\documentclass{aa}  

\newcommand{\todo}{\ifmmode \text{\color{red}\Huge{\(\bullet\)}} \else {\color{red}{\Huge$\bullet$}}\fi}
\newcommand{\tido}{\ifmmode {{\color{red}\bullet}} \else {\color{red}$\bullet$}\fi}

\newcommand{\E        }[1]{\ifmmode 10^{#1} \else $10^{#1}$\fi}
\newcommand{\tE        }[1]{\ifmmode \times10^{#1} \else $\times10^{#1}$\fi}
\newcommand{\pc}	{\ifmmode {\rm pc} \else pc\fi}
\newcommand{\kpc}	{\ifmmode {\rm kpc} \else kpc\fi}
\newcommand{\ld}	{\ifmmode {\rm l.d.} \else l.d.\fi}
\newcommand{\kms}	{\ifmmode {\rm km\,s}^{-1} \else km\,s$^{-1}$\fi}
\newcommand{\cc}	{\ifmmode {\rm cm}^{-3}    \else cm$^{-3}$\fi}
\newcommand{\cmii}	{\ifmmode {\rm cm}^{-2}    \else cm$^{-2}$\fi}
\newcommand{\ergss}	{\ifmmode {\rm erg\,s}^{-1} \else erg s$^{-1}$\fi}
\newcommand{\ergcms}	{\ifmmode {\rm erg\,cm}^{-2}\,{\rm s}^{-1} \else erg\,cm$^{-2}$\,s$^{-1}$\fi}
\newcommand{\ergcmsA}	{\ifmmode {\rm erg\,cm}^{-2}\,{\rm s}^{-1}\,{\rm \AA}^{-1}
\else erg\,cm$^{-2}$\,s$^{-1}$\,\AA$^{-1}$\fi}
\newcommand{  \ergcmsHz  }{\ifmmode{\rm erg\,cm}^{-2}\,{\rm s}^{-1}\,{\rm Hz}^{-1}
                       \else ergs\,cm$^{-2}$\,s$^{-1}$\,Hz$^{-1}$\fi}
\newcommand{\kev}	{\ifmmode {\rm keV} \else keV\fi}

\newcommand{\mic}	{\ifmmode {\rm \mu m} \else $\mu$m\fi}
\newcommand{\vFWHM}	{\ifmmode v_{\mbox{\tiny FWHM}} \else $v_{\mbox{\tiny FWHM}}$\fi}
\newcommand{\vBLR}	{\ifmmode v_{\mbox{\tiny BLR}} \else $v_{\mbox{\tiny BLR}}$\fi}
\newcommand{\sigBLR}	{\ifmmode \sigma_{\mbox{\tiny BLR}} \else $\sigma_{\mbox{\tiny BLR}}$\fi}
\newcommand{\vNLR}	{\ifmmode v_{\mbox{\tiny NLR}} \else $v_{\mbox{\tiny NLR}}$\fi}
\newcommand{\tauBLR}	{\ifmmode \tau_{\mbox{\tiny BLR}} \else $\tau_{\mbox{\tiny BLR}}$\fi}

\newcommand{\Hubble}	{\ifmmode {\rm km\,s}^{-1}\,{\rm Mpc}^{-1} \else km\,s$^{-1}$\,Mpc$^{-1}$\fi}
\newcommand{\NDunit}	{\ifmmode {\rm Mpc}^{-3} \else Mpc$^{-3}$\fi}
\newcommand{\LFunit}	{\ifmmode {\rm Mpc}^{-3}\,{\rm mag}^{-1} \else Mpc$^{-3}$\,mag$^{-1}$\fi}
\newcommand{\MFunit}	{\ifmmode {\rm Mpc}^{-3}\,{\rm dex}^{-1} \else Mpc$^{-3}$\,dex$^{-1}$\fi}

\newcommand{\Msun}{\ifmmode M_{\odot} \else $M_{\odot}$\fi}
\newcommand{\Lsun}{\ifmmode L_{\odot} \else $L_{\odot}$\fi}
\newcommand{\Zsun}{\ifmmode Z_{\odot} \else $Z_{\odot}$\fi}
\newcommand{\mpyr}{\ifmmode \Msun\,{\rm yr}^{-1} \else $\Msun\,{\rm yr}^{-1}$\fi}

\newcommand{\qnote}{\ifmmode q_{0} \else $q_{0}$\fi}
\newcommand{\Hnote}{\ifmmode H_{0} \else $H_{0}$\fi}
\newcommand{\hnote}{\ifmmode h_{0} \else $h_{0}$\fi}
\newcommand{\anote}{\ifmmode a_{0} \else $a_{0}$\fi}
\newcommand{\tnote}{\ifmmode t_{0} \else $t_{0}$\fi}

\def\gsim{\;\rlap{\lower 2.5pt \hbox{$\sim$}}\raise 1.5pt\hbox{$>$}\;}
\def\lsim{\;\rlap{\lower 2.5pt \hbox{$\sim$}}\raise 1.5pt\hbox{$<$}\;}

\newcommand{  \Halpha   }{\ifmmode {\rm H}\alpha \else H$\alpha$\fi}

\newcommand{  \ha       }{\Halpha}
\newcommand{  \Hbeta    }{\ifmmode {\rm H}\beta \else H$\beta$\fi}

\newcommand{  \hb       }{\Hbeta}
\newcommand{  \Hgamma   }{\ifmmode {\rm H}\gamma \else H$\gamma$\fi}
\newcommand{  \Hdelta   }{\ifmmode {\rm H}\delta \else H$\delta$\fi}
\newcommand{  \Lya      }{\ifmmode {\rm Ly}\alpha \else Ly$\alpha$\fi}
\newcommand{  \Lyb      }{\ifmmode {\rm Ly}\beta \else Ly$\beta$\fi}
\newcommand{  \Pa       }{\ifmmode {\rm P}\alpha \else P$\alpha$\fi}
\newcommand{  \Pb       }{\ifmmode {\rm P}\beta \else P$\beta$\fi}
\newcommand{  \Bra      }{\ifmmode {\rm Br}\alpha \else Br$\alpha$\fi}
\newcommand{  \Brg      }{\ifmmode {\rm Br}\gamma \else Br$\gamma$\fi}
\newcommand{  \hii      }{\ifmmode {\rm H}\,\textsc{ii} \else H\,\textsc{ii}\fi}
\newcommand{  \hei      }{\ifmmode {\rm He}\,\textsc{i} \else He\,\textsc{i}\fi}
\newcommand{  \heii     }{\ifmmode {\rm He}\,\textsc{ii} \else He\,\textsc{ii}\fi}
\newcommand{  \HeIIuv   }{\ifmmode {\rm He}\,\textsc{ii}\,\lambda1640 \else He\,\textsc{ii}\,$\lambda1640$\fi}
\newcommand{  \HeIIop   }{\ifmmode {\rm He}\,\textsc{ii}\,\lambda4686 \else He\,\textsc{ii}\,$\lambda4686$\fi}
\newcommand{  \CII	}{\ifmmode \left[{\rm C}\,\textsc{ii}\right]\,\lambda157.74\,\mu{\rm m} \else [C\,{\sc ii}]\ $\lambda157.74\,\mu{\rm m}$\fi}
\newcommand{  \cii	}{\ifmmode \left[{\rm C}\,\textsc{ii}\right] \else [C\,{\sc ii}]\fi}

\newcommand{  \ciii     }{\ifmmode {\rm C}\,\textsc{iii}\right] \else C\,\textsc{iii}]\fi}
\newcommand{  \CIII     }{\ifmmode {\rm C}\,\textsc{iii}\right]\,\lambda1909 \else C\,\textsc{iii}]\,$\lambda1909$\fi}
\newcommand{  \civ      }{\ifmmode {\rm C}\,\textsc{iv}  \else C\,\textsc{iv}\fi}
\newcommand{  \CIV      }{\ifmmode {\rm C}\,\textsc{iv}\,\lambda1549 \else C\,\textsc{iv}\,$\lambda1549$\fi}

\newcommand{\COIItoI}{\ifmmode ^{12}{\rm CO}\,\mbox{(2--1)} \else  $^{12}{\rm CO}$\,\mbox{(2--1)}}
\newcommand{  \NIIopt   }{\ifmmode \left[{\rm N}\,\textsc{ii}\right]\,\lambda6584 \else [N\,\textsc{ii}]\,$\lambda6584$\fi}
\newcommand{  \nii      }{\ifmmode \left[{\rm N}\,\textsc{ii}\right]  \else [N\,\textsc{ii}]\fi}
\newcommand{  \niii     }{\ifmmode {\rm N}\,\textsc{iii} \else N\,\textsc{iii}\fi}
\newcommand{  \NIII     }{\ifmmode {\rm N}\,\textsc{iii}\,\lambda4640 \else N\,\textsc{iii}\,$\lambda4640$\fi}
\newcommand{  \niv      }{\ifmmode {\rm N}\,\textsc{iv}  \else N\,\textsc{iv}\fi}
\newcommand{  \NIVuv    }{\ifmmode {\rm N}\,\textsc{iv}\,\lambda1486 \else N\,\textsc{iv}\,$\lambda1486$\fi}
\newcommand{  \nv       }{\ifmmode {\rm N}\,\textsc{v}   \else N\,\textsc{v}\fi}
\newcommand{\oi}{\ifmmode \left[{\rm O}\,\textsc{i}\right] \else [O\,{\sc i}]\fi}
\newcommand{\OI}{\ifmmode \left[{\rm O}\,\textsc{i}\right]\,\lambda6300 \else [O\,{\sc i}]$\,\lambda6300$\fi}
\newcommand{\oii}{\ifmmode \left[{\rm O}\,\textsc{ii}\right] \else [O\,{\sc ii}]\fi}
\newcommand{\OII}{\ifmmode \left[{\rm O}\,\textsc{ii}\right]\,\lambda3727 \else [O\,{\sc ii}]\,$\lambda3727$\fi}
\newcommand{\oiii}{\ifmmode \left[{\rm O}\,\textsc{iii}\right] \else [O\,{\sc iii}]\fi}
\newcommand{\OIII}{\ifmmode \left[{\rm O}\,\textsc{iii}\right]\,\lambda5007 \else [O\,{\sc iii}]\,$\lambda5007$\fi}
\newcommand{  \OIIIbf   }{\ifmmode {\rm O}\,\textsc{iii}\,\lambda3133 \else O\,\textsc{iii}\,$\lambda3133$\fi}
\newcommand{  \OIIIuv   }{\ifmmode {\rm O}\,\textsc{iii}\,\lambda1663 \else O\,\textsc{iii}\,$\lambda1663$\fi}
\newcommand{  \oiv      }{\ifmmode {\rm O}\,\textsc{iv}  \else O\,\textsc{iv}\fi}
\newcommand{  \OIVuv    }{\ifmmode {\rm O}\,\textsc{iv}\,\lambda1402  \else O\,\textsc{iv}\,$\lambda1402$\fi}
\newcommand{  \OIVIR    }{\ifmmode {\rm O}\,\textsc{iv}\,25.9\,\mu {\rm m} \else O\,\textsc{iv}\,$25.9\,\mu$m\fi}
\newcommand{  \ovi      }{\ifmmode {\rm O}\,\textsc{vi}   \else O\,\textsc{vi}\fi}
\newcommand{  \Ovi      }{\ifmmode {\rm O}\,\textsc{vi}\,\lambda1035 \else O\,\textsc{vi}\,$\lambda1035$\fi}
\newcommand{  \nei      }{\ifmmode {\rm Ne}\,\textsc{i}   \else Ne\,\textsc{i}\fi}
\newcommand{  \neii     }{\ifmmode {\rm Ne}\,\textsc{ii}  \else Ne\,\textsc{ii}\fi}
\newcommand{  \NeiiIR   }{\ifmmode {\rm Ne}\,\textsc{ii}\,12.8\,\mu {\rm m} \else Ne\,\textsc{ii}\,$12.8\,\mu$m\fi}
\newcommand{  \neiii    }{\ifmmode {\rm Ne}\,\textsc{iii} \else Ne\,\textsc{iii}\fi}
\newcommand{  \neiv     }{\ifmmode {\rm Ne}\,\textsc{iv}  \else Ne\,\textsc{iv}\fi}
\newcommand{  \nev      }{\ifmmode {\rm Ne}\,\textsc{v}   \else Ne\,\textsc{v}\fi}
\newcommand{  \NevIR    }{\ifmmode {\rm Ne}\,\textsc{v}\,24.3\,\mu {\rm m} \else Ne\,\textsc{v}\,$24.3\,\mu$m\fi}
\newcommand{  \nevi     }{\ifmmode {\rm Ne}\,\textsc{vi}  \else Ne\,\textsc{vi}\fi}
\newcommand{  \mgi      }{\ifmmode {\rm Mg}\,\textsc{i} \else Mg\,\textsc{i}\fi}
\newcommand{  \mgii     }{\ifmmode {\rm Mg}\,\textsc{ii} \else Mg\,\textsc{ii}\fi}
\newcommand{  \MgII     }{\ifmmode {\rm Mg}\,\textsc{ii}\,\lambda2798 \else Mg\,\textsc{ii}\,$\lambda2798$\fi}
\newcommand{  \sii      }{\ifmmode {\rm S}\,\textsc{ii} \else S\,\textsc{ii}\fi}
\newcommand{  \siii     }{\ifmmode {\rm S}\,\textsc{iii} \else S\,\textsc{iii}\fi}
\newcommand{  \siv      }{\ifmmode {\rm S}\,\textsc{iv} \else S\,\textsc{iv}\fi}
\newcommand{  \sili     }{\ifmmode {\rm Si}\,\textsc{i}   \else Si\,\textsc{i}\fi}
\newcommand{  \silii    }{\ifmmode {\rm Si}\,\textsc{ii}  \else Si\,\textsc{ii}\fi}
\newcommand{  \Siliv    }{\ifmmode {\rm Si}\,\textsc{iv}  \else Si\,\textsc{iv}\fi}
\newcommand{  \SilIVuv  }{\ifmmode {\rm Si}\,\textsc{iv}\,\lambda1400  \else Si\,\textsc{iv}\,$\lambda1400$\fi}
\newcommand{  \AlIII   }{\ifmmode {\rm Al}\,\textsc{iii}\,\lambda1857 \else Al\,\textsc{iii}\,$\lambda1857$\fi}
\newcommand{  \Aliii   }{\ifmmode {\rm Al}\,\textsc{iii} \else Al\,\textsc{iii}\fi}
\newcommand{  \caii     }{\ifmmode {\rm Ca}\,\textsc{ii} \else Ca\,\textsc{ii}\fi}
\newcommand{  \feii     }{\ifmmode {\rm Fe}\,\textsc{ii} \else Fe\,\textsc{ii}\fi}
\newcommand{  \feiii    }{\ifmmode {\rm Fe}\,\textsc{iii} \else Fe\,\textsc{iii}\fi}
\newcommand{  \Kalpha   }{\ifmmode {\rm K}\alpha \else K$\alpha$\fi}

\newcommand{ \Lhb   }{\ifmmode L_{\hb} \else $L_{\hb}$\fi}
\newcommand{ \Lha   }{\ifmmode L_{\ha} \else $L_{\ha}$\fi}
\newcommand{ \fwhb  }{\ifmmode {\rm FWHM}\left(\hb\right) \else FWHM(\hb)\fi}
\newcommand{\sighb  }{\ifmmode \sigma\left(\hb\right) \else $\sigma\left(\hb\right)$\fi}
\newcommand{ \ewhb  }{\ifmmode {\rm EW}\left(\hb\right) \else EW(\hb)\fi}
\newcommand{ \fwha  }{\ifmmode {\rm FWHM}\left(\ha\right) \else FWHM(\ha)\fi}
\newcommand{ \ewha  }{\ifmmode {\rm EW}\left(\ha\right) \else EW(\ha)\fi}
\newcommand{ \Lmg   }{\ifmmode L\left(\mgii\right) \else $L\left(\mgii\right)$\fi}
\newcommand{ \fwmg  }{\ifmmode {\rm FWHM}\left(\mgii\right) \else FWHM(\mgii)\fi}
\newcommand{ \Lciv  }{\ifmmode L\left(\civ\right) \else $L\left(\civ\right)$\fi}
\newcommand{ \fwciv }{\ifmmode {\rm FWHM}\left(\civ\right) \else FWHM(\civ)\fi}
\newcommand{ \fwhm  }{\ifmmode {\rm FWHM} \else FWHM\fi} 
\newcommand{ \voff  }{\ifmmode v_{\rm off} \else $v_{\rm off}$\fi} 
\newcommand{ \vmax  }{\ifmmode v_{\rm max} \else $v_{\rm max}$\fi} 

\newcommand{ \mumg  }{\ifmmode \mu\left(\mgii\right) \else $\mu\left(\mgii\right)$\fi}
\newcommand{ \fmg   }{\ifmmode f\left(\mgii\right) \else $f\left(\mgii\right)$\fi}
\newcommand{ \muciv }{\ifmmode \mu\left(\civ\right) \else $\mu\left(\civ\right)$\fi}
\newcommand{ \fciv  }{\ifmmode f\left(\civ\right) \else $f\left(\civ\right)$\fi}
\newcommand{  \auvo     }{\ifmmode \alpha_{\nu,{\rm UVO}} \else $\alpha_{\nu,{\rm UVO}}$\fi}
\newcommand{  \Ledd     }{\ifmmode L_{\rm Edd} \else $L_{\rm Edd}$\fi}
\newcommand{  \lamLlam  }{\ifmmode \lambda L_{\lambda} \else $\lambda L_{\lambda}$\fi}
\newcommand{  \lLl      }{\ifmmode \lambda L_{\lambda} \else $\lambda L_{\lambda}$\fi}
\newcommand{  \nuLnu    }{\ifmmode \nu L_{\nu} \else $\nu L_{\nu}$\fi}
\newcommand{  \nLn      }{\ifmmode \nu L_{\nu} \else $\nu L_{\nu}$\fi}
\newcommand{  \Luv      }{\ifmmode L_{1450} \else $L_{1450}$\fi}
\newcommand{  \Lop      }{\ifmmode L_{5100} \else $L_{5100}$\fi}
\newcommand{  \lLop     }{\ifmmode \log\left(\Lop/\ergs\right) \else $\log\left(\Lop/\ergs\right)$\fi}
\newcommand{  \Lthree   }{\ifmmode L_{3000} \else $L_{3000}$\fi}
\newcommand{  \lLthree  }{\ifmmode \log\left(\Lthree/\ergs\right) \else $\log\left(\Lthree/\ergs\right)$\fi}
\newcommand{  \Lsix      }{\ifmmode L_{6200} \else $L_{6200}$\fi}
\newcommand{  \lLisx     }{\ifmmode \log\left(\Lop/\ergs\right) \else $\log\left(\Lop/\ergs\right)$\fi}
\newcommand{  \Lxray    }{\ifmmode L_{\rm X} \else $L_{\rm X}$\fi}

\newcommand{  \Lhard    }{\ifmmode L_{\rm 2-10} \else $L_{\rm 2-10}$\fi}
\newcommand{  \Lsoft    }{\ifmmode L_{\rm 0.5-2} \else $L_{\rm 0.5-2}$\fi}

\newcommand{\Fthree}{\ifmmode F_{3000} \else $F_{3000}$\fi}
\newcommand{\fuv}{\ifmmode f_{\lambda}\left(1450{\rm \AA}\right) \else $f_{\lambda}\left(1450 {\rm \AA}\right)$\fi}
\newcommand{\fthree}{\ifmmode f_{\lambda}\left(3000{\rm \AA}\right) \else $f_{\lambda}\left(3000{\rm \AA}\right)$\fi}
\newcommand{\fH}{\ifmmode f_{\lambda}\left(1.65\micron\right) \else
$f_{\lambda}\left(1.65\micron\right)$\fi}

\newcommand{\fbol}{\ifmmode f_{\rm bol} \else $f_{\rm bol}$\fi}
\newcommand{\fbolwv}{\ifmmode f_{\rm bol}\left(\lambda\right) \else $f_{\rm bol}\left(\lambda\right)$\fi}
\newcommand{\fbolopt}{\ifmmode f_{\rm bol}\left(5100{\rm \AA}\right) \else $f_{\rm bol}\left(5100{\rm \AA}\right)$\fi}
\newcommand{\fbolthree}{\ifmmode f_{\rm bol}\left(3000{\rm \AA}\right) \else $f_{\rm bol}\left(3000{\rm \AA}\right)$\fi}
\newcommand{\fboluv}{\ifmmode f_{\rm bol}\left(1450{\rm \AA}\right) \else $f_{\rm bol}\left(1450{\rm \AA}\right)$\fi}

\newcommand{\fbolbat}{\ifmmode f_{\rm bol}\left(14-150\,\kev\right) \else $f_{\rm bol}\left(14-150\,\kev\right)$\fi}
\newcommand{\fbolhard}{\ifmmode f_{\rm bol}\left(2-10\,\kev\right) \else $f_{\rm bol}\left(2-10\,\kev\right)$\fi}

\newcommand{\fobs}{\ifmmode f_{\rm obs} \else $f_{\rm obs}$\fi}

\newcommand{  \mbh      }{\ifmmode M_{\rm BH} \else $M_{\rm BH}$\fi}
\newcommand{  \lmbh     }{\ifmmode \log\left(\mbh/\Msun\right) \else $\log\left(\mbh/\Msun\right)$\fi} 
\newcommand{  \lledd    }{\ifmmode L/L_{\rm Edd} \else $L/L_{\rm Edd}$\fi}
\newcommand{  \mmedd    }{\ifmmode \dot{m}/\dot{m}_{\rm \,Edd} \else $\dot{m}/\dot{m}_{\rm \,Edd}$\fi}
\newcommand{  \Lbol     }{\ifmmode L_{\rm bol} \else $L_{\rm bol}$\fi}
\newcommand{  \lbol     }{\ifmmode L_{\rm bol} \else $L_{\rm bol}$\fi}
\newcommand{  \lLbol    }{\ifmmode \log\left(\Lbol/\ergs\right) \else $\log\left(\Lbol/\ergs\right)$\fi} 
\newcommand{  \Lagn     }{\ifmmode L_{\rm AGN} \else $L_{\rm AGN}$\fi}
\newcommand{  \lagn     }{\ifmmode L_{\rm AGN} \else $L_{\rm AGN}$\fi}

\newcommand{  \tgrow     }{\ifmmode t_{\rm growth} \else $t_{\rm growth}$\fi}
\newcommand{  \tAD     }{\ifmmode t_{\rm acc} \else $t_{\rm acc}$\fi}
\newcommand{  \tacc    }{\ifmmode t_{\rm acc} \else $t_{\rm acc}$\fi}
\newcommand{  \tUni      }{\ifmmode t_{\rm Universe} \else $t_{\rm Universe}$\fi}

\newcommand{  \Mdotin	}{\ifmmode \dot{M}_{\rm infall} \else $\dot{M}_{\rm infall}$\fi}
\newcommand{  \Mdotbh	}{\ifmmode \dot{M}_{\rm BH} \else $\dot{M}_{\rm BH}$\fi}
\newcommand{  \Mdotad	}{\ifmmode \dot{M}_{\rm AD} \else $\dot{M}_{\rm AD}$\fi}
\newcommand{  \Mdotacc	}{\ifmmode \dot{M}_{\rm acc} \else $\dot{M}_{\rm acc}$\fi}
\newcommand{  \Mdotthin	}{\ifmmode \dot{M}_{\rm thin} \else $\dot{M}_{\rm thin}$\fi}
\newcommand{  \Mdotdisk	}{\ifmmode \dot{M}_{\rm disk} \else $\dot{M}_{\rm disk}$\fi}

\newcommand{  \Mindot	}{\ifmmode \dot{M}_{\rm infall} \else $\dot{M}_{\rm infall}$\fi}
\newcommand{  \Mbhdot	}{\ifmmode \dot{M}_{\rm BH} \else $\dot{M}_{\rm BH}$\fi}
\newcommand{  \Maddot	}{\ifmmode \dot{M}_{\rm AD} \else $\dot{M}_{\rm AD}$\fi}
\newcommand{  \Maccdot	}{\ifmmode \dot{M}_{\rm acc} \else $\dot{M}_{\rm acc}$\fi}
\newcommand{  \Mthdot	}{\ifmmode \dot{M}_{\rm thin} \else $\dot{M}_{\rm thin}$\fi}
\newcommand{  \Mdsdot	}{\ifmmode \dot{M}_{\rm disk} \else $\dot{M}_{\rm disk}$\fi}

\newcommand{  \as	}{\ifmmode a_{\rm *} \else $a_{\rm *}$\fi}
\newcommand{  \avec	}{\ifmmode \vec{a}_{\rm *} \else $\vec{a}_{\rm *}$\fi}
\newcommand{  \re	}{\ifmmode \eta      	 \else $\eta$\fi}
\newcommand{  \RISCO	}{\ifmmode R_{\rm ISCO}  \else $R_{\rm ISCO}$\fi}

\newcommand{  \mseed    }{\ifmmode M_{\rm seed} \else $M_{\rm seed}$\fi}
\newcommand{  \mbul     }{\ifmmode M_{\rm bulge} \else $M_{\rm bulge}$\fi} 
\newcommand{  \mstar    }{\ifmmode M_{*} \else $M_{*}$\fi} 
\newcommand{  \mgal     }{\ifmmode M_{*} \else $M_{*}$\fi} 
\newcommand{  \mhost    }{\ifmmode M_{\rm host} \else $M_{\rm host}$\fi}
\newcommand{  \mmsmall  }{\ifmmode M_{\rm BH}/M_{*} \else $M_{\rm BH}/M_{*}$\fi}
\newcommand{  \mmlarge  }{\ifmmode M_{*}/M_{\rm BH} \else $M_{*}/M_{\rm BH}$\fi}

\newcommand{  \mmdotlarge}{\ifmmode \dot{M}_*/\Mbhdot \else $\dot{M}_*/\Mbhdot$\fi}
\newcommand{  \mmdotsmall}{\ifmmode \Mbhdot/\dot{M}_* \else $\Mbhdot/\dot{M}_*$\fi}

\newcommand{  \mmwp     }{\ifmmode \left(M_{*}/M_{\rm BH}\right) \else $\left(M_{*}/M_{\rm BH}\right)$\fi}
\newcommand{  \ml       }{\ifmmode M_{*}/L_{*} \else $M_{*}/L_{*}$\fi}
\newcommand{  \mlwp     }{\ifmmode \left(M_{*}/L\right) \else $\left(M_{*}/L\right)$\fi}
\newcommand{  \mlk      }{\ifmmode \left(M_{*}/L_{K}\right) \else $\left(M_{*}/L_{K}\right)$\fi}
\newcommand{  \sigs     }{\ifmmode \sigma_{*} \else $\sigma_{*}$\fi}
\newcommand{  \Reff     }{\ifmmode R_{\rm e} \else $R_{\rm e}$\fi}
\newcommand{  \Rvir     }{\ifmmode R_{\rm vir} \else $R_{\rm vir}$\fi}
\newcommand{  \Rtwo     }{\ifmmode R_{200} \else $R_{200}$\fi}
\newcommand{  \Rfive    }{\ifmmode R_{500} \else $R_{500}$\fi}
\newcommand{  \Rgrp     }{\ifmmode R_{\rm grp} \else $R_{\rm grp}$\fi}
\newcommand{  \nser     }{\ifmmode n_{\rm s} \else $n_{\rm s}$\fi}
\newcommand{  \LSF      }{\ifmmode L_{\rm SF}  \else $L_{\rm SF}$\fi}
\newcommand{  \LFIR     }{\ifmmode L_{\rm FIR} \else $L_{\rm FIR}$\fi}
\newcommand{  \Lfir     }{\ifmmode L_{\rm FIR} \else $L_{\rm FIR}$\fi}
\newcommand{  \LTIR     }{\ifmmode L_{\rm TIR} \else $L_{\rm TIR}$\fi}
\newcommand{  \Ltir     }{\ifmmode L_{\rm TIR} \else $L_{\rm TIR}$\fi}

\newcommand{  \mdyn     }{\ifmmode M_{\rm dyn} \else $M_{\rm dyn}$\fi} 
\newcommand{  \mgas     }{\ifmmode M_{\rm gas} \else $M_{\rm gas}$\fi} 
\newcommand{  \mh       }{\ifmmode M_{\rm h} \else $M_{\rm h}$\fi}
\newcommand{  \mhalo    }{\ifmmode M_{\rm halo} \else $M_{\rm halo}$\fi}
\newcommand{  \sfr      }{\ifmmode {\rm SFR} \else SFR\fi}

\newcommand{ \Lcii     }{\ifmmode L_{\cii} \else $L_{\cii}$\fi}
\newcommand{ \fwcii  }{\ifmmode {\rm FWHM}\cii \else FWHM\cii\fi}

\newcommand{\bj}{\ifmmode b_{\rm J} \else $b_{\rm J}$\fi}

\newcommand{\iab}{\ifmmode i_{\rm AB} \else $i_{\rm AB}$\fi}

\newcommand{\jab}{\ifmmode J_{\rm AB} \else $J_{\rm AB}$\fi}
\newcommand{\hab}{\ifmmode H_{\rm AB} \else $H_{\rm AB}$\fi}
\newcommand{\kab}{\ifmmode K_{\rm AB} \else $K_{\rm AB}$\fi}

\newcommand{\jveg}{\ifmmode J_{\rm Vega} \else $J_{\rm Vega}$\fi}
\newcommand{\hveg}{\ifmmode H_{\rm Vega} \else $H_{\rm Vega}$\fi}
\newcommand{\kveg}{\ifmmode K_{\rm Vega} \else $K_{\rm Vega}$\fi}

\def\arcmin{\hbox{$^\prime$}}
\def\arcsec{\hbox{$^{\prime\prime}$}}

\newcommand{  \Chisq    }{\ifmmode \chi^{2} \else $\chi^{2}$}
\newcommand{  \nelec    }{\ifmmode n_{e} \else $n_{e}$\fi}     % electron density
\newcommand{  \nh       }{\ifmmode n_{\rm H} \else $n_{\rm H}$\fi}     % hydrogen density
\newcommand{  \Ncol     }{\ifmmode N_{\rm col} \else $N_{\rm col}$\fi} % column density
\newcommand{  \NH       }{\ifmmode N_{\rm H} \else $N_{\rm H}$\fi}     % column density

\def\arcmin{\hbox{$^\prime$}}
\def\arcsec{\hbox{$^{\prime\prime}$}}
\def\ion#1#2{#1$\;${\small\rm\@Roman{#2}}\relax}

\newcommand{\SiX}{\ifmmode \left[{\rm Si}\,\textsc{x}\right]\,\lambda14300 \else [Si\,{\sc x}]\,$\lambda14300$\fi}
\newcommand{\SiVI}{\ifmmode \left[{\rm Si}\,\textsc{vi}\right]\,\lambda19640 \else [Si\,{\sc vi}]\,$\lambda19640$\fi}
\newcommand{\SXI}{\ifmmode \left[{\rm S}\,\textsc{xi}\right]\,\lambda19196 \else [S\,{\sc xi}]\,$\lambda19196$\fi}
\newcommand{\SVIII}{\ifmmode \left[{\rm S}\,\textsc{viii}\right]\,\lambda9915 \else [S\,{\sc viii}]\,$\lambda9915$\fi}
\newcommand{\SIX}{\ifmmode \left[{\rm S}\,\textsc{ix}\right]\,\lambda12520 \else [S\,{\sc ix}]\,$\lambda12520$\fi}
\newcommand{\FeXIII}{\ifmmode \left[{\rm Fe}\,\textsc{xiii}\right]\,\lambda10747 \else [Fe\,{\sc xiii}]\,$\lambda10747$\fi}

\newcommand{  \hi       }{\ifmmode {\rm H}\,\textsc{i} \else H\,\textsc{i}\fi}

\newcommand{  \oxab       }{\ifmmode 12+\log_{10}{\rm (O/H)} \else $12+\log_{10}{\rm (O/H)}$ \fi}

\newcommand{  \aco     }{\ifmmode \alpha_{\rm ^{12}CO(1-0)} \else $\alpha_{\rm ^{12}CO(1-0)}$\fi}
\newcommand{  \acoSpec     }{\ifmmode \alpha_{\rm ^{12}CO(1-0)} \else $\alpha_{\rm ^{12}CO(1-0)}$\fi}

\newcommand{\acoRes}{\ifmmode \langle\aco\rangle=4.4{\pm}0.9 \else $\langle\aco\rangle=4.4{\pm}0.9 $\fi}

\newcommand{\acoResSpec}{\ifmmode \langle\acoSpec\rangle=4.4{\pm}0.9 \else $\langle\acoSpec\rangle=4.4{\pm}0.9 $\fi}
\usepackage{graphicx}	% Including figure files
\usepackage{amsmath}	% Advanced maths commands
\usepackage{amssymb}	% Extra maths symbols
\usepackage{bm}	        % Bold math symbols
\usepackage{csvsimple}
\usepackage{ragged2e}
\usepackage{multirow}
\usepackage{siunitx} % for \num{} notation
\usepackage[utf8]{inputenc}
\usepackage{dirtytalk}
\usepackage{xparse}
\usepackage{enumitem} 
\usepackage{array}
\usepackage{CJKutf8}
\usepackage[flushleft]{threeparttable}
\usepackage{txfonts}
\newcommand{\orcid}[1]{\href{https://orcid.org/#1}{\includegraphics[width=10pt]{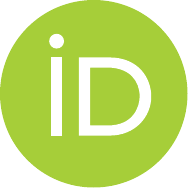}}}

\usepackage[pdfpagelabels=false]{hyperref}	% Hyperlinks
\hypersetup{colorlinks=true,linkcolor=blue,citecolor=blue,filecolor=blue,urlcolor=blue,}

\usepackage[nameinlink,capitalize]{cleveref}
\begin{document}

\renewcommand{\figureautorefname}{Fig.} % Fig. 1 but Figs 1 and 2
\renewcommand{\equationautorefname}{Eq.} % or equation (1) ?
\renewcommand{\sectionautorefname}{Section} % or Sec. or Sect. ?
\renewcommand{\subsectionautorefname}{Section}
\renewcommand{\subsubsectionautorefname}{Section}
\renewcommand{\appendixautorefname}{Appendix}

% workaround: \hyperref[sec:errcontr]{Appendix~\ref{sec:errcontr}}

% Homogeneous typesetting of molecular species and their transitions
% + Works in text and math mode (always in upright font shape).
% + Works in normal and bold font series.
%-- Examples:
% + molecular species: \chem{^{12}CO} or \species{^{12}CO} --> $\mathrm{^{12}CO}$
% + molecular species and transition: \chem{^{12}CO}{10} --> $\mathrm{^{12}CO}\,(1{-}0)$
% + transition: \trans{10} --> $(1{-}0)$

%-- Required packages
% \usepackage{amsmath}
% \usepackage{xparse}
% \usepackage{bm}

%-- Command for species

\newcommand{  \Hi      }{\ifmmode {\rm H}\,\textsc{i} \else H\,\textsc{i}\fi}

\newcommand*{\species}[1]{%
    \ensuremath{\IfBold{\bm{\mathrm{#1}}}{\mathrm{#1}}}%
}

%-- Command for transition

\newcommand*{\trans}[1]{%
    \ensuremath{\IfBold{\bm{\mytrans{#1}}}{\mytrans{#1}}}%
}

\newcommand{\delim}{{-}}
\newcommand*{\mytrans}[1]{%
    \ifnum#1=10 (1\delim0)\else%
    \ifnum#1=21 (2\delim1)\else%
    \ifnum#1=32 (3\delim2)\else%
    \ifnum#1=43 (4\delim3)\else%
    \ifnum#1=54 (5\delim4)\else%
    \ifnum#1=65 (6\delim5)\else%
    \ifnum#1=76 (7\delim6)\else%
    \ifnum#1=87 (8\delim7)\else%
    \ifnum#1=98 (9\delim8)\else%
    \ifnum#1=109 (10\delim9)\else#1%
    \fi\fi\fi\fi\fi\fi\fi\fi\fi\fi%
}

%-- IfBold Command
\makeatletter
\newcommand*{\IfBold}{%
  \ifx\f@series\my@test@bx
    \expandafter\@firstoftwo
  \else
    \expandafter\@secondoftwo
  \fi
}
\newcommand*{\my@test@bx}{bx}
\makeatother

%-- Command for chemical species and transition: \chem{^{12}CO}{10} or \chem{^{12}CO}
% Using xparse package
\DeclareDocumentCommand{\chem}{ m g }{%
    {\species{#1}%
        \IfNoValueF {#2} {\,\trans{#2}}%
    }%
}

%-----------------------------------------------------------------

   \title{Wide-field CO isotopologue emission and the CO-to-H$_2$ factor across the nearby spiral galaxy M101}
   %\subtitle{}

   \author{Jakob S. den Brok \inst{1,2}\fnmsep\thanks{\email{jakob.denbrok@gmail.com}} \orcid{0000-0002-8760-6157}
          \and
    Frank Bigiel\inst{1}\orcid{0000-0003-0166-9745}
        \and 
    Jérémy Chastenet\inst{3}\orcid{0000-0002-5235-5589}
        \and 
    Karin Sandstrom\inst{4}\orcid{0000-0002-5235-5589}
        \and 
    Adam Leroy\inst{5}\orcid{0000-0002-2545-1700}
        \and 
    Antonio Usero\inst{6}\orcid{0000-0003-1242-505X}
        \and 
    Eva Schinnerer\inst{7}\orcid{0000-0002-3933-7677}
        \and 
    Erik W. Rosolowsky\inst{8}\orcid{0000-0002-5204-2259}
        \and 
    Eric W. Koch\inst{2}\orcid{0000-0003-2551-7148}
      \and 
     I-Da Chiang (\begin{CJK*}{UTF8}{bkai}江宜達 \end{CJK*})
     \inst{9}\orcid{0000-0003-2551-7148}
        \and
    Ashley T. Barnes\inst{1,10}\orcid{0000-0003-0410-4504}
        \and 
    Johannes Puschnig\inst{1}\orcid{0000-0003-1111-3951}
        \and 
    Toshiki Saito\inst{11}\orcid{0000-0002-2501-9328}
        \and 
    Ivana~Be\v{s}li\'c\inst{1}\orcid{0000-0003-0783-0157}
        \and 
    Melanie Chevance\inst{12, 13}\orcid{0000-0002-5635-5180}
        \and 
    Daniel~A.~Dale\inst{14}\orcid{0000-0002-5782-9093}
        \and 
    Cosima Eibensteiner\inst{1}\orcid{0000-0002-1185-2810}
        \and 
    Simon Glover\inst{12}\orcid{0000-0001-6708-1317}
        \and 
    Mar\'ia~J.~Jim\'enez-Donaire\inst{6}\orcid{0000-0002-9165-8080}
        \and 
    Yu-Hsuan Teng\inst{4}\orcid{0000-0003-4209-1599}
        \and
    Thomas G. Williams\inst{15} \orcid{0000-0002-0012-2142}
    }

   \institute{
    Argelander-Institut für Astronomie, Universität Bonn, Auf dem Hügel 71, 53121 Bonn, Germany
    \and 
    Center for Astrophysics $\mid$ Harvard \& Smithsonian, 60 Garden St., 02138 Cambridge, MA, USA
    \and 
    Sterrenkundig Observatorium, Universiteit Gent, Krijgslaan 281 S9, B-9000 Gent, Belgium
    \and
    Center for Astrophysics \& Space Sciences, Department of Physics, University of California San Diego, 9500 Gilman Drive, La Jolla,
CA 92093, USA
    \and
    Department of Astronomy, The Ohio State University, 140 West 18th Ave, Columbus, OH 43210, USA
    \and
    Observatorio Astron\'omico Nacional (IGN), C/ Alfonso XII, 3, E-28014 Madrid, Spain
    \and 
    Max Planck Institute for Astronomy, Königstuhl 17, 69117 Heidelberg, Germany
    \and
    4-183 CCIS, University of Alberta, Edmonton, Alberta, T6G 2E1, Canada
    \and
    Institute of Astronomy and Astrophysics, Academia Sinica, No. 1, Sec. 4, Roosevelt Road, Taipei 10617
    \and
    European Southern Observatory, Karl-Schwarzschild-Straße 2, 85748 Garching, Germany
    \and
    National Astronomical Observatory of Japan, 2-21-1 Osawa, Mitaka, Tokyo, 181-8588, Japan
    \and 
    Instit\"ut  f\"{u}r Theoretische Astrophysik, Zentrum f\"{u}r Astronomie der Universit\"{a}t Heidelberg, Albert-Ueberle-Strasse 2, 69120 Heidelberg, Germany
    \and
    Cosmic Origins Of Life (COOL) Research DAO, coolresearch.io
    \and
    Department of Physics \& Astronomy, University of Wyoming, Laramie, WY 82071 
    \and 
    %Universit\"{a}t Heidelberg, Zentrum f\"{u}r Astronomie, Institut f\"{u}r Theoretische Astrophysik, Albert-Ueberle-Str 2, D-69120 Heidelberg, Germany
    %\and 
    Sub-department of Astrophysics, Department of Physics, University of Oxford, Keble Road, Oxford OX1 3RH, UK
             }

   \date{Received September XX; accepted YY}

  \abstract{ Carbon monoxide (CO) emission constitutes the most widely used tracer of the bulk molecular gas in the interstellar medium (ISM) in extragalactic studies.  The CO-to-H$_2$ conversion factor, \aco, links the observed CO emission to the total molecular gas mass. However, no single prescription perfectly describes the variation of \aco\ across all environments within and across galaxies as a function of metallicity, {molecular} gas opacity, line excitation, and other factors. Using spectral line observations of CO and its isotopologues mapped across a nearby galaxy, we can constrain the molecular gas conditions and link them to a variation in \aco. Here, we present new, wide-field ($10\times10\rm\,arcmin^2$) IRAM 30-m telescope 1mm and 3mm line observations of \chem{^{12}CO}, \chem{^{13}CO}, and \chem{C^{18}O} across the nearby, grand-design, spiral galaxy M101. From the CO isotopologue line ratio analysis alone, we find that selective nucleosynthesis and changes in the opacity are the main drivers of the variation in the line emission across the galaxy. In a further analysis step, we estimated \aco\ using different approaches, including (i) via the dust mass surface density derived from far-IR emission as an independent tracer of the total gas surface density and (ii) local thermal equilibrium (LTE) based measurements using the optically thin \chem{^{13}CO}{10} intensity. We find an average value of $\acoResSpec\rm\,M_\odot\,pc^{-2}(K\,km\,s^{-1})^{-1}$ across the disk of the galaxy, with a decrease by a factor of 10 toward the $2\rm\,kpc$ central region.  In contrast, we find LTE-based \aco\ values are lower by a factor of $2-3$ across the disk relative to the dust-based result. Accounting for \aco\ variations, we found significantly reduced molecular gas depletion time by a factor 10 in the galaxy's center.  In conclusion, our result suggests implications for commonly derived scaling relations, such as an underestimation of the slope of the Kennicutt Schmidt law, if \aco\ variations are not accounted for.  
  }
  
   \keywords{galaxies: ISM -- ISM: molecules -- radio lines: galaxies}

   \maketitle
%
%-------------------------------------------------------------------
%%%%%%%%%%%%%%%%%%%%%%%%%%%%%%%%%%%%%%%%%%%%%%%%%%%%%%%%%%%%%%%%%%%%%%%%%%%%%%%%%%%%%%%%%%%%%%%%%%%%

%%%%%%%%%%%%%%%%% INTRODUCTION  %%%%%%%%%%%%%%%%%%

\section{Introduction}
\label{sec:intro}
The low-$J$ rotational transitions of carbon monoxide (CO) are key tracers of the bulk molecular gas mass in the interstellar medium (ISM) within and across galaxies. The $^{12}$CO molecule constitutes the second most abundant molecule after molecular hydrogen, H$_2$. It has a permanent dipole moment and a much higher moment of inertia than H$_2$. Consequently, $^{12}$CO has low energy rotational transitions, leading to excitation and detectable emission at low temperatures -- unlike the lowest H$_2$ rotational lines which require ${\gtrsim}$100 K to excite. Hence, in particular, at low temperatures ($T{\sim}10\,$ K) and number densities above $n_{\rm H}{\sim}10^2 \: {\rm cm^{-3}}$, CO is regularly used as an effective tracer of the molecular ISM.
The conversion from $^{12}$CO emission to the amount of molecular hydrogen relies on the application of an appropriate CO–to–H$_2$ conversion factor which corresponds to a light-to-mass ratio \citep[see the review by][]{Bolatto2013}. We note that H$_2$ column densities, $N_{\rm H_2}$ [cm$^{-2}$], are generally derived from the low-$J$ $^{12}$CO(1-0) integrated intensity, $W_{\rm ^{12}CO(1-0)}$ [$\rm K\,km\,s^{-1}$], using the conversion factor X$_{\rm ^{12}CO(1-0)}$ [$\rm cm^{-2}\,(K\,km\,s^{-1})^{-1}$]:
\begin{equation}
    N_{\rm H_2} = X_{\rm CO}\times W_{\rm ^{12}CO(1-0)}.
\end{equation}
Equivalent to the factor X$_{\rm CO}$, but in different units, \aco\ [$\rm M_\odot\,pc^{-2}\,(K\,km\,s^{-1})^{-1}$] converts the integrated intensity into the total molecular gas mass surface density (including the contribution of elements heavier than hydrogen), $\Sigma_{\rm mol}\,[M_\odot\,\rm pc^{-2}]$, via:
\begin{equation}
    \Sigma_{\rm mol} = \aco\times W_{\rm ^{12}CO(1-0)}.
\end{equation}
The value of \aco\ varies with the ISM environment. In low-metallicity regions, for example, a significant fraction of the molecular gas becomes CO-dark since dust shielding against photodissociation of CO is reduced \citep{Maloney1988,Israel1997,Leroy2007, Wolfire2010, Glover2011, Leroy2011, Bolatto2013,  Schruba2017, Williams2019}. In addition, previous studies find that \aco\ tends to decrease toward the centers of galaxies \citep{Sandstrom2013, Cormier2018, Israel2020}. {Changes in} temperature and gas turbulence \citep[e.g.,][]{Israel2020, Sun2020, Teng2022},{ which both affect CO emissivity and hence the conversion factor,} could explain the observed decrease in \aco. Given that CO is so straightforwardly observable, a concrete prescription for \aco\ as a function of local ISM properties poses a longstanding goal.

Obtaining robust \aco\ calibrations is challenging since the molecular gas mass must be measured independently of CO emission. One commonly used technique consists of using dust emission to trace the {combined atomic and molecular (i.e., {total})} gas distribution in the ISM \citep[e.g.,][]{Thronson1988,Israel1997, Leroy2011, Planck2011_XXI, Sandstrom2013}. From an empirical standpoint in the Milky Way, dust seems to be well mixed with the {total} gas at the kiloparsec-scales \citep{Planck2011_XXI}. In addition, the dust emission remains optically thin across most nearby spiral galaxies. Using IR or (sub)millimeter emission, one can model the dust spectral energy distribution and obtain an estimate of the dust mass surface density. We can translate the dust mass to a total gas column or mass surface density using a metallicity-dependent dust-to-gas ratio (DGR). The DGR can, however, be environmentally dependent and vary across a galaxy \citep{RomanDuval2014}. Since the ionized gas is only expected to contribute a small fraction of the column density of gas mixed with dust \citep{Planck2011_XXI}, we can reasonably consider this dust-based column density to reflect the sum of atomic gas \hi, and molecular gas. 
Using \hi\ emission observations, we can separate the total gas into its two components and separate out the amount of molecular gas. By comparing it to the measured CO intensity, we can derive an estimate for \aco. 

We can also use CO isotopologue emission to infer the temperature, density, and opacity of molecular clouds in nearby galaxies  \citep[e.g.,][]{Davis2014,Alatalo2015,RomanDuval2016, Cormier2018,  Israel2020, Teng2022}. 
The low-$J$ \chem{^{12}CO} transitions usually remain optically thick, whereas \chem{^{13}CO} and \chem{C^{18}O}  are optically thinner. Consequently, comparing optically thin \chem{^{13}CO} and \chem{C^{18}O} lines to the optically thick \chem{^{12}CO} lines gives insights into the optical depth. Moreover, contrasting two optically thin lines offers an understanding of changes in relative abundances of the different isotopologue species \citep{Davis2014, Zhang2018, Brown2019}. 
For instance, the various C and O isotopes { and the CO isotopologue species abundances vary with}  processes, such as nucleosynthetic and chemical processes \citep{Henkel1994, Timmes1995, Prantzos1996}. Hence, studying the emission of several CO isotopologues can provide insight into the chemical enrichment of the molecular gas. Due to lower abundances, the emission of these CO isotopologues is, however, fainter by $1-2$ orders of magnitude than the \chem{^{12}CO} emission \citep[e.g.][]{denBrokClaws}.

\begin{table}
    \begin{center}
    \caption{Properties of M101.} 
    \label{tab:target_descr}
    \begin{tabular}{l l}
    \hline \hline
     Property    & Value  \\ \hline
     Other Names         & NGC\,5457, PGC\,50063  \\
     Right Ascension (J2000)$^{\rm (a)}$ & 14$^{\rm h}$\,03$^{\rm m}$\,12$^{\rm s}$.6 \\
     Declination (J2000)$^{\rm (a)}$ & 54$^{\circ}$\,20\arcmin\,57\arcsec \\
     Inclination, $i^{\rm (b)}$ & $18^\circ$ \\
     Position Angle$^{\rm (b)}$ & $39^\circ$ \\
     Radius, $r_{25}^{\rm (a)}$ & $12.0^\prime$ \\
     Distance, $d^{\rm (a)}$ & $6.65$\,Mpc \\
     Systemic Velocity, $V_{\rm hel}^{\rm (a)}$ & $237\,$km\,s$^{-1}$\\
     Morphology$^{\rm (a)}$ & SABc \\
     ${\rm SFR}^{\rm (c)}$  & $3.4$\, M$_\odot$\,yr$^{-1}$\\
     $\log_{10}(M_\star/\mathrm{M}_\odot)^{\rm (c)}$ & 10.39 \\ \hline
    \end{tabular}
     \end{center}
    \raggedright{ {\bf Notes:}\\
    (a) \cite{Anand2021};\\ 
    (b) \cite{Sofue1999};\\
    (c) \cite{Leroy2019}.}
\end{table}

\begin{figure}
    \centering
    \includegraphics[width=\columnwidth]{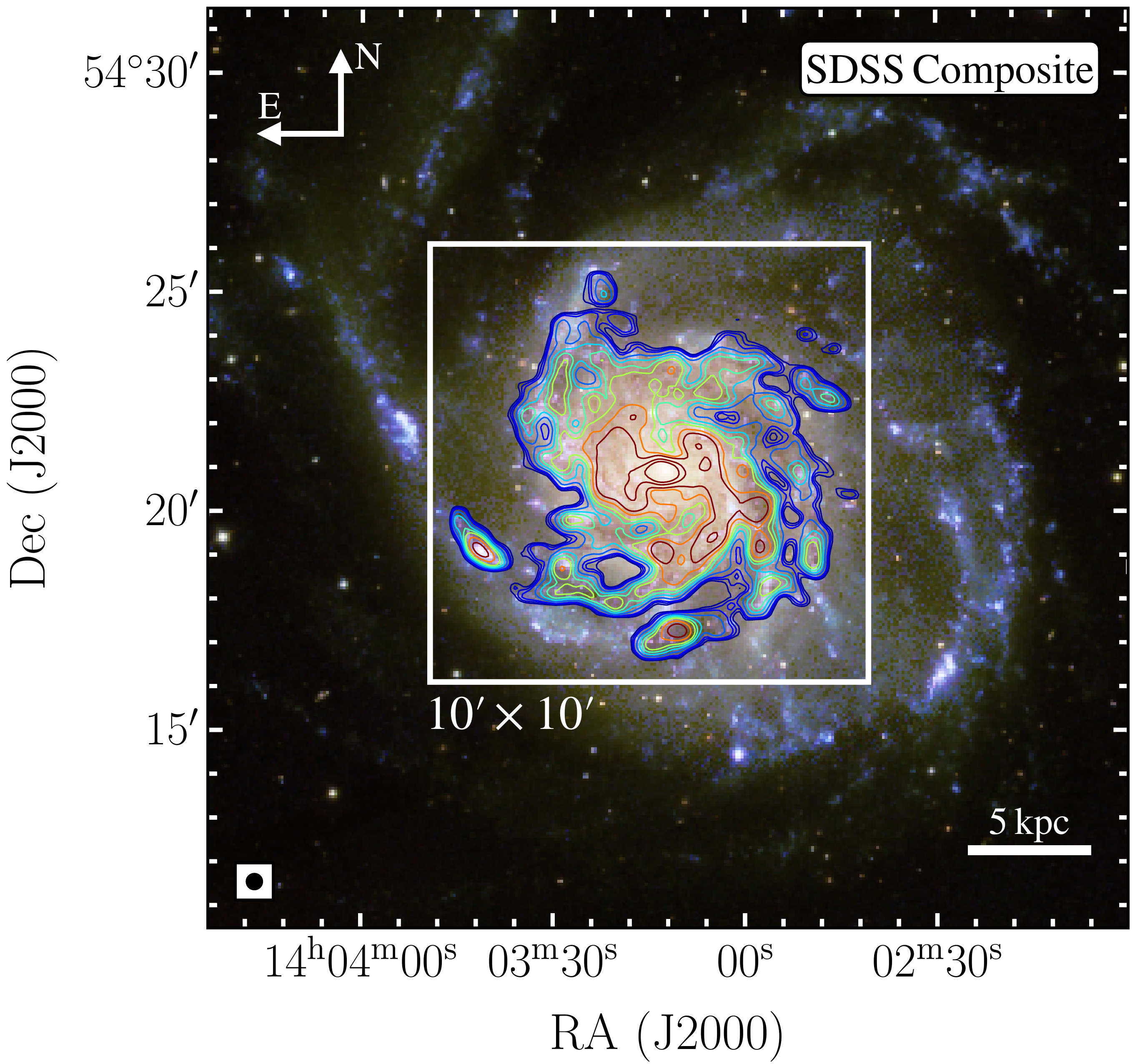}
    \caption{{\bf SDSS composite RGB image with \chem{^{12}CO}{10} overlay}.  Color image using public SDSS data from the 16th data release \citep{Ahumada2020}. We combined the  u, g, and r filter bands. Contours illustrate the IRAM \mbox{30m} \chem{^{12}CO}{10} integrated intensities. The mm observations have a resolution of $23$\arcsec (${\sim}800\,\rm pc$), and are indicated by the black circle in the lower left. The $10\arcmin\times 10\arcmin$ field-of-view of our IRAM \mbox{30m} observation is indicated by the white rectangular outline. Contours are drawn at arbitrary intervals between $0.5-10$\,K\, km\,s$^{-1}$ to highlight the structure of the galaxy.}
    \label{fig:hubble_rgb}
\end{figure}

Both estimating the CO-to-H$_2$ conversion factor via the dust mass surface density and studying the molecular gas conditions using CO isotopologues require high-sensitivity observations of CO. As a result, studies that resolve these diagnostics across large parts of galaxies are still rare. Here, we present IRAM 30m telescope observations of the $J{=}1{\rightarrow}0$ rotational transition of \chem{^{12}CO}, \chem{^{13}CO}, and \chem{C^{18}O} for the galaxy M101.
It is a well-studied, massive, face-on, nearby ($D= 6.65$\, Mpc; \citealt{Anand2021}), star-forming spiral galaxy in the northern hemisphere. In addition to its proximity, the galaxy has a low inclination ($i=18^\circ$), which allows for well-resolved, extended studies across the full galactic disk. M101 has a considerable apparent size across the sky with an extent of the disk in the optical of ${\sim}20\arcmin\times20\arcmin$ \citep{Paturel2003}. It is tidally interacting \citep{Waller1997} with nearby companion galaxies. Furthermore, M101 is of particular interest due to its well-documented metallicity gradient \citep{Kennicutt2003, Croxall2016, Berg2020} based on auroral line measurements. The gradient is stronger than in other nearby spiral galaxies \citep[{f}or M101, the gradient is $-1.1\rm dex/r_{25}$;][]{Berg2020}. In \autoref{fig:hubble_rgb}, we show an optical composite image using observations from the \textit{Sloan Digital Sky Survey} \citep[SSDS; ][]{Blanton2017}. In addition, we show the \chem{^{12}CO}{10} line map presented in this paper using overlayed contours. The galaxy has a wealth of ancillary data across all wavelength regimes. As part of the IRAM \mbox{30m} large program HERACLES \citep{Leroy2009}, wide-field \chem{^{12}CO}{21} observations exist, which complement our observations of the 3mm CO $J=1\rightarrow$0 lines. In addition, there exists a dust surface density map \citep{Chastenet2021} and a  \hi\ map from THINGS \citep{Walter2008} that allow resolved application of the dust-based modeling technique. \autoref{tab:target_descr} lists key properties of the galaxy derived from previous surveys and studies.

The IRAM \mbox{30m} wide-field ${\sim}$kpc multi-CO line observations of M101 complement the IRAM \mbox{30m} large program CLAWS \citep{denBrokClaws}, which obtained deep multi-CO kpc-scale observations of the galaxy M51. In combination, we can investigate differences and similarities in molecular gas conditions traced by CO emission between these two massive, star-forming spiral galaxies.  
Moreover, since we have the same suite of data for M51 as for M101, we can systematically assess the CO integrated intensity ratio (hereafter referred to as simply ``{CO line ratio}") and \aco\ conversion factor trends across different nearby galaxies.

This paper is organized as follows: in \autoref{sec:obs_data} we present and describe the IRAM \mbox{30m} observations as well as the ancillary data that are used in this paper. 
The main results of the paper, which includes results from the CO isotopologue analysis and the \aco\ variation across the galaxy, are presented in  \autoref{sec:result} and \autoref{sec:result2}. Finally, \autoref{sec:diskuss} diskusses the implications of the CO line ratio and \aco\ variation on commonly derived molecular ISM scaling relations and provides a parameterization of the conversion factor in terms of commonly observed parameters. We conclude in \autoref{sec:conclusion}.

%%%%%%%%%%%%%%%%%%%%%%%%%%%%%%%%%%%%%%%%%%%%%%%%%%%%%%%%%%%%%%%%%%%%%%%%%%%%%%%%%%%%%%%%%%%%%%%%%%%%

%%%%%%%%%%%%%%%%% OBSERVATIONS %%%%%%%%%%%%%%%%%%
\section{Observations and data reduction}
\label{sec:obs_data}
\subsection{Observations}
As part of an IRAM \mbox{30m} observing program (\#160-20, PI: den Brok), we used the EMIR receivers to map line emission in the 1\,mm (220\,GHz) and 3\,mm (100\,GHz) windows in dual polarization from the disc of M101 for a total of ${\sim}$80\,h (${\sim}$65\,h on-source time) in the time period of January to March 2021. The receiver bandwidth was 15.6\,GHz per polarization. We carried out the observations simultaneously in the E90 and E230 bands using both the upper-inner (UI) and upper-outer (UO) bands.
We used the Fast Fourier Transform spectrometers with $195$\,kHz spectral resolution (FTS200). The spectrometer yielded a spectral resolution of ${\sim}0.5$\,km\,s$^{-1}$ for the E090 and ${\sim}0.2$\,km\,s$^{-1}$ for the E230 band. \hyperref[tab:obs_lines]{Table~\ref*{tab:obs_lines}} lists the key lines we targeted. 

\begin{table*}
    \begin{center}
    \caption{Summary of the lines targeted as part of the IRAM 30m observing program. Several observational parameters and key characteristics of the extracted data products are included.}
    \label{tab:obs_lines}
    \begin{tabular}{c c c c c c c c c c}
         \hline \hline 
          Band & Line & $\nu_{\rm rest}$ & \multicolumn{2}{c}{Beam size} & $\langle \rm rms\rangle$& On-source time & $\langle T_{\rm sys}\rangle$ & $\langle$pwv$\rangle$\\
         && [GHz] &[$\arcsec$]&[pc]&[mK]& [hr] & [K]&[mm]\\
         &&&(1)&(1)&(2)&(3)&(4)&(5)\\\hline
         \multirow{3}{*}{E0 (3\,mm)}&\chem{^{12}CO}{10}&115.271&25.6&830&13.7&\multirow{3}{*}{65.4}&\multirow{3}{*}{217}&\multirow{3}{*}{4.6}\\
         &\chem{^{13}CO}{10}&110.201&26.8&860&7.4\\
         &\chem{C^{18}O}{10}&109.782&26.9&870&7.3\\ \hline
         \multirow{1}{*}{E2 (1.3\,mm)}&\chem{^{12}CO}{21}&230.538&12.8&410&21.8&\multirow{1}{*}{37.4}&\multirow{1}{*}{211}&\multirow{1}{*}{1.1}\\
         \hline
    \end{tabular}
    \end{center}
    {\raggedright {\bf Notes:} (1) Beam size of the final data cube after reduction. (2) Average rms measured for a 4\,km\,s$^{-1}$ channel width. (3) Total on-source time, including only the subset of data finally used to generate the cubes after reduction. The scanning speed was 8$''$/sec. While we simultaneously observed bands E0 and E2 {to target the $J {=} 1 {\rightarrow} 0$ and $J{=}2\rightarrow1$ transitions}, the on-source time for the \chem{^{12}CO}{21} is shorter because we also dedicated time to target the $J=2\rightarrow1$ transitions of \chem{^{13}CO} and \chem{C^{18}O}, which required another tuning. However, we do not detect any \trans{21} emission of these CO isotopologues in the 1mm regime. (4) Average system temperature (for a subset of data used for the final cube). (5) Average precipitable water vapor (pwv) during observations (for a subset of data used for the final cube).}
\end{table*}

For the mapping, we used a similar approach to the one from the EMPIRE survey \citep[see][]{Donaire2019}. Using the on-the-fly and position switching (OTF-PSW) mode, we mapped a field of $10\,\mathrm{arcmin} \times 10\,\mathrm{arcmin}$ (corresponding to ${\sim}20\,\rm kpc\times20\,\rm kpc$ or $0.83\,r_{25}\times0.4\,r_{25}$).  In addition, we included two emission-free reference positions (OFF position) offset by 300$''$ toward the north and east of M101's center. We scanned the field in RA and DEC directions using multiple straight paths that are each offset by $8''$ from each other. After an iteration over the full field, we shifted the scanned box by {$\sqrt{2} \times N$, with $N = 2\arcsec,4\arcsec,6\arcsec$}, along the position angle PA${=}+45^{\circ}$. This guarantees that, in the end, we cover M101 with a much finer, $2''$, instead of $8''$, grid {along the $x$ and $y$ direction}. We set the read-out dump time to $0.5$\,s, and the final spacing between data points reach $4''$.
A typical observation session had a length of $6{-}9$\,h during the night, with 11 sessions in total. The telescope's pointing and focus were determined at the beginning of each session using observations of a bright quasar. We corrected the focus after 4\,h of observing, and the pointing of the telescope was adjusted every $1{-}1.5$\,h using a nearby quasar. To ensure a proper antenna temperature ($T_\mathrm{a}^\star$) calibration, we did a chopper-wheel calibration every $10{-}15$ minutes using hot-/\linebreak[0]{}cold-load absorber and sky measurements. Finally, to achieve accurate flux calibration, we observed line calibrators (IRC+10216 or W3OH) at the beginning or end of each observing session.

\subsection{Data reduction}

The following steps summarize the data processing and reduction. For these individual routines, we employ the scripts used for the HERACLES and EMPIRE pipeline \citep[see description in][]{Donaire2019} and basic calibration steps by \texttt{MRTCAL}\footnote{\url{ https://www.iram-institute.org/medias/uploads/mrtcal-check.pdf}}. 
\begin{enumerate}
    \item First, we convert the spectrum to the corrected antenna temperature scale ($T_\mathrm{a}^\star$) by scaling each science scan using the most recent previous calibration scan.
    \item We then subtract the most recent OFF measurement from the calibrated spectrum. This concludes the most basic calibration steps.
    \item Next, using the \textit{Continuum and Line Analysis Single-dish Software} (\texttt{CLASS}\footnote{\url{https://www.iram.fr/IRAMFR/GILDAS/doc/html/class-html/class.html}}), we extract the target lines  and create the velocity axis given the rest frequency of the relevant line. 
    \item To subtract the baseline, we perform a constant linear fit. For the fit, we account for the systemic velocity of M101. We omit the range of 100 to 400\,km\,s$^{-1}$ around the center of the line (which corresponds to the velocity range of the galaxy). 
    \item Finally, we regrid the spectra to have a 4\,km\,s$^{-1}$ channel width across the full bandpass. 
    Such a spectral resolution is sufficient to sample the line profile, as shown by previous observations and IRAM 30m surveys, such as HERACLES, EMPIRE and CLAWS. The spectra are then saved as a FITS file.
\end{enumerate}

To estimate the flux calibration stability, we observed the spectra of line calibrators (e.g. IRC+10216) on several nights. We find a maximum day-to-day variation in amplitude of ${\sim}5$\% across all observations, which is consistent with the more extended analysis of the stability of the line calibrators in \cite{Cormier2018} done for the EMPIRE survey. The average actual noise in the cube data is listed in \autoref{tab:obs_lines}.

We performed a more sophisticated final data reduction using an \texttt{IDL} routine, which is based on the HERACLES data reduction pipeline \citep{Leroy2009}. With this {routine}, we can remove bad scans and problematic spectra. Furthermore, the routine performs a \emph{platforming correction} at the edges of the FTS units. This ensures that the various sub-band continua are at a common
level. We note that the receiver's tuning was chosen so that no target line is affected by potential offsets due to platforming.
After the platforming correction, we perform a baseline fitting again. We start by excluding a generous line window using the \chem{^{12}CO}{10} line emission. We place a window extending in both spectral directions around the mean \chem{^{12}CO}{10} velocity. The window's full width for each pixel depends on the specific velocity range of the galaxy's emission derived from HERACLES CO(2-1) data. It ranges between 50 and 300\,km\,s$^{-1}$ for each pixel. We place two windows of the same width adjacent to the central window on both sides. The pipeline then fits a second-order polynomial to the baseline in these windows. The routine finally subtracts the resulting baseline from the full spectrum.

Bad scans and spectra are removed by sorting the remaining spectra by their rms. The pipeline determines the channel-rms from line-free windows after the baseline subtraction. We remove the spectra in the highest tenth percentile. 

For the following analysis, we use the main beam temperature ($T_{\rm mb}$). The main beam temperature is connected to the corrected antenna temperature scale ($T_\mathrm{a}^\star$) via
\begin{equation}
    T_{\rm mb} = \frac{F_{\rm eff}}{B_{\rm eff}} T_\mathrm{a}^\star~,
\end{equation}
with the forward ($F_{\rm eff}$) and beam ($B_{\rm eff}$) efficiencies, which depend on the observed frequency. We determined the value of the efficiencies using a cubic interpolation of the efficiencies listed in the IRAM documentation\footnote{ \url{http://www.iram.es/IRAMES/mainWiki/Iram30mEfficiencies}}.
{Adopting these values, we find a $F_{\rm eff}/B_{\rm eff}$ ratio of 1.2 at 115\,GHz and 1.6 at 230\,GHz}.

Finally, we generated science-ready data cubes by gridding the spectra onto a $2''$ spaced Cartesian grid. The final beam of each data cube, given in \autoref{tab:obs_lines} is coarser than the telescope beam, {because we performed a further convolution of the OTF data (at telescope beam resolution) with a Gaussian beam that has a width corresponding to two-thirds of the FWHM of the telescope beam. Such a gridding kernel is needed as we translate from the data sampled on the OTF grid to a regular grid \citep{Mangum2007}. Our choice of a gridding kernel equal to two-thirds of the FWHM of the telescope beam reflects a trade-off between  signal-to-noise and resolution. The noise is sampled on the scale of the data dumps (every 0.5\,s) while the telescope samples the sky with the PSF of the telescope.} The average noise in the cube data is listed in \autoref{tab:obs_lines}.

This work does not account for flux contamination due to error beam contribution. We note that M101 shows no strong arm-interarm contrast in CO emission (as opposed to other similar spiral galaxies, such as, for example, M51). Therefore, the magnitude of the error beam contribution is expected to be minor. 
In \cite{denBrokClaws}, the effect of error beam contributions is discussed in detail. In particular, in the presence of strong contrast between bright and faint regions, the faint region can suffer from significant error beam contributions. The exact contribution is difficult to quantify as the exact shape of the error beam of a single-dish telescope fluctuates depending on the telescope's elevation. That is why only first-order estimates on the extent of the contribution can be made. IRAM provides estimates of the full 30m telescope beam pattern in their reports \citep[e.g.][]{Kramer2013}. The 1\,mm regime is more strongly affected by such error beam contributions, since the telescope's main beam efficiency is lower ($B_{\rm eff}^{3\,\rm mm}=78\%$ and $B_{\rm eff}^{1.3\,\rm mm}=59\%$) and the beam size is smaller. While \cite{denBrokClaws} find in general contributions to be ${<}10\%$ in M51, it can in certain interarm regions reach up to 40\%. In particular, regions with strong contrast are affected. For M101, we do not expect the error beam to contribute more than 10\%, given the overall low contrast across its disk.

\subsection{Ancillary data and measurements}
For a complete analysis, we use archival and ancillary data sets. In this section, we provide a brief description of the additional data sets used in the analysis. For our \aco\ estimation approach, we particularly require robust dust mass surface density and atomic gas mass surface density maps.

\subsubsection{Dust mass surface density maps}
The dust surface density maps are the products of emission spectral energy distribution (SED) fitting following the procedure by \citet{Chastenet2021}. They used a total of 16 photometric bands, combining mid- and far-IR maps-
This includes the 3.4, 4.6, 12, 22\,$\mu$m from the Wide-field Infrared Survey Explorer \citep[WISE;][]{Wright2010}, 3.6, 4.5, 5.8, 8, 24, 70, 160\,$\mu$m from Spitzer \citep[][]{Fazio2004, Rieke2004, Werner2004}, and 70, 100, 250, 350, and 500\,$\mu$m from Herschel \cite[][]{Griffin2010, Pilbratt2010, Poglitsch2010}.
For the M101 dust mass map they relied on Herschel data from KINGFISH \citep{Kennicutt2011}, Spitzer data from \cite{Dale2009}, and WISE maps from the z0mgs survey \citep{Leroy2019}. 
The angular resolution would be ${\sim}36''$, if we include up to 500\,$\mu$m. For our analysis, we employ a resolution of ${\sim}21''$ by only using up to $250\,\mu$m. The fitted dust masses up to $250\,\mu$m. is consistent with the one up to $500\,\mu$m. (priv. comm. with Jeremy Chastenet).
\citet{Chastenet2021} used the \citet{Draine2007} physical dust model to fit the data, with the {\tt DustBFF} fitting tool \citep{Gordon2014}. The free parameters for dust continuum emission fitting are the minimum radiation field heating the dust, $U_{\rm min}$, the fraction of dust grains heated by a combination of radiation fields at various intensities, $\gamma$, the total dust surface density, $\Sigma_{\rm dust}$, the fraction of grains with less than $10^3$ carbon atoms, $q_{\rm PAH}$, and a scaling factor for stellar surface brightness, $\Omega_*$. We note that we correct the dust mass surface density with a normalization factor of 3.1 \citep{Chastenet2021}. The renormalization is necessary so that the dust mass estimates agree with predictions based on the metal content \citep[e.g.,][]{Planck2014,Planck2015,Dalcanton2015}. { \citet{Chastenet2021} derived the value 3.1 by fitting the dust model to a common MW diffuse emission spectrum and comparing to other dust models using the same abundance constraints.} The uncertainty is set pixelwise as 10\% of the dust mass surface density value. 
Details on IR image preparation, fitting procedure, and results can be found in \citet{Chastenet2021}. 

\subsubsection{Radial metallicity gradients}
We employ radial metallicity gradient measurements from \cite{Berg2020}. They derive the chemical abundances from optical auroral line measurements in \hii\ regions across M101 and M51. Their observations are part of the CHemical Abundances Of Spirals (CHAOS) project \citep{Berg2015}. We use the slope and intercept of the gradient provided by \cite{Berg2020} (see Table 2 therein, we correct the slope since we use an updated value for M101's $r_{25}$):
\begin{equation}
    12+\log(\rm O/H)=\begin{cases}(8.78{\pm}0.04)-(1.10{\pm}0.07)R_{g}[r_{25}] & \text{for M101} \\ (8.75{\pm}0.09)-(0.27{\pm}0.15)R_{g}[r_{25}] & \text{for M51}\end{cases}.
\end{equation}

Often, the metallicity is also expressed in terms of solar metallicity fraction, $Z$. We assume a solar abundance of $\oxab_\odot=8.73$ \citep{Lodders2010} and convert the oxygen abundance to a metallicity ($Z=\Sigma_{\rm metal}/\Sigma_{\rm gas}$, where $\Sigma_{\rm gas}$ includes the mass of He as well). The following equation relates the fractional metallicity, $Z$, to the oxygen abundance:
\begin{equation}
\label{eq:Z}
    Z = \frac{1}{M_{\rm O}/M_{\rm metal}}\frac{m_{\rm O}}{1.36\,m_{\rm H}}10^{(\oxab)-8.73}.
\end{equation}
We assume a fixed oxygen-to-metals ratio, $M_{\rm O}/M_{\rm metal}=0.51$ \citep{Lodders2003}. The atomic masses for oxygen and hydrogen are indicated by $m_{\rm O}$ and $m_{\rm H}$, respectively. The factor $1.36$ is used to include Helium.

\subsubsection{Atomic gas surface density}
To estimate the atomic gas surface density ($\Sigma_{\rm atom}$), we use archival \hi\, 21\,cm line emission data from \textit{The \hi\ Nearby Galaxy Survey} \citep[THINGS;][]{Walter2008}. The data were observed with the Very Large Array (VLA) in B, C, and D configurations. We use the natural weighted data. These have an angular resolution of ${\sim}11$\arcsec (${\sim}350$\,pc) and a spectral resolution of ${\sim}5$\,km\,s$^{-1}$.
We note that the THINGS M101 data suffer from a negative baseline level due to missing zero-spacings. To improve the data, we feathered the interferometric VLA data using an Effelsberg single dish observation from \textit{The Effelsberg-Bonn H I Survey} \citep[EBHIS;][]{Winkel2016}. We use \texttt{uvcombine}\footnote{\url{uvcombine.readthedocs.io}} and the CASA version 5.6.1 \texttt{feather} function and determine a single dish factor of $1.7$. We convert the \hi\ line emission ($I_{\hi}$) to atomic gas surface density via \citep{Walter2008}:
\begin{align}
    \Sigma_{\rm \hi} &[M_\odot\,{\rm pc}^{-2}] \nonumber\\ &= 1.36\times (8.86\times 10^3)\times \left(\frac{I_{\hi} [{\rm Jy\,beam^{-1}\,km\,s^{-1}}]}{{\rm B_{\rm maj}[\arcsec]\times B_{\rm min}[\arcsec]}}\right),
\end{align}
where the factor $1.36$ accounts for the mass of helium and heavy elements and assumes optically thin 21-cm emission. $\rm B_{\rm max}$ and $\rm B_{\rm min}$ are the FWHM of the major and minor axes of the main beam mentioned above. We provide further details on the feathering and how it affects the subsequent \hi\ measurements in \autoref{sec:hi}.

\subsubsection{Stellar mass and SFR data}
We employ stellar mass and SFR surface density maps from the z0mgs survey \citep{Leroy2019}. The SFR surface density is estimated using a combination of ultraviolet observations from the Galaxy Evolution Explorer \citep[GALEX;][]{Martin2005} and mid-infrared data from the Wide-field
Infrared Survey Explorer \citep[WISE;][]{Wright2010}. We use the SFR maps with the combination FUV (from GALEX; at 150\,nm wavelength)+ WISE4 (from WISE; at  22\,$\mu$m).

We use the stellar mass surface density maps computed with the technique utilized for sources in the PHANGS-ALMA survey \citep{Leroy2021_PHANGS}. In short, the $\Sigma_{\star}$ estimate is based on near-infrared emission observations at 3.6\,$\mu$m (IRAC1 on Spitzer) or 3.4\,$\mu$m (WISE1). The final stellar mass is then derived from the NIR emission using an SFR-dependent mass-to-light ratio.

\begin{figure*}
    \centering
    \includegraphics[width=0.95\textwidth]{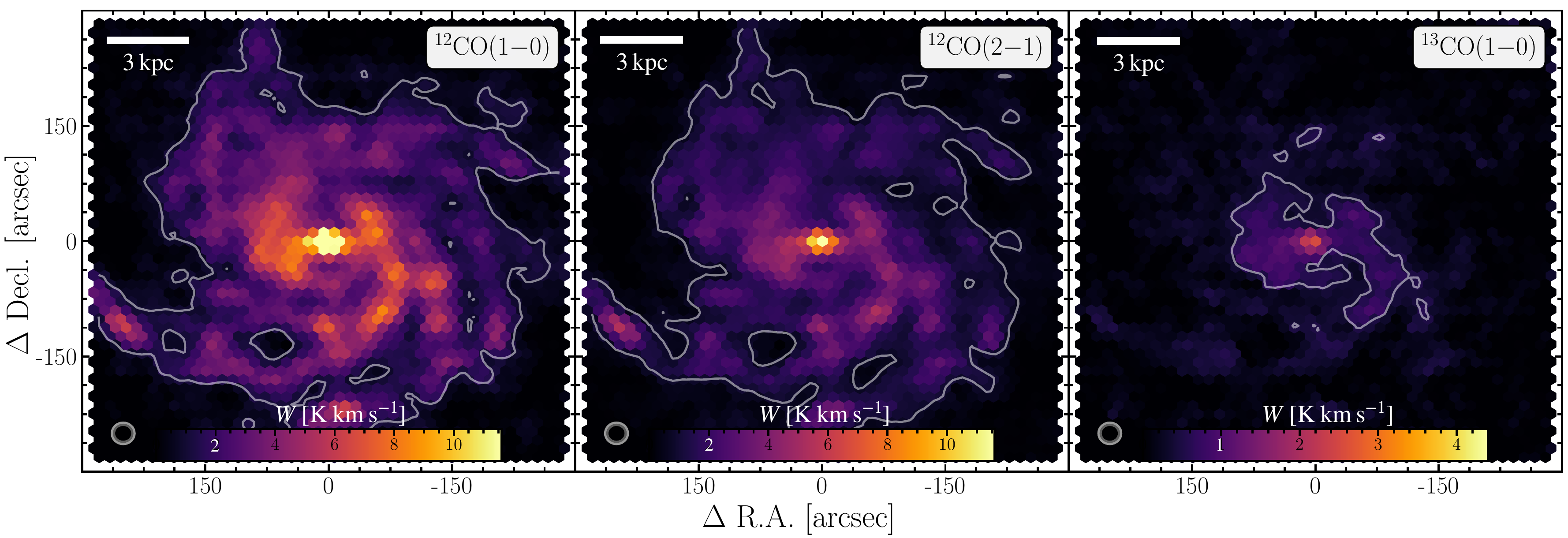}
    \caption{{\bf Integrated intensities}. All maps have been convolved to a common beamsize of 27\arcsec\ (the beamsize is indicated by the circle in the lower left corner). Color scale in units ${\rm K\, km\,s^{-1}}$. Contour indicates S/N=5 of the respective CO isotopologue transition. We do not provide the \chem{C^{18}O}{10} emission line map since we do not detect significant emission across the galaxy. Coordinates are relative to the center coordinates in \autoref{tab:target_descr}.}
    \label{fig:maps101}
\end{figure*}

\subsection{Final data product}
\label{sec:final_datap}
For the analysis in this paper, we homogenize the resolution of the data. We convolve all observations to a common angular resolution of 27\arcsec (${=}840$\,pc), adopting a Gaussian 2D kernel. We regrid all data onto a hexagonal grid where the points are separated by half the beam size (13\arcsec). We perform these steps using a modified pipeline, which has been utilized for IRAM \mbox{30m} large programs before (EMPIRE, \citealt{Donaire2019};  CLAWS, \citealt{denBrokClaws}).

We use the HERACLES/EMPIRE pipeline to determine the integrated intensity for the individual pixels in the regridded cube for each line, including \hi. The goal is to create a signal mask that helps optimize the S/N of the derived integrated intensities. The masked region over which to integrate is determined using a bright emission line. Since \hi\ is faint in the center, we use the \chem{^{12}CO}{10} line for the mask determination for pixels with a galactocentric radius $r\le 0.23\times r_{25}$. We select the factor $0.23$, because, based on observations of star-forming galaxies, the CO surface brightness drops, on average, by a factor of $1/e$ at this radius \citep{Puschnig2020}. This ensures that \chem{^{12}CO} is still detected significantly relative to the \hi\ emission line. For lines of sight with a larger galactocentric radius, the routine employs the \hi\ emission line to determine the relevant spectral range. 
We make a 3D mask where emission is detected at ${\rm S/N}>4$ and then expand the resulting mask into regions with ${\rm S/N}>2$ detections. Finally, we pad the mask {along the spectral axis} by $\pm2$ channels in velocity.
The integrated intensity is then computed by integrating over the channels within the mask. Indicating the number of channels within the mask by $n_{\rm chan}$, the routine computes as follows:
\begin{equation}
    W_{\rm line}[{\rm K\,km\,s^{-1}}] = \sum^{n_{\rm chan}} T_\mathrm{mb}(v)[{\rm K}] \cdot \Delta v_{\rm chan}[{\rm km\,s^{-1}}] 
\end{equation}
where $T_{\rm mb}$ is the surface brightness temperature of a given channel and $\Delta v_{\rm chan}$ is the channel width. \hyperref[fig:aCOMap]{Figure~\ref*{fig:maps101}} shows the integrated intensity for the \chem{^{12}CO}{10}, \chem{^{12}CO}{21}, and \chem{^{13}CO}{10} emission lines. The uncertainty of the integrated intensity for each sightline is computed using the final convolved and regridded cubes with the following equation:
\begin{equation}
    \sigma_W[{\rm K\,km\,s^{-1}}] = \sqrt{n_{\rm chan}}\cdot  \sigma_{\rm rms}[{\rm K}] \cdot \Delta v_{\rm chan}[{\rm km\,s^{-1}}] 
\end{equation}
We indicate the position-dependent $1\sigma$ root-mean-squared (rms) value of the noise per channel with $\sigma_{\rm rms}$. Our approach does not assume any variation of the noise with frequency for each target line. To determine the channel noise, the routine computes the median absolute deviation across the signal-free part of the spectrum scaled by a factor of $1.4826$ (to convert to a standard deviation equivalent).

%%%%%%%%%%%%%%%%%%%%%%%%%%%%%%%%%%%%%%%%%%%%%%%%%%%%%%%%%%%%%%%%%%%%%%%%%%%%%%%%%%%%%%%%%%%%%%%%%%%%

%%%%%%%%%%%%%%%%% ANALYSIS %%%%%%%%%%%%%%%%%%
%\section{Analysis}
%\label{sec:analysis}
%\input{Sections/03_analysis}

%%%%%%%%%%%%%%%%%%%%%%%%%%%%%%%%%%%%%%%%%%%%%%%%%%%%%%%%%%%%%%%%%%%%%%%%%%%%%%%%%%%%%%%%%%%%%%%%%%%%

%%%%%%%%%%%%%%%%% RESULTS %%%%%%%%%%%%%%%%%%
\section{Results: CO isotopologue line emission}
\label{sec:result}
\subsection{CO emission across M101}
\label{sec:COemission}

In \autoref{fig:maps101}, we show the moment-0 maps of \chem{^{12}CO}{10} and \trans{21}, and \chem{^{13}CO}{10}. We detect significant \chem{^{12}CO}{10} and \trans{21} integrated intensities across the full $10\arcmin\times10\arcmin$ field-of-view. We see elevated emission tracing the galaxy's bar and spiral arms. We also find higher integrated intensity values relative to the surroundings at the eastern tip of the southern spiral arm. We find significant \chem{^{13}CO}{10} integrated intensities within $r_{\rm gal}\lesssim5$\,kpc. The \chem{C^{18}O}{10} is too faint, and we do not detect any integrated intensity at ${\rm S/N{>}3}$.

\begin{figure}
    \centering
    \includegraphics[width=0.85\columnwidth]{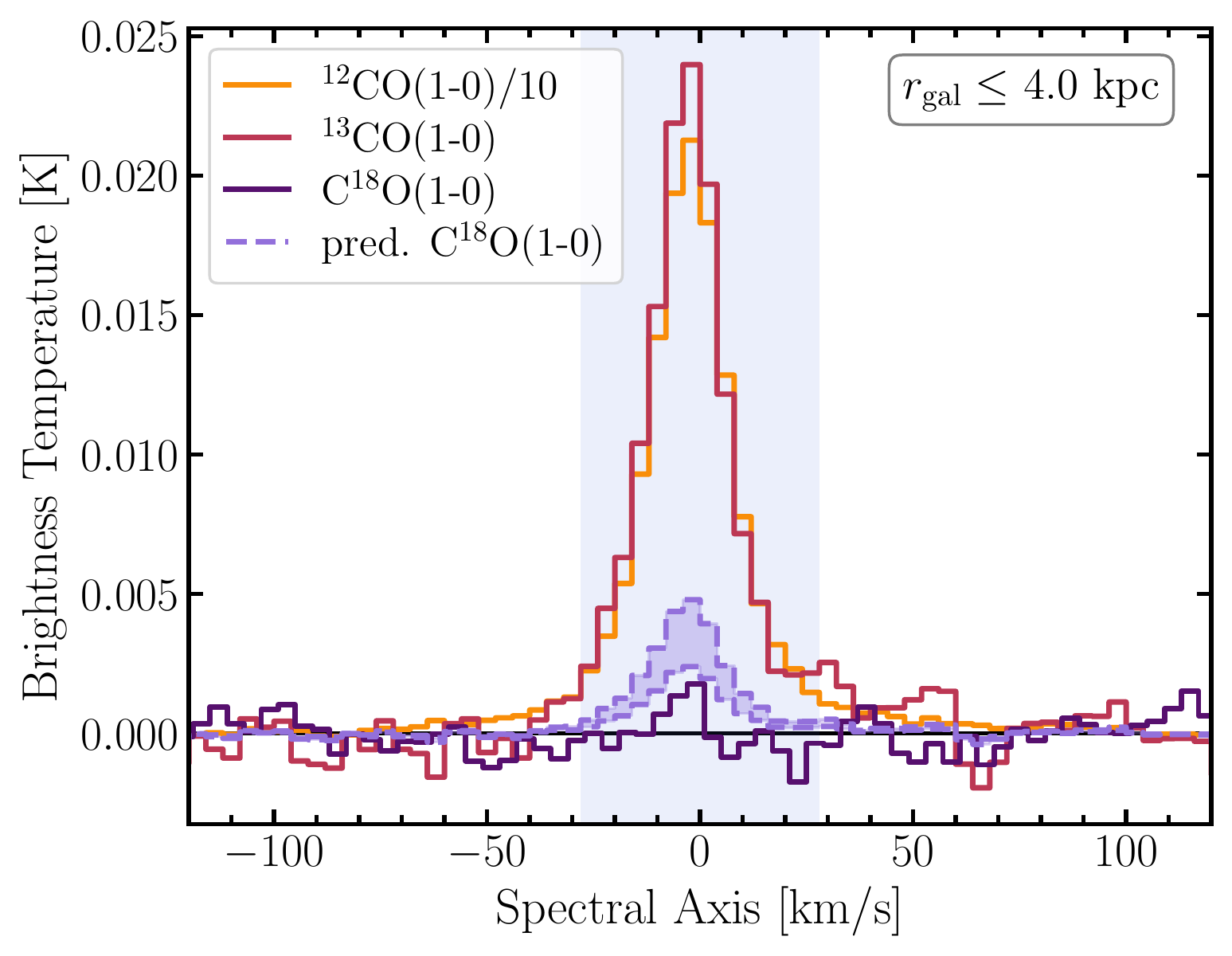}
    \caption{{\bf Radially stacked CO spectra for $r_{\rm gal}\le4\,{\rm kpc}$.} We stack over the central $4\,{\rm kpc}$. Furthermore, the predicted \chem{C^{18}O}{10} emission line is shown, based on the \chem{^{13}CO}{10} emission line and assuming a line ratio of $0.2>R_{18/13}>0.1$ \citep{Donaire2017}. The \chem{C^{18}O}{10} in M101 seems to be fainter than we would expect based on values from EMPIRE. The blue-shaded background indicates the line mask over which we integrate the spectrum.}
    \label{fig:stacks18co}
\end{figure}

To improve the S/N, we stack the spectra by binning sightlines according to various parameters. The procedure is described in \autoref{app:stacking}. For a full reference, \autoref{fig:stacks13co} in \autoref{app:stacking} shows the radially stacked spectra of the \chem{^{12}CO}{10} and \chem{^{13}CO}{10} integrated intensity. We stack the spectra in radial bins with a step size of 1.25\,kpc out to 10\,kpc. Thanks to the improved ${\rm S/N}$ in the stacked spectra, we do find significant (${\rm S/N}>3$) \chem{^{13}CO}{10} integrated intensity out to $r_{\rm gal}\le8\,{\rm kpc}$. 
However, \chem{C^{18}O}{10} emission remains undetected for our 1.25\,kpc radial bins and even when stacking all central 4\,kpc sightlines (\autoref{fig:stacks18co}). In \autoref{fig:stacks18co}, we show for comparison the expected range of integrated intensities based on the \chem{^{13}CO}{10} integrated intensity and the assumption of a \chem{C^{18}O}/\chem{^{13}CO}{10}${\equiv}R_{18/13}$ line ratio commonly found in spiral galaxies of $0.2>R_{18/13}>0.1$ \citep{Langer1993,Donaire2017}. 
The integrated intensity is lower by a factor ${\sim}$2 from the predicted range (we find a 3$\sigma$ upper limit $W^{\rm ul}=0.1\,{\rm K\,km\,s^{-1}}$ and predicted based ratio derived in nearby galaxies $W^{\rm pred.}=0.15-0.2\,{\rm K\,km\,s^{-1}}$ with an average uncertainty of $0.01\,{\rm K\,km\,s^{-1}}$). For comparison, ratios commonly found in the literature range from $R_{18/13}>1$ in ULIRGs \citep{Brown2019}, to $R_{18/13}{\sim}0.3$ in starburst \citep{Tan2011}, $R_{18/13}{\sim}0.1$ in the Milky Way \citep{Langer1993}, and $R_{18/13}{\sim}0.15$ for nearby spiral galaxies (EMPIRE; \citealt{Donaire2017}).

\begin{figure}
    \centering
    \includegraphics[width=\columnwidth]{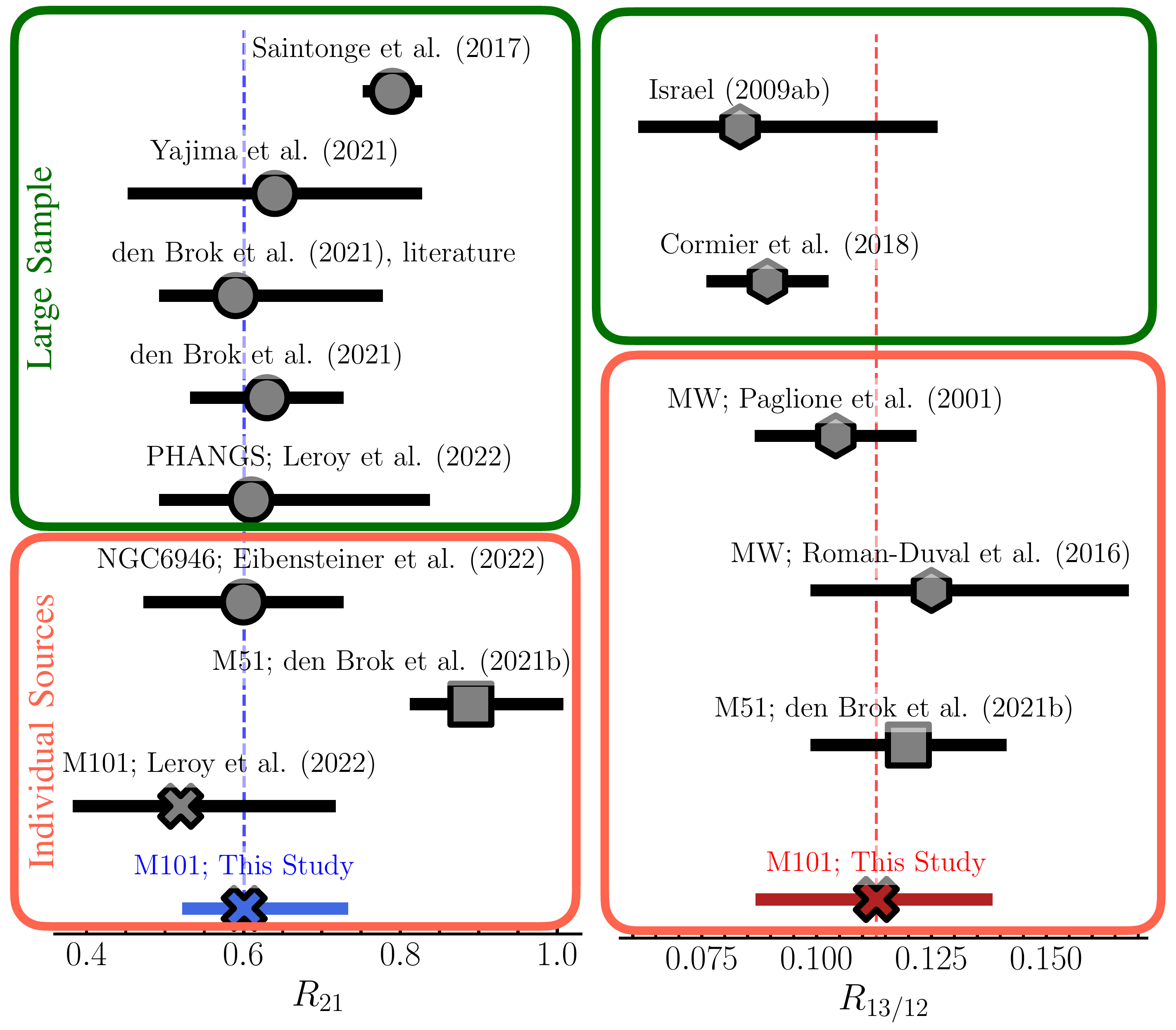}
    \caption{{\bf CO line ratio comparison to literature values.} We compare the average $R_{21}$ and $R_{13/12}$ values estimated from the distribution of the M101 data points to literature values. Errorbars indicate the 1$\sigma$ distribution of sample values. If the literature value corresponds to the value for a specific galaxy, the source's name is provided. Measurement for M101 indicated by the cross. The square symbol indicates the result from M51. (\textit{Left}) Collection of $R_{21}$ distributions. (\textit{Right}) The $R_{13/12}$ distribution is shown. Our measurement agrees well with results for M51 and the Milky Way.}
    \label{fig:ratio_lit_comp}
\end{figure}

\begin{figure*}
    \centering
    \includegraphics[width=0.9\textwidth]{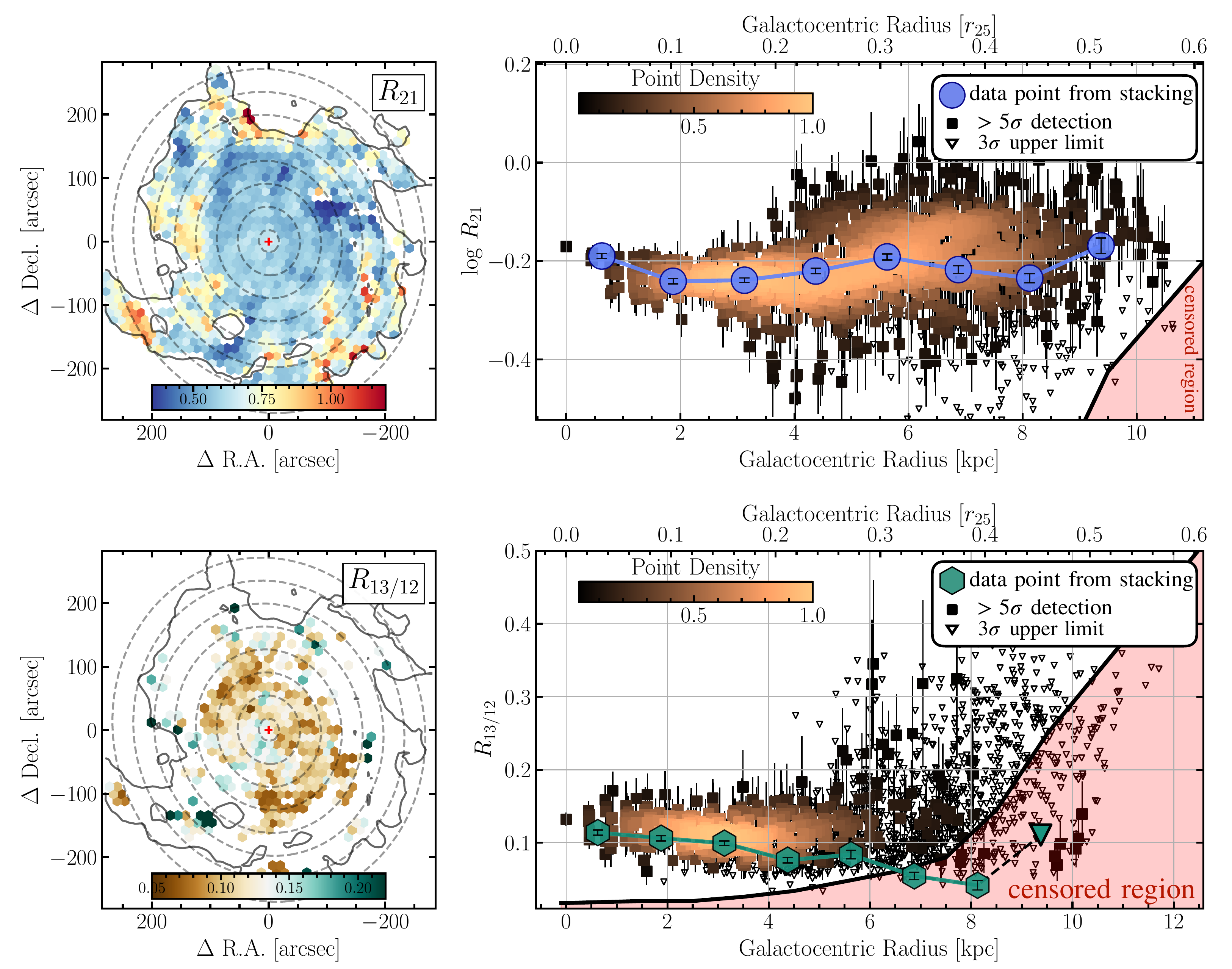}
    \vspace{2mm}
    \includegraphics[width=0.89\textwidth]{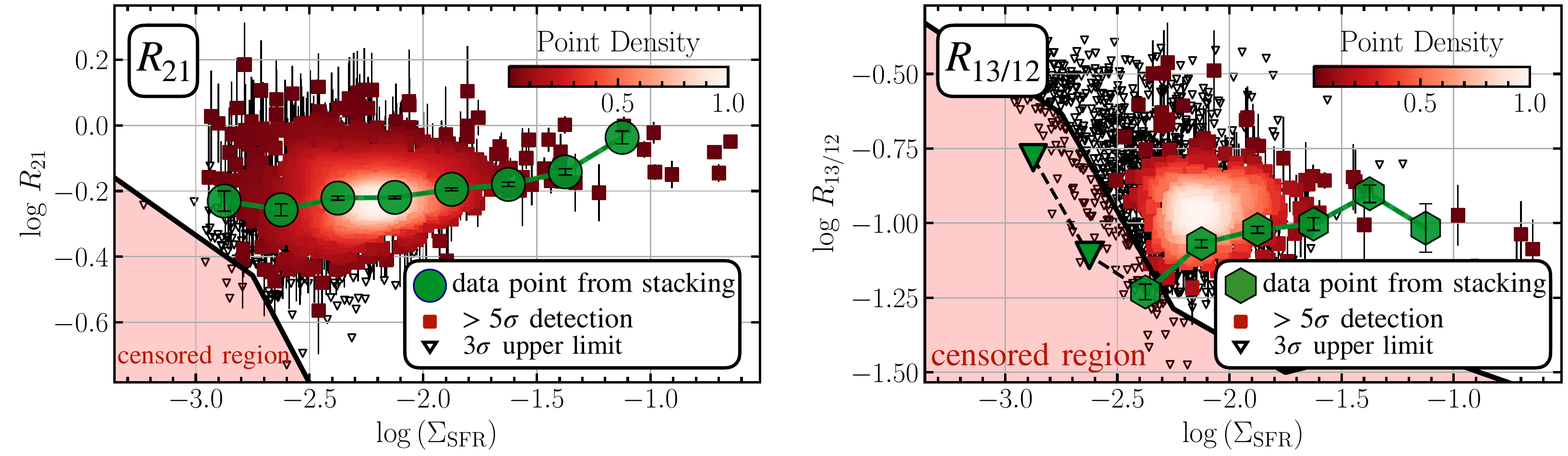}
    \caption{{\bf Spatial and radial variation of the CO line ratio.} Top row shows the $R_{21}$ line ratio while the central row shows the $R_{13/12}$ line ratio. The maps (left top and middle panels) show the spatial distribution of the line ratios. The colored points show sightlines with 5$\sigma$ in both integrated intensities. The 5$\sigma$ contour of the \chem{^{12}CO}{10} integrated intensity is shown by the solid contour.  The dashed circles indicate the radial bins used for the stacking.  The radial plots (right panels) show the radial trends of the line ratios. The panels in the bottom row show trends of the \chem{^{12}CO}{21}/\trans{10} ($R_{21}$) ratio (left) and the \chem{^{13}CO}/\chem{^{12}CO}{10} ratio ($R_{13/12}$) (right) with the SFR surface density. The ratio derived by the stacked line brightness is indicated by the larger blue or green symbols (see \autoref{app:stacking} for a brief description of the stacking technique). We note that because the S/N \chem{^{13}CO}{10} is significantly lower than for \chem{^{12}CO}{10}, and more lines of sights do not show a significant detection, the stacked points  yield a lower $R_{13/12}$.  Triangles indicate 3$\sigma$ upper limits. The uncertainty of the points is indicated, but it is generally smaller than the point size. The censored region applying to the individual lines of sight is  shown by the red (1$\sigma$) shaded region. The region indicates where, due to the lower average sensitivity of one observation set, we do not expect to find significantly detected data points (i.e., the region where only one dataset will be significantly detected).}
    \label{fig:ratiomaps101}
\end{figure*}

\subsection{CO line ratios}

\begingroup
\renewcommand{\arraystretch}{1.5} 
\begin{table*}
    \centering
    \begin{threeparttable}
    
    \caption{{\bf Mean values and Kendall's~$\bm{\tau}$ rank correlation coefficient ($\bm{p}$-value given in parenthesis).} Measured for the line ratios of stacked spectra as function of galactocentric radius and SFR surface density (see \autoref{fig:ratiomaps101}).% and \autoref{fig:ratioOther}). 
    }
    \label{tab:summaryratio}
    \begin{tabular}{l c c | c c } \hline
    &&& \multicolumn{2}{c}{Kendall's~$\tau$ rank correlation coefficient}\\
    Line Ratio&$\langle R \rangle$&$\langle R \rangle^{\rm equal}$& Radius &$\Sigma_{\rm SFR}$  \\
    &(1)&(2)& \\ \hline \hline
    $R_{21}$&$0.60^{+0.07}_{-0.11}$&$0.62^{+0.08}_{-0.14}$&$0.36\ (0.3)$&$0.92\,(4\times10^{-4})$\\
    $R_{13/12}$&$0.11^{+0.03}_{-0.02}$&$0.12^{+0.03}_{-0.03}$&$-0.90\ (0.003)$ &$0.73\,(0.06)$  \\
    \hline
    \end{tabular}
    
    \begin{tablenotes}
      \small
      \item {\bf Notes:} The value in parentheses indicates Kendall's $\tau$ $p$-value. We consider any correlation with $p\le0.05$ significant. (1) $\langle R \rangle$ indicates the average line ratio weighted by \chem{^{12}CO}{21} integrated intensity. The uncertainty for each line ratio is given by the weighted 16th and 84th percentile range. (2) The \chem{^{12}CO}(1-0) median line ratio and 16th and 84th percentiles (since all pixels have the same size, this corresponds to weighing all points equally).
    \end{tablenotes}
    \end{threeparttable}
\end{table*}
\endgroup

We reiterate that we refer to the integrated intensity ratio between two lines simply as \textit{line ratio}. We investigate the line ratio distribution across M101 and compare it to literature values from previous studies. The \chem{^{12}CO}{10} and \trans{21}, as well as the \chem{^{13}CO}{10} emission, is bright enough so that we can investigate its variation across the field-of-view. In particular, the following line ratios are of interest:
\begin{align}
    R_{21} &\equiv \frac{W_{\chem{^{12}CO}{21}}}{W_{\chem{^{12}CO}{10}}} \\
    R_{13/12} &\equiv \frac{W_{\chem{^{13}CO}{10}}}{W_{\chem{^{12}CO}{10}}}
    \\
    R_{18/13} &\equiv \frac{W_{\chem{C^{18}O}{10}}}{W_{\chem{^{13}CO}{10}}}
\end{align}
In \autoref{fig:ratio_lit_comp}, we compare the average line ratio value $R_{21}$ and $R_{13/12}$ which we determined across the full field-of-view (see \autoref{tab:summaryratio}) with values measured in the literature.
We illustrate the spatial variation of these two line ratios and their radial trends in \autoref{fig:ratiomaps101}. We show the line ratio of the individual sightlines as well as the radially stacked ones discussed in \autoref{sec:COemission}, which have a radial bin size of $1.25\,{\rm kpc}$. Furthermore, we illustrate the censored region in the line ratio parameter space (see \autoref{fig:ratiomaps101}). This indicates the region in the parameter space where at least one of the lines is not detected with more than 1$\sigma$ significance (see \autoref{sec:CensoredReg} for a description of the censored region). In addition, we compare the line ratio to $\Sigma_{\rm SFR}$, which traces changes in temperature and density of the gas \citep{Narayanan2012}. We note that previous studies found a trend of $R_{21}$ with the SFR surface density, which would make it a potential tracer of line ratio variation \citep[e.g.,][]{Sawada2001,Yajima2021, Leroy2022}. 

\subsubsection{\texorpdfstring{$R_{21}$}{Lg} line ratio}

We compare the intensity-weighted mean $R_{21}$ value in \autoref{fig:ratio_lit_comp} to the line ratio distribution within and across other sources and samples. Regarding individual sources (orange box in \autoref{fig:ratio_lit_comp}), our result agrees well to within 1$\sigma$ with the ratio of $\langle R^{\rm Leroy2022}_{21} \rangle = 0.52^{+0.19}_{-0.13}$ reported for this galaxy by \citep{Leroy2022}, based on IRAM \mbox{30m} HERA and NRO data. There is only a mild increase of the ratio within the central region ($r_{\rm gal}<1$\,kpc), with a line ratio of $0.69$ for the central sightline. Also, the center of NGC 6946 shows a similar dynamical range of $\langle R_{21}^{\rm NGC\,6946}\rangle=0.6^{+0.1}_{-0.1}$\citep{Eibensteiner2022}. M51 with ratio $\langle R_{21}^{\rm M51}\rangle=0.89^{+0.11}_{-0.07}$ remains an outlier to all these studies as already noted by \citet{denBrokClaws}. Additionally, we compare it to the overall ratio distribution within a sample of galaxies (green box in \autoref{fig:ratio_lit_comp}). When contrasting our average result of M101 to the full EMPIRE survey, which consists of nine nearby spiral galaxies, we find an  almost  identical  median  value: \cite{denbrok2021} report $\langle R_{21}^{\rm EMPIRE}\rangle=0.63^{+0.09}_{-0.09}$. In addition, our value agrees well with the average line ratio for a set of literature single-pointing measurements of nearby spiral galaxies, namely $\langle R_{21}^{\rm literature}\rangle=0.59^{+0.18}_{-0.09}$, which \cite{denbrok2021} have compiled. \cite{Yajima2021} find an average $\langle R_{21}^{\rm Yajima}\rangle=0.64^{+0.18}_{-0.18}$, which agrees with our finding in M101 within the error margins. Recently, \citet{Leroy2022} investigated $R_{21}$ on kpc-scales for a large sample of CO maps of nearby galaxies. They report a median line ratio across all galaxies studied of $\langle R_{21}^{\rm PHANGS}\rangle=0.61^{+0.21}_{-0.11}$. Finally, we find that the average value derived from \mbox{xCOLD GASS} measurements \citep{Saintonge2017} is slightly higher with $\langle R_{21}^{\rm xCOLD GASS}\rangle=0.79^{+0.03}_{-0.03}$ than the value we find. We note that the \mbox{xCOLD GASS} includes galaxies with high star formation rates, which could be associated with enhanced $R_{21}$. Overall, we see that our average value found in M101 agrees well with those derived from a larger set of nearby star-forming spiral galaxies.

Regarding internal variation of the $R_{21}$ line ratio across M101, we find no radial trend for the individual lines of sight as well as the stacked values (see the top right panel in \autoref{fig:ratiomaps101}). Also, Kendall's $\tau$ correlation coefficient does not indicate any significant correlation (see \autoref{tab:summaryratio}). We do not find any significant azimuthal variation of $R_{21}$ across the galaxy. This is qualitatively seen in the map in the top left panel of  \autoref{fig:ratiomaps101}. Neither the bar ends nor the spiral arm or interarm regions show a significant difference in the line ratio. Also, when we bin by spiral phase, a method to quantify the difference between arm and interarm regions, we do not see any clear trend (see \autoref{app:stacking}).  Across the full galaxy, we find a \chem{^{12}CO}{10} brightness weighted mean ratio of $\langle R_{21} \rangle = 0.60^{+0.07}_{-0.11}$.
Comparing to the nine galaxies of the EMPIRE sample, \citet{denbrok2021} generally find a significant increase of $R_{21}$ toward the center by 10-20\% in the galaxies that have a barred structure. In contrast, M101 does not seem to conform to this trend. The fact that the line ratio stays constant across the galaxy, despite apparent environmental differences in the molecular gas condition (such as center or disk, arm or interarm), puts constraints on the connection of $R_{21}$ to the environmental temperature and density variation.

Past studies describe a way to parametrize $R_{21}$ variation using the SFR surface density, $\Sigma_{\rm SFR}$ \citep{denbrok2021, Leroy2022}. Understanding ways to parameterize $R_{21}$ is particularly crucial for studies that rely on \chem{^{12}CO}{21} as opposed to \chem{^{12}CO}{10} observations to derive molecular gas parameters and hence need an accurately calibrated $R_{21}$ to predict the \chem{^{12}CO}{10} brightness from other $J$ lines. The bottom row of \autoref{fig:ratiomaps101} shows the distribution of the line ratios for the individual sightlines with the SFR surface density. We also show the stacked line ratio to better illustrate the trends. When looking at the stacked points, we find a significant ($p=4\times10^{-4}$) positive ($\tau = 0.92$) correlation for $R_{21}$ with the SFR surface density, $\Sigma_{\rm SFR}$. A positive correlation with SFR surface density is also reported by \citet{Leroy2022} who, studying the PHANGS-ALMA sample, found a Spearman's rank correlation coefficient of $\rho=0.55$ for the galaxy-wide, normalized binned $R_{21}$ to the normalized SFR surface density. Comparing the slope of the correlation in logarithmic space, we find a slightly shallower slope of $m=0.10\pm0.2$, compared to $m=0.13$ found by \citet{Leroy2022}. So despite the overall flat $R_{21}$ trend across M101, we can still recover the trend with $\Sigma_{\rm SFR}$ for the stacked data points, which is in agreement with previous studies. Regarding the individual lines of sight, the scatter of ${>}$0.2\,dex still dominates over the degree of variation of $R_{21}$ expected from the dynamical range in $\Sigma_{\rm SFR}$ of 2\,dex.

\subsubsection{\texorpdfstring{$R_{13/12}$}{Lg}  line ratio} 

We compare the intensity weighted mean $R_{13/12}$ line ratio distribution of M101 to findings of various previous studies in \autoref{fig:ratio_lit_comp} (right panel). The average ratio of $\langle R_{13/12}^{\rm M51}\rangle=0.12^{+0.02}_{-0.072}$ found in M51 \citep{denBrokClaws} is consistent within the error margin with the average ratio we find in this study, however, its scatter is slightly larger. \cite{Cormier2018} studied the \chem{^{12}CO}-to-\chem{^{13}CO} line ratio (i.e., the inverse of the ratio we investigate) for the nine EMPIRE galaxies. Converting their finding to $R_{13/12}$, they obtain  $\langle R_{13/12}^{\rm EMPIRE}\rangle=0.09^{+0.01}_{-0.01}$, again consistent with our finding. Similarly, studying the central ${\sim}20''$ of around ten nearby galaxies, including AGN and central starbursts, a range of $0.06<R_{13/12}<0.13$ is found by \cite{Israel2009a,Israel2009b}. For comparison, we also show measurements from the Milky Way \citep{Paglione2001}. For galactic radii larger than 2\,kpc, they find an average value of $\langle R_{13/12}^{\rm MW}\rangle=0.10^{+0.02}_{-0.02}$.

Regarding resolved line ratios within a galaxy, we find for $R_{13/12}$ a negative radial trend when looking at the stacked data points (shown in the bottom right panel in \autoref{fig:ratiomaps101}) with a Kendall's coefficient of $\tau = -0.90$ and a $p$-value of $p=0.003$ (we consider a correlation with a $p$-value below 0.05 to be significant). We note that using the stacked data points, we can actually sample the censored region, which applies to the individual lines of sight, as we have significant \chem{^{13}CO}{10} integrated intensities out to ${\sim}8$\,kpc (see \autoref{sec:COemission} and \autoref{fig:stacks13co}). The trend is less evident when looking at individual sightlines, as the scatter seems significantly larger than the radial trend, and we are limited by the censored region. We find an average line ratio of $\langle R_{13/12} \rangle = 0.11^{+0.03}_{-0.02}$. The stacked integrated intensities decreases from $R_{13/12}|_{r_{\rm gal}=0} = 0.113\pm0.004$ down to $R_{13/12}|_{r_{\rm gal}=8\,{\rm kpc}} = 0.055\pm0.005$ further out. Such a radial decrease is also present in M51 \citep{denBrokClaws}. For comparison, studying this ratio in the Milky Way, \cite{RomanDuval2016} find a radial gradient of the ratio decreasing from $R_{13/12}=0.16$ at 4\,kpc to $R_{13/12}=0.1$ at 8\,kpc radial distance. This Milky Way finding agrees well with our finding in M101. The map at the middle left in \autoref{fig:ratiomaps101} does also not show any azimuthal variation of the line ratio. However, we note that the significant sightlines are mainly from the center, bar ends, and spiral arm regions, while the \chem{^{13}CO}{10} emission within the interarm regions is too faint. The variation of this particular line ratio is due to a combined effect of variation of the optical depth of \chem{^{12}CO} line emission and differences in the relative abundance of \chem{^{13}CO} and \chem{^{12}CO} (under the assumption that \chem{^{13}CO} remains optically thin on kpc scales; see \autoref{sec:disc_isoratio}).

Finally, we investigate the distribution of the CO line ratio across the disk of the galaxy with respect to the SFR surface density (see bottom right panel of \autoref{fig:ratiomaps101}). The stacked $R_{13/12}$ data points show only a mild positive trend ($\tau=0.73$) with the SFR surface density ($p=0.06$), with a scatter of ${\sim}0.25$\,dex for the individual sightlines. However, we note that M101 shows only a narrow dynamical range of SFR surface densities. For comparison, M51 covers ${>}2$\,dex in SFR surface densities, while M101 shows approximately $1\,$dex.   We note  that a similar mild positive trend is observed within individual nearby galaxies \citep{Cao2017, Cormier2018} with respect to the SFR surface density.

\begin{figure*}
    \centering
    \includegraphics[width=\textwidth]{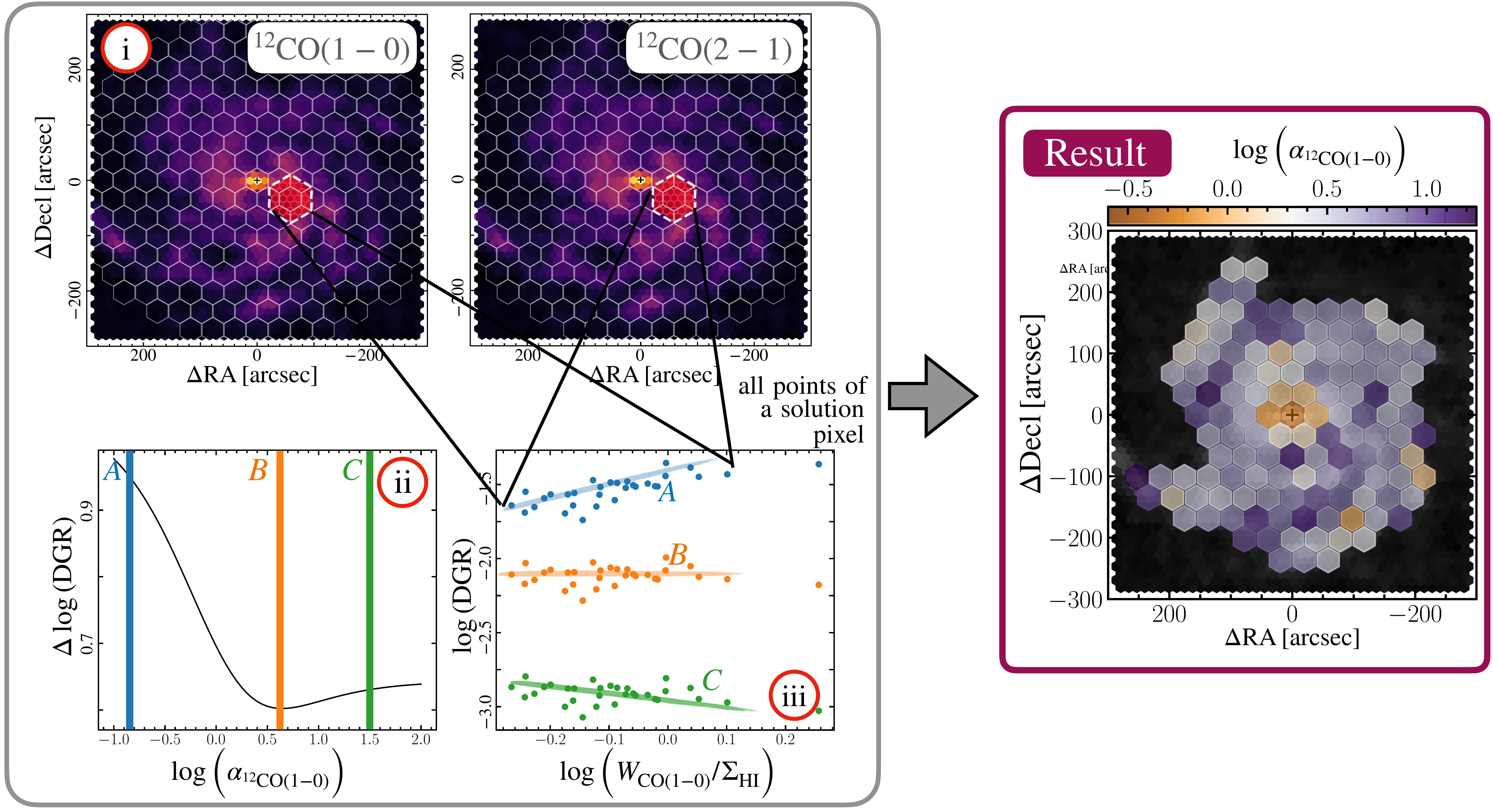}
    \caption{{\bf Solution pixel approach to estimate \aco.} (\textit{Left}) From the scatter minimization approach, described by \cite{Leroy2011} and \cite{Sandstrom2013}, we obtain estimates of \aco\ and $\rm DGR$. The top panels (\textit{i}) show the hexagons that illustrate the individual solution pixels for both \chem{^{12}CO}{10} and \trans{21} transmission. The solution pixels consist of 37 underlying, half-beam spaced lines of sight. We note that the underlying hexagon tiling is meant to show the results in each solution pixel (the actual solution pixels have 40\% overlap). %In the sake of clarity, we have reduced the solution pixel size in the figure to obtain a clear 2D map. 
    In the maps, we highlight an individual solution pixel. We vary \aco\ and compute the $\rm DGR$ following  equation (\ref{eq:dust-gas}). We select the value for which the variation in $\rm DGR$ is minimal. The bottom left panel (\textit{ii}) shows the variation of the $\rm DGR$ as a function of different \aco.  The variation for the selected solution pixel is minimal for \aco\ labeled $B$. We perform this analysis for each solution pixel. The bottom right panel (\textit{iii}) illustrates why the variance differs when changing \aco. Here we combine all significant points from the solution pixel indicated in panel (i) from both CO lines. The black lines point to the solution pixel where the individual lines of sight are drawn from. We correct the \chem{^{12}CO}{21} data with the average line ratio of the solution pixel. The panel illustrates the differences in $\rm DGR$ for three selected \aco\ (labeled $A$, $B$, and $C$). Based on the selection of \aco\, the DGR values will be positively or negatively correlated (as illustrated by the colored line, which is drawn schematically to guide the eye). (\textit{Right}) The resulting \aco\ value for each solution pixel based on the combined \chem{^{12}CO}{10} and \chem{^{12}CO}{21} integrated intensities.  } 
    \label{fig:aCOMap}
\end{figure*}

\section{Results: CO-to-H2 conversion factor}
\label{sec:result2}
\subsection{\texorpdfstring{\aco}{Lg} estimation}
\label{sec:aco_presc}
Under the assumption that dust and gas are well mixed on the scales we probe, the following relation connects the dust mass and the total gas surface density (both in units of $M_\odot\,\rm pc^{-2}$) via the dust-to-gas ratio ($\rm DGR$):
\begin{equation}
\label{eq:dust-gas}
    \frac{\Sigma_{\rm dust}}{\rm DGR}=\Sigma_{\hi} + \Sigma_{\rm H_2} = \Sigma_{\hi} + \alpha_{\rm ^{12}CO(1-0)}\times W_{^{12}\rm CO(1-0)},
\end{equation}
where  \aco\ is the CO-to-H$_2$ conversion factor {in units of $\left[M_\odot\,\rm pc^{-2}(K\,km\,s^{-1})^{-1}\right]$}, which converts the CO-integrated intensity into a molecular gas mass surface density.
There are, however, two unknown quantities in \cref{eq:dust-gas}: The key parameter of interest, \aco\, and the $\rm DGR$ value. Both parameters are expected to vary with the galactic environment and are likely also linked to each other. To estimate both parameters, we introduce some modifications to the  so-called scatter minimization technique developed in \cite{Leroy2011} and \cite{Sandstrom2013}. The idea is to solve simultaneously for \aco\ and DGR. In essence, we find and select a value for \aco\ which -- given a set of measurements of $\Sigma_{\hi}$, $\Sigma_{\rm dust}$ and  $W_{^{12}\rm CO(1-0)}$ -- yields the most uniform distribution of ${\rm DGR}$ values over a certain (${\sim}3$kpc size) area.
The approach consists of the following steps:
\begin{enumerate}
    \item We split the galaxy into so-called \emph{solution pixels}, which are hexagonal regions containing 37 half-beam sampled data points. The solution pixels are separated center-to-center by 1.5 times the beam size (panel (i) in \autoref{fig:aCOMap} illustrates a solution pixel in red).
    \item Using \autoref{eq:dust-gas}, we compute the $\rm DGR$ for each solution pixel with the underlying pixel using a range of \aco\ values. For \aco\, we vary the value from 0.01 to 10~$M_\odot\,\rm pc(K\,km\,s^{-1})^{-1}$ in steps of 0.1 dex (panel (iii) in \autoref{fig:aCOMap} shows the resulting $\rm DGR$ values using three different \aco\ values $A$, $B$, and $C$). The scatter in the resulting $\rm DGR$ values for each solution pixel will vary with the choice of \aco.
    
    \item In addition to obtaining 37 $\rm DGR$ data points per solution pixel from \chem{^{12}CO}{10}, we obtain an additional 37 measurements by using the \chem{^{12}CO}{21} integrated intensity measurements. We convert these to a \chem{^{12}CO}{10} integrated intensity using the average $R_{21}$ of the solution pixel.
    \item The \aco\ value of the solution pixel is chosen such that the scatter of the $\rm DGR$ values of the combined 74 data points is minimal.
\end{enumerate}

For a more detailed description of the implementation, we refer to Section 3 in \cite{Sandstrom2013}. We note that the solution pixels overlap (they share ${\sim}40\%$ of the area with the neighboring solution pixels). Consequently, they are not fully independent from each other. We illustrate the solution pixel in \autoref{fig:aCOMap} (The pixel colored in red illustrates the full extent of a solution pixel).

With this approach, we have now constraints on the values of \aco\ and ${\rm DGR}$. The approach makes the following assumptions:
\begin{enumerate}
    \item There is a dynamical range in the $W_{^{12}\rm CO(1-0)}/\Sigma_{\hi}$ ratio ($x$ axis of the panel (iii) in \autoref{fig:aCOMap}) beyond statistical scatter. Otherwise, there is no leverage by varying \aco\ to find the minimum variation in the ${\rm DGR}$ values.  We test for any potential degeneracies  of the scatter minimization solution in \autoref{sec:discrep} in case the dynamical range is limited.
    \item Regarding the $\rm DGR$ value: we assume that the total gas and dust are well mixed on ${\sim}$kpc scales. This ensures that \cref{eq:dust-gas} is valid. Furthermore, we assume that ${\rm DGR}$ remains constant on ${\sim}3$kpc scales, ${\rm DGR}$ does not change with varying  atomic and molecular phase balance, and a negligible fraction of dust is present in the ionized gas phase.
    \item $R_{21}$ remains constant over the scales of a solution pixel. This is justified given the generally flat line ratio trends found across other nearby galaxies, with only mild increases of 10\% toward some galaxy centers \citep{denbrok2021}.
\end{enumerate}

\begin{figure*}
    \centering
    \includegraphics[width=0.95\columnwidth]{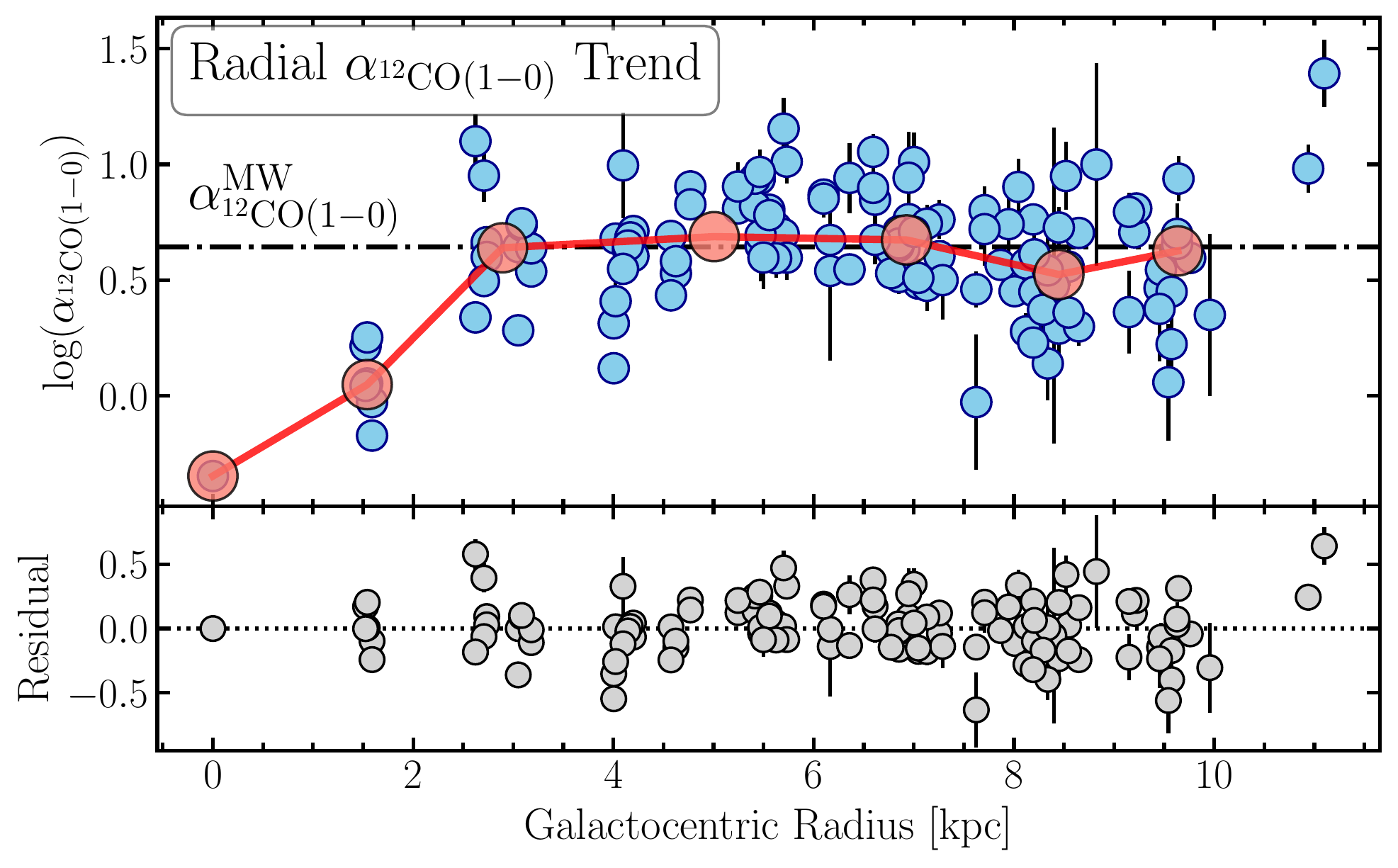}
    \includegraphics[width=0.95\columnwidth]{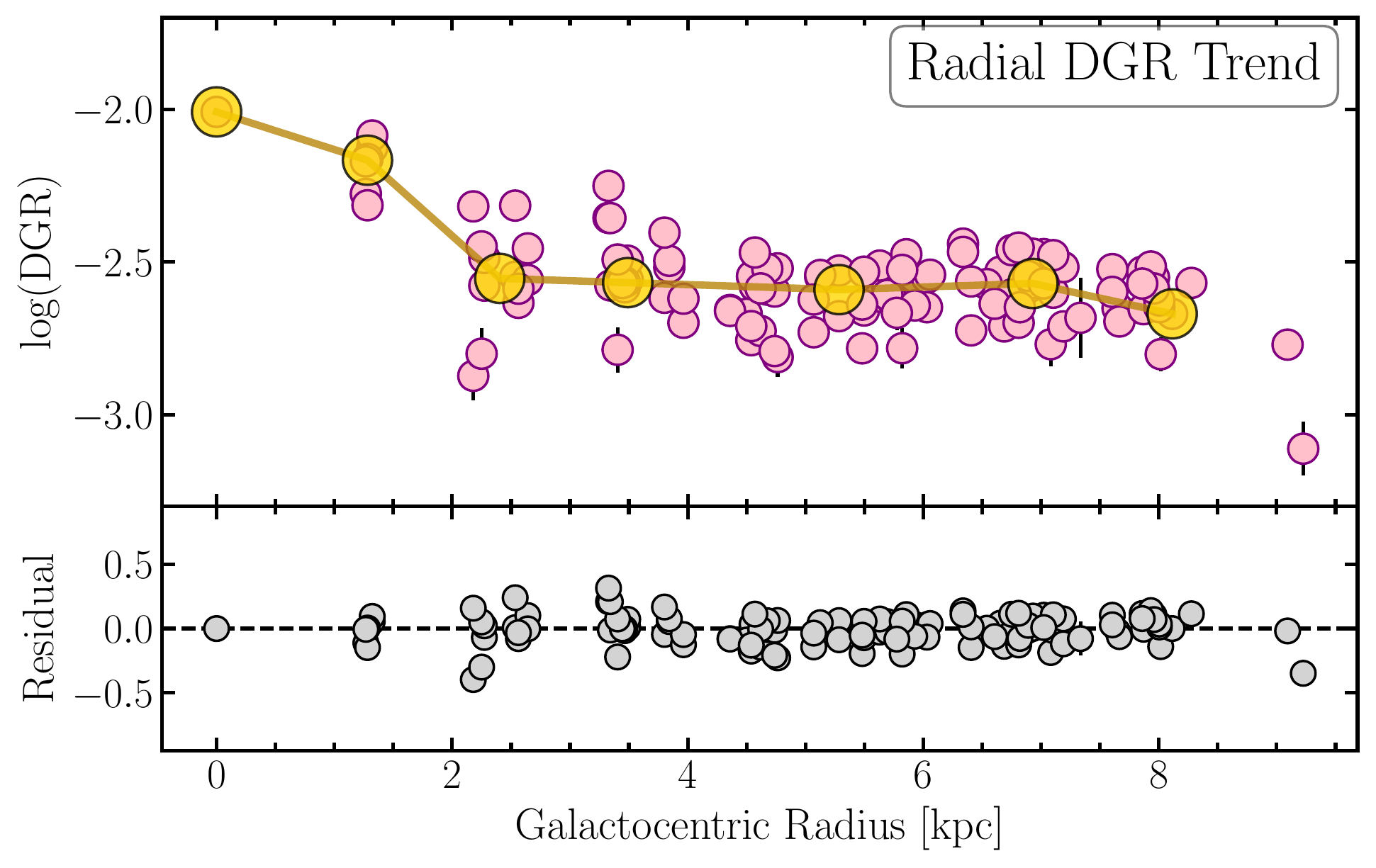}
    \caption{{\bf Radial \aco\ and $\rm DGR$ trend in M101.} The left panel shows radial \aco\ trend, and the right panel illustrates radial $\rm DGR$ dependency. (\textit{Top}) Smaller blue (pink) points show the individual \aco\ ($\rm DGR$) measurements for the various solution pixels. Larger red (yellow) points show the derived trend based on binning the data. (\textit{Bottom}) Residual \aco\ or $\rm DGR$ values after subtracting the radial trend based on linearly interpolating the binned data trend (solid red /yellow line). 
    }
    \label{fig:radial_sub}
\end{figure*}
 
We estimate the uncertainty of the \aco\ value by performing a Monte Carlo test. For each measurement ($\Sigma_{\hi}$, $\Sigma_{\rm dust}$ and $W_{^{12}\rm CO(1-0)}$) we add random noise drawn from a normal distribution with the width  corresponding to their measurement errors. We repeat this resampling 100 times. Our final \aco\ value and corresponding uncertainty are determined via bootstrapping. Iterating with $n_{\rm iter}=1000$, we draw $n_{\rm sample}=1000$ samples from the Monte Carlo iterations and take the mean and standard deviation.

\begingroup
\renewcommand{\arraystretch}{1.5} % Default value: 1
\begin{table*}
    \centering
    \begin{threeparttable}
    
    \caption{{\bf Median \aco\ values for M101 and M51.}  }
    \label{tab:summaryaco}
    \begin{tabular}{l c c c | c  c c } \hline
    &\multicolumn{3}{c}{M101}&\multicolumn{3}{c}{M51} \\
    \aco$^{\rm a}$:& binned$^{\rm b}$ & num. weighted$^{\rm c}$ & lum. weighted$^{\rm d}$& binned$^{\rm b}$ & num. weighted$^{\rm c}$ & lum. weighted$^{\rm d}$ \\ \hline \hline
    All &$4.3^{+0.9}_{-0.9}$ &$4.4^{+3.1}_{-2.2}$ &$4.1^{+2.7}_{-3.0}$  & $3.3^{+0.6}_{-0.6}$ & $3.2^{+3.0}_{-1.5}$ & $3.1^{+0.4}_{-1.4}$  \\
    Center$^{\rm e}$ & \multicolumn{3}{c|}{$0.43$} & \multicolumn{3}{c}{$3.1$} \\
    Disk &$4.4^{+0.9}_{-0.9}$& $4.5^{+3.2}_{-1.8}$ & $4.5^{+3.2}_{-1.6}$ & $3.7^{+0.6}_{-0.6}$&$3.5^{+3.3}_{-1.8}$&$3.1^{+1.6}_{-1.5}$ \\
    \hline
    \end{tabular}
    
    \begin{tablenotes}
      \small
      \item {\bf Notes:} (a) Conversion factor in units $M_\odot\,\rm pc^{-2}\,(K\,km\,s^{-1})$, (b) binning together all the datapoints. Uncertainty represents the binned propagated uncertainty. (c) Median with 16$^{\rm th}$ and 84$^{\rm th}$ percentile scatter (d) \chem{^{12}CO}{10} intensity weighted median with 16$^{\rm th}$ and 84$^{\rm th}$ percentile scatter. (e) Center consists only of one solution pixel.
    \end{tablenotes}
    \end{threeparttable}
\end{table*}
\endgroup

We note as a caveat that we do not account for systematic uncertainties in dust mass measurements. Phase-dependent depletion is observed, and the DGR is likely higher in dense, molecular regions \citep{Jenkins2009}. On the other hand, the dust appears to emit more effectively in dense regions \citep{Dwek1998, Paradis2009,Kohler2015}. These effects are discussed in detail in \cite{Leroy2011,Sandstrom2013}. They find that variation in ${\rm DGR}$ and dust emissivity could lead to a bias of \aco\ towards higher values (by a factor of ${<}$2).  
Further systematic uncertainties could be introduced by the variation of the dust-to-metals ratio, the emissivity calibration, or  dust absorption coefficient \citep[e.g.][]{Clark2016,Chiang2018, Clark2019, Chastenet2021}.
We note that such a trend is systematic and cannot explain any galaxy-internal variation (such as a radial trend) we find in M101. Overall, such effects could be considered by updates to the scatter minimization technique in future work.
{We also do not account for changes in the conversion factor due to CO freeze-out, which occurs predominantly in the densest regions of molecular clouds \citep[$n_{\rm H_2}{>}10^{5}\rm\,cm^{-3}$; e.g., ][]{Whitworth2018}. Since the low-$J$ CO emission is optically thick, we do not expect a significant impact on the observed CO integrated intensity (hence leading to a change in the conversion factor). This is further supported by simulations from \citet{Glover2016}, who find that in molecular clouds at solar neighborhood metallicity CO freeze-out affects the derived CO-to-H$_2$ conversion factor by only $2-3\%$. }

\subsection{Trends in \texorpdfstring{\aco}{Lg} distribution}
\label{sec:corr_aco}

The panel on the right-hand side in \autoref{fig:aCOMap} shows the spatial distribution of the estimated \aco. From a qualitative assessment, we find a decrease in \aco\ and an increase of the $\rm DGR$ toward the center of the galaxy.  \hyperref[fig:radial_sub]{Figure~\ref*{fig:radial_sub}}  shows the radial trend of $\aco$ as well as the residual. 
The result illustrates the lower \aco\ values towards the center, while it has a relatively constant value inside the disk ($r>2\,\rm kpc$). For the central solution pixel, we have $\aco^{\rm center}=(0.43\pm0.03)\,\rm M_\odot\,{\rm pc^{-2}\,(K\,km\,s^{-1})^{-1}}$, while the average value in the disk amounts to $\langle\aco\rangle|_{\rm disk }= (4.4{\pm}0.9)\,\rm M_\odot\,{\rm pc^{-2}\,(K\,km\,s^{-1})^{-1}}$. 
However, we find a large $1\sigma$ point-to-point scatter in \aco\ inside the disk of ${\sim}0.3$ dex. Based on our Monte Carlo implementation of iteratively computing \aco, we find that the propagated uncertainty of \aco\ is ${\sim}0.1\,$dex. \hyperref[tab:summaryaco]{Table~\ref*{tab:summaryaco}} lists the \aco\ values using different binnings.

Our finding of low \aco\ values toward the center is consistent with other studies targeting larger samples of galaxies. They find conversion factors $5{-}10$ times lower than the average MW factor in the center of nearby spiral galaxies \citep{Israel1997,Sandstrom2013, Israel2020}. For reference, past studies also found such low values, for example, for LIRGs \citep[e.g.][]{Downes1998,Kamenetzky2014, Sliwa2017a}, likely due to more excited or turbulent gas similar to conditions in galaxy centers.
We also note that, in particular, the low conversion factor value we find for the center of M101 is consistent with the optically thin \chem{^{12}CO} emission limit. In the presence of highly turbulent gas motions or large gas velocity dispersion, it is possible that the low-$J$ \chem{^{12}CO} emission turns less optically thick. In fact, the $R_{13/12}$ line ratio gives us a potential way to assess whether \chem{^{12}CO} becomes optically thin toward the center. The middle right panel in \autoref{fig:ratiomaps101} shows a decreasing radial trend of $R_{13/12}$. If the trend is only due to optical depth changes of \chem{^{12}CO}, we would expect an opposite trend with decreasing $R_{12/13}$ toward the center. Hence, if the \chem{^{12}CO} emission is indeed less optically thick {in the center, the observed trend in $R_{12/13}$ } implies that the relative abundance of \chem{^{13}CO} has to increase toward the center of M101, and we can make a prediction of \aco.
Under representative molecular ISM conditions with an excitation temperature of $T_{\rm ex}=30\,\rm K$, a canonical CO abundance of $\left[\rm CO/H_2\right]=10^{-4}$, and assuming local thermal equilibrium (LTE), we expect $\aco^{\rm opt.\,thin}\approx0.34\,\rm M_\odot\,pc^{-2}(K\,km\,s^{-1})^{-1}$ \citep{Bolatto2013}, which is very close to the value we find for the center of M101.

We note that M101 is also included in the sample investigated by \cite{Sandstrom2013}. They find a central \aco value of $\aco^{\rm center}=0.35\pm0.1$, which lies within the margin of error of the value we find ($\aco^{\rm center}=0.43\pm0.05$). However, they find a galaxy-wide average value of $\langle\aco^{\rm S13}\rangle=2.3$, which is a factor 2 lower than the value we find in this study. To test the impact of different datasets, we repeat the \aco\ estimation using a different combination of \chem{^{12}CO}{21} (CLAWS and HERACLES) and \hi\ (non-feathered and feathered) datasets. This way, we can assess how the difference in datasets affects the resulting \aco\ values. For details on the comparison, we refer to \autoref{sec:aco_datasets}. The discrepancy between the median \aco\ value measured here and that from \cite{Sandstrom2013} can be traced back to the fact that \cite{Sandstrom2013} relied on \chem{^{12}CO}{21} observations from IRAM \mbox{30m}/HERA, used a constant $R_{21}=0.7$ ratio to convert between the $J=2\rightarrow1$ and $J=1\rightarrow0$ transition and used THINGS \hi\ data that have not been short-spacing corrected. On the one hand, we find from our analysis that substituting the CLAWS data with the HERACLES \chem{^{12}CO}{21} observations does not significantly affect the average \aco\ distribution. On the other hand, using the non-feathered \hi\ data lowers the \aco\ measurements by 0.1\,dex. We also find that using a constant $R_{21}$ and only relying on the \chem{^{12}CO}{21} observations, will further systematically lower \aco\ by 0.2\, dex, hence reproducing the results from \citet{Sandstrom2013}.

Contrasting our finding to results from studies using another \aco\ estimation approach, we find that our median \aco\ value for the disk of M101 is, in fact, consistent with virial mass measurements. For example, \cite{Rebolledo2015} studied the conversion factor in certain brighter regions of M101 and found, on average, values close to the MW average.

\subsection{\texorpdfstring{$\rm \aco$}{Lg} based on multi-line modeling}

Using the \chem{^{13}CO}{10} emission line, we can perform a simple LTE modeling attempt to obtain an additional, independent estimate of \aco, which we refer to hereafter as $\aco^{\rm LTE}$. Assuming LTE, we can calculate the conversion factor using the following equation
\begin{equation}
    \label{eq:aco13}
    \aco^{\rm LTE} = \left[\frac{\rm H_2}{\rm ^{13}CO}\right]\times \frac{\eta_{\rm 12}}{\eta_{\rm 13}}\times\frac{6.5\times10^{-6}}{1-\exp(-5.29/T_{\rm exc})}\times R_{13/12}. 
\end{equation}
In this formula, the CO isotopologue line ratio $R_{13/12}$ traces the optical depth, $T_{\rm exc}$ indicates the excitation temperature of \chem{^{13}CO}, $\left[\frac{\rm H_2}{\rm ^{13}CO}\right]$ describes the relative \chem{^{13}CO} abundance, and $\eta$ is the beam filling factor of the \chem{^{12}CO}{10} and \chem{^{13}CO}{10} emission respectively. We refer to \citet{Donaire2017} for a more detailed derivation of the equation.

\begin{figure}
    \centering
    \includegraphics[width=\columnwidth]{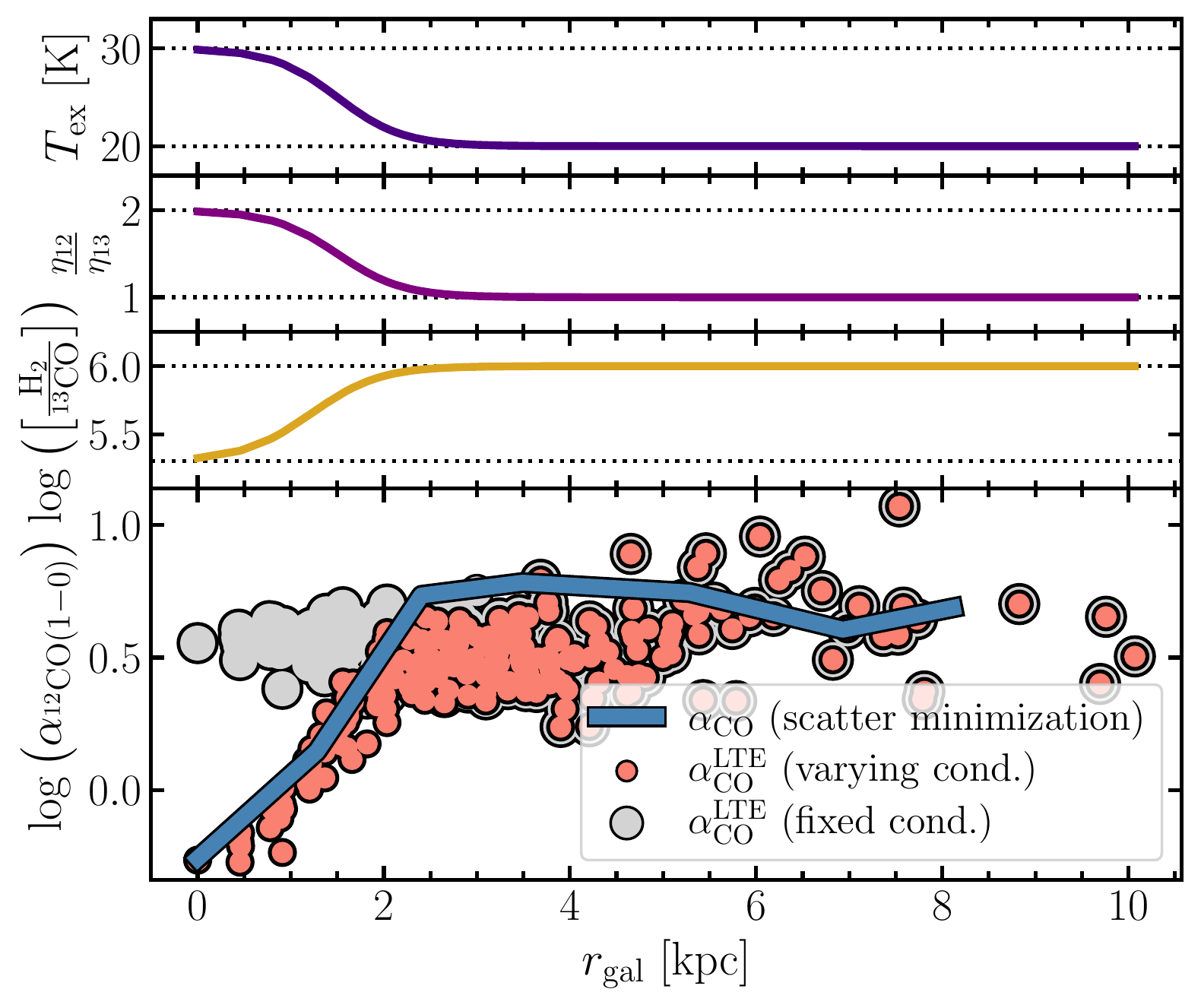}
    \caption{{\bf \chem{^{13}CO} derived $\aco^{\rm LTE}$} We estimate the conversion factor under LTE assumptions using the \chem{^{12}CO}{10} and \chem{^{13}CO}{10} emission. We perform two iterations: (i) keeping the conditions fixed across the galaxy apart from the $R_{13/12}$ ratio and (ii) varying the \chem{^{13}CO} excitation temperature,$T_{\rm ex}$, the beam filling factor ratio, $\eta_{12}/\eta_{13}$, and the \chem{^{13}CO} abundance between disk and center using a sigmoid function. (\textit{Top three panels}) Variation of input parameters for \aco\ derivation.  (\textit{Bottom Panel}) Radial Trend in $\alpha_{\rm CO}^{\rm LTE}$. Grey points indicate measurements under fixed conditions. Red points indicate $\alpha_{\rm CO}^{\rm LTE}$ assuming a variation of the input parameters as shown in the top three panels. The blue line shows the radial trend derived from the scatter minimization technique.   }
    \label{fig:aco_13_trend}
\end{figure}
\begin{figure*}
    \centering
    \includegraphics[width=0.9\textwidth]{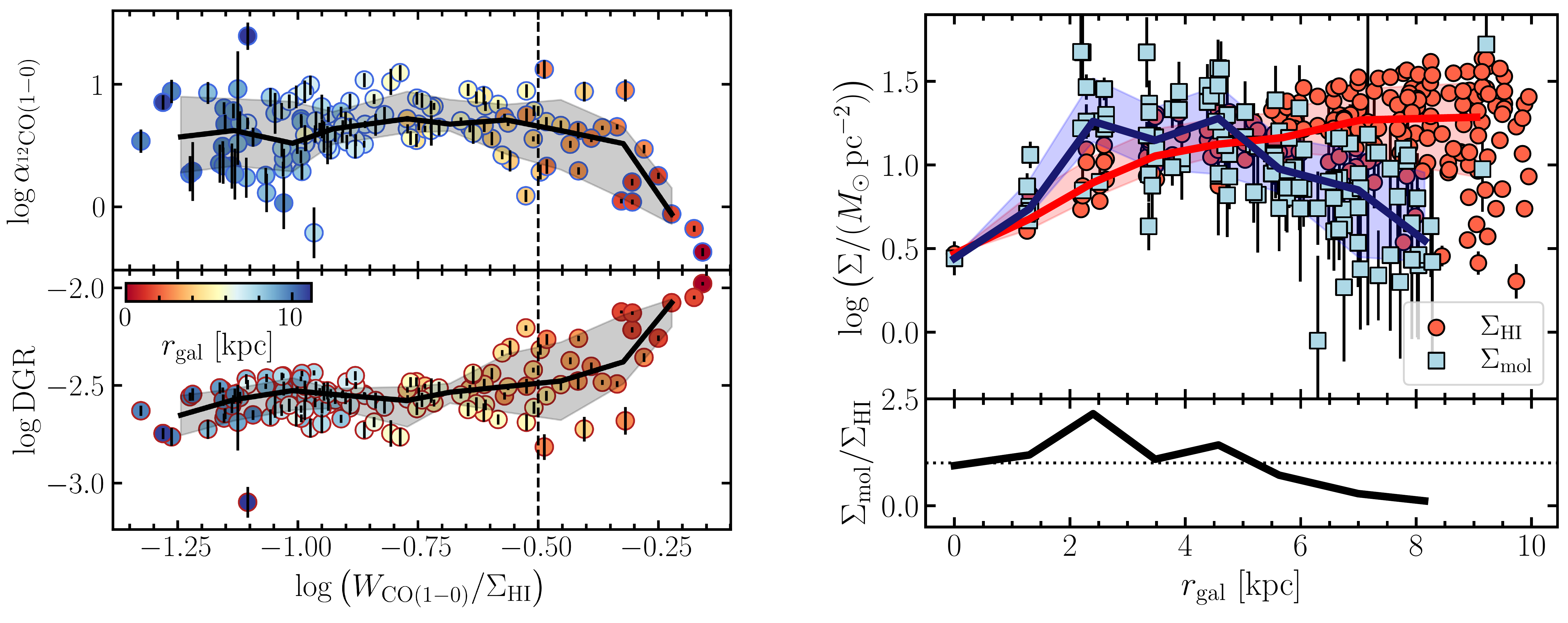}
    \caption{{\bf DGR and \aco\ trends and atomic/molecular gas profiles} (Left) The trend of \aco\ and $\rm DGR$ as a function of the $\log(W_{\rm CO}/\Sigma_{\hi})$, which roughly translates to the molecular-to-atomic gas fraction. The points indicate the individual solution pixels. The color indicates the galactocentric radius. The black line indicates the binned trend, and the dark-shaded region shows the $1\sigma$ scatter. The vertical dashed line is arbitrarily drawn and shows approximately beyond where the two parameters start to deviate from a flat trend.  (Right) Radial profiles of $\Sigma_{\hi}$ and $\Sigma_{\rm mol}$ at solution pixel scale resolution ($\sim 2$\,kpc). The line indicates the radially binned values. The bottom panel shows the ratio of molecular to atomic gas mass surface density. {The horizontal dotted line indicates unity between the molecular and atomic gas surface density.}}
    \label{fig:CO_to_HI}
\end{figure*}

\hyperref[fig:aco_13_trend]{Figure~\ref*{fig:aco_13_trend}}  shows the derived $\aco^{\rm LTE}$ values as a function of the galactocentric radius. We use two different approaches to estimate the input parameters (besides $R_{13/12}$) in \autoref{eq:aco13}:

(i) We assume constant LTE conditions so that the lines are thermalized across M101 following values provided in \citet{Cormier2018}. In particular, we fix the excitation temperature $T_{\rm ex}=20\,\rm K$, the beam filling factor ratio $\eta_{12}/\eta_{13}=1$, and the \chem{^{13}CO} abundance $[\rm H_2/\chem{^{13}CO}]=1\times10^{6}$.  
These values are adopted from \cite{Cormier2018}.
The result is indicated by the grey points in \autoref{fig:aco_13_trend}. We find a relatively flat trend with $\langle\aco^{\rm LTE}\rangle = 3.5^{+0.7}_{-0.9}$. 

(ii) Because the molecular gas conditions are likely not constant across the galaxy, we perform the $\aco^{\rm LTE}$ calculation again. This time, we simultaneously vary the excitation temperature, beam filling factor ratio, and abundance ratio between the center and the disk, thus mimicking a more realistic two-phase model than assuming constant conditions throughout the galaxy. Upon varying the parameters, the beam filling factor and the abundance ratio affect the resulting $\aco^{\rm LTE}$ value directly linearly, while the excitation temperature is exponentially linked to the conversion factor.  We use a convenient sigmoid function\footnote{We use a sigmoid function with an arbitrary width to vary the conditions smoothly between disk and center. We do this to simulate a more realistic transition between the two phases. In our case, we set the width of the sigmoid to the width of a solution pixel (${\sim}2$\,kpc).} to allow for a smooth variation of the parameters between the disk and center limit as a function of galactocentric radius. We use the limit values used in \citet{Cormier2018} as input. We vary the \chem{^{13}CO} excitation temperature, $T_{\rm ex}$, between 20\,K (disk) and 30\,K (center). Such values align with findings in the Milky Way \citep{Roueff2021}. The increase of the abundance towards the center by a factor 5 is motivated by our finding that $R_{13/12}$ is enhanced towards the center (see \autoref{sec:disc_isoratio} for further discussion). Finally, we also vary the beam filling factor ratio $\eta_{12}/\eta_{13}$ between a value of 1 (disk) and 2 (center). The measurements are shown as red points in \autoref{fig:aco_13_trend}. 
The top panels of \autoref{fig:aco_13_trend} show the radial trend for the individual parameter we use as combined input for \autoref{eq:aco13}. 
Using this approach, we can reproduce the depression of the conversion factor toward the center of the galaxy. For the disk ($r>2\,\rm kpc$), we find $\langle\aco^{\rm LTE, \rm disk}\rangle=2.8^{+1.1}_{-0.7}$, while in the center, we find $\langle\aco^{\rm LTE, \rm center}\rangle=0.6^{+0.2}_{-0.1}$. We stress that this exercise does not constrain the degree of variation of the individual input parameters. With this approach, we investigate whether the observed radial variation of \aco\ is reproducible when applying changing input parameters that agree with regular findings from the center and disk  region of nearby galaxies.

We note that with our \chem{^{13}CO} approach, we obtain \aco\ values in the disk that are systematically lower by a factor of 1.6 than the values we find with the scatter minimization approach (for comparison, the average \aco\ value of the disk derived from the scatter minimization technique is indicated in \autoref{tab:summaryaco}). Such a finding of systematically lower \aco\ values based on \chem{^{13}CO} is consistent with previous studies \citep[e.g.,][]{Meier2001,Meier2004,Heiderman2010, Cormier2018}. Similarly, \citet{Szucs2016} show by using numerical simulation of realistic molecular clouds that total molecular mass predictions based on \chem{^{13}CO} are systematically lower by up to a factor of 2–3 due to uncertainties related to chemical and optical depth effects. \citet{Cormier2018} conclude that the systematic offset between \chem{^{12}CO} and \chem{^{13}CO} based \aco\ estimates likely derive from the simplifying assumption of a similar beam filling factor of the two lines across the disk. Such a difference could be explained by the fact that \chem{^{12}CO} is tracing the diffuse molecular gas phase, while \chem{^{13}CO} is likely more confined to the somewhat denser molecular gas phase. The fact that for the depression of \aco\ both estimates agree likely also reflects that our simplified assumptions of the variation of the parameters to the center reflect the actual physical {molecular} gas conditions more properly. To robustly and quantitatively constrain the parameters, such as the excitation temperature and abundance, observations of other \chem{^{13}CO} rotational transitions would be necessary.

In principle, we could match both prescriptions with just slightly different parameter profiles for the LTE-based \aco\ estimation. So far, for instance, we have adopted a MW-based \chem{^{13}CO} abundance in the disk. If we assume that abundance values in the disk are larger by a factor of 2 in M101 than in the MW, we would recover the same \aco\ trend from both prescriptions. However, further observations of other $J$ \chem{^{13}CO} transitions are needed to constrain the underlying \chem{^{13}CO} abundance in M101. 

Our LTE-based \aco\ estimates offer valuable qualitative insight into potential drivers of the CO-to-H$_2$ conversion factor variation. Quantitatively assessing the \aco\ values is difficult due to the underlying assumptions that need to be made for the input parameters (excitation temperature, beam filling factor, and \chem{^{13}CO} abundance). By allowing variation of the parameters toward the center, the depression of \aco\ can be accurately described. 

\subsection{The \texorpdfstring{$\rm DGR$}{Lg} across M101}

Based on our scatter minimization approach, we also derive estimates of the $\rm DGR$ for the individual solution pixels. The right panel in \autoref{fig:radial_sub} shows the radial trend in $\rm DGR$. Similarly to \aco\ we find a clear difference of the value towards the center (larger values by $0.5\,$dex), while the disk shows a relatively flat trend of $\rm \log DGR = (-0.003^{+0.005}_{-0.002})$. Furthermore, the disk shows a relatively small  point-to-point scatter of only $0.2\,\rm dex$. The values we find for the $\rm DGR$ are significantly lower than the average Milky Way solar neighborhood ($\rm DGR^{\rm MW}=0.01$, which differs by $0.5$\,\rm dex; \citealt{Frisch2003}) and nearby spiral galaxies ($\rm DGR^{\rm spiral}=0.014$, which is differed by $0.6$\,\rm dex; \citealt{Sandstrom2013}). 

In contrast, in their comprehensive study of the DGR in M101, \citet{Chiang2018} find values in agreement with our DGR results. They find a power law metallicity dependence of the DGR, with values ranging from $10^{-3}$ (at $12+\log(\rm O/H)=8.3$) to $10^{-2}$ (at $12+\log(\rm O/H)=8.6$).  We cover a dynamical range in metallicity ($12+\log(\rm O/H)$) of 0.3\,dex between the center and disk of M101. Using the relation between metallicity and the DGR found by \citet{Chiang2018} in M101, we would expect to find a 0.6\,dex variation of $\rm DGR$. This is close to the actual 0.5\,dex we find. We note that potential causes for the difference could be that \citet{Chiang2018} (i) applied a constant \aco\ value, (ii) used a modified black body model approach to fit the dust mass surface density, and (iii) did not apply a short spacing correction for the THINGS \hi\ data.

\begin{figure*}
    \centering
    \includegraphics[width=\textwidth]{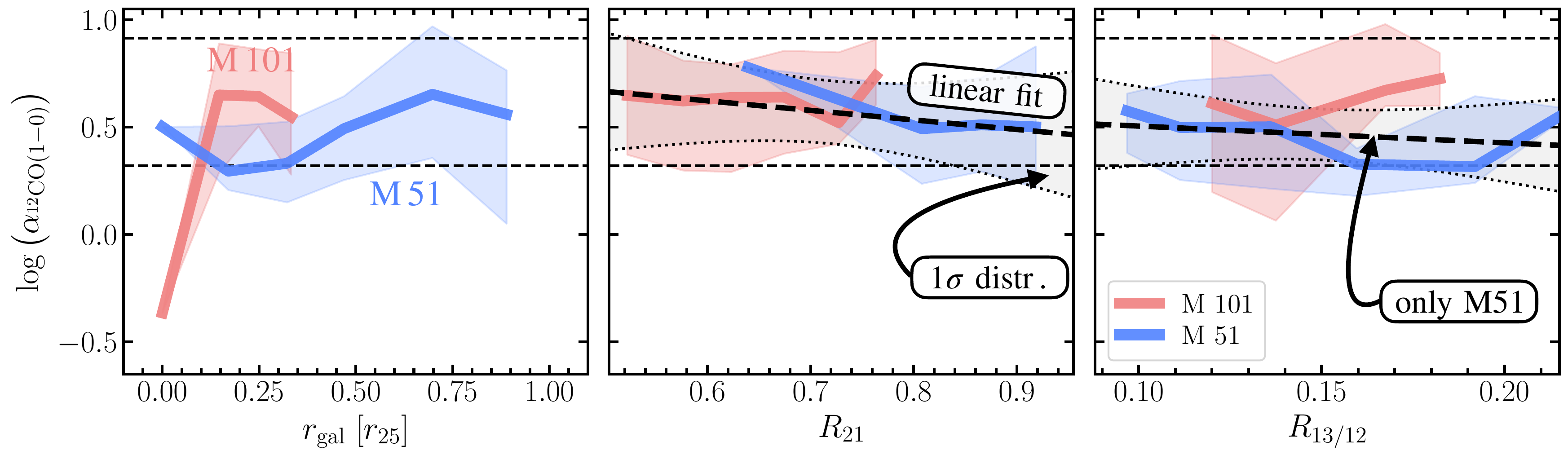}
    \caption{{\bf Comparing \aco\ Trends in M101 and M51.} Comparison of radial \aco\ trend in M101(red) and M51(blue). Each panel shows the trend line for both galaxies separately. The trend is determined by binning the \aco\ values of the individual line of sights. The two dashed lines show the 16$^{\rm th}$ and 84$^{\rm th}$ percentile distribution of the combined M51 and M101 \aco\ dataset. The shaded region around the trend line shows the respective 1$\sigma$ scatter of the respective trend line fit. (\textit{Left}) The trend of \aco\ with galactocentric radius normalized by $r_{25}$.  (\textit{Center}) $^{12}$CO line ratio $R_{21}$ correlation. The thick-dashed line shows a linear regression to the stacked data points (combining M51 and M101). The grey-shaded region between the dotted curves indicates the 1$\sigma$ confidence interval of the fit. (\textit{Right}) Trend with the $^{13}$CO-to-$^{12}$CO line ratio, $R_{13/12}$. The linear regression only fits the trend for M51 since no clear trend is seen for M101.}
    \label{fig:M51_M101_comp}
\end{figure*}

\subsection{The \texorpdfstring{\hi}{Lg}-to-\texorpdfstring{H$_2$}{Lg} ratio}

Besides the radial trend, we check the trend of \aco\ and $\rm DGR$ with the CO-to-\hi\ intensity ratio. While we expect the molecular surface density to increase toward the center, the atomic gas surface density is expected to stay flat in the disk \citep[e.g.,][]{Casasola2017,Mok2017}. The left panel of \autoref{fig:CO_to_HI} shows the variation of \aco\ and $\rm DGR$ as function of $\log W_{\chem{^{12}CO}}/\Sigma_{\hi}$. We see that both the conversion factor and the dust-to-gas ratio remain constant across different solution pixels for  $\log \left(W_{\chem{^{12}CO}}/\Sigma_{\hi}\right)<-0.5$ (we note that we only require that the \aco\ and ${\rm DGR}$ value remain constant on the solution pixel level).  Only at $\log\left( W_{\chem{^{12}CO}}/\Sigma_{\hi}\right)>-0.5$, which corresponds to more central solution pixels, we see a systematic deviation, with 1\,dex lower \aco\ values, and an increase of ${\sim}0.5$\,dex for the $\rm DGR$. Because the parameter $\log\left( W_{\chem{^{12}CO}}/\Sigma_{\hi}\right)$ correlates with radius (as seen by the clear color gradient in the panel), we find an equivalent trend as the radial trends shown in \autoref{fig:radial_sub}.

The right panel in \autoref{fig:CO_to_HI} illustrates the radial surface density profiles of the atomic and molecular gas mass. The individual points represent the solution pixels and the colored line indicates the respective radially binned trend. We see that the atomic mass surface density, $\Sigma_{\hi}$, decreases by ${\sim}0.7$ dex toward the central kpc region of the galaxy. For the derivation of the molecular gas mass, $\Sigma_{\rm mol}$, we account here for the variation in \aco\ derived from the scatter minimization technique. We see that the molecular gas mass surface density decreases radially outward by ${\sim}1$dex from 2\,kpc to 8\,kpc. However, when accounting for \aco\ variation, we also see a depression of the molecular gas mass surface density toward the center by again ${\sim}$1\,dex. (from 2\,kpc inward to 0\,kpc). The black trend at the bottom of {right panel of \autoref{fig:CO_to_HI}} shows the molecular-to-atomic gas mass ratio. We see that the outer regions are more \hi-dominated, while H$_2$ becomes increasingly relevant radially inward up to ${\sim}2$\,kpc. Toward the center of the galaxy, the dominance of H$_2$ over \hi\ seems to decrease again.

\subsection{Comparison of \texorpdfstring{$\rm DGR$}{Lg} and \texorpdfstring{\aco}{Lg} trends in M51 and M101}
\label{sec:M51_aco}

For comparison, we investigate \aco\ trends in the nearby massive star-forming galaxy M51 (NGC 5194).
We estimate the conversion factor in M51 identically to the approach used for M101 as described in \autoref{sec:aco_presc}. We use \chem{CO}{10} data from PAWS \citep{Pety2013}, \chem{CO}{21} from CLAWS \citep{denBrokClaws}, \hi\ observations from THINGS \citep{Walter2008} and dust mass maps using the \cite{Draine2007} model. We note that we do not perform short-spacing corrections for the \hi\ data since, upon visual inspection of the spectra, we find that M51 is less affected by negative bowling issues than the M101 observations. Nonetheless, we caution that we could miss a fraction of the total flux by relying only on short-spacing correction. This would mainly lead to a systematic offset of the \aco\ values and not affect the overall trend  (as discussed in \autoref{sec:aco_datasets}).  \hyperref[fig:M51_M101_comp]{Figure~\ref*{fig:M51_M101_comp}} shows the trend of \aco\ determined from the scatter minimization technique for both M101 and M51 as a function of galactocentric radius, $R_{21}$, and $R_{13/12}$. For reference, we show the \aco\ map for M51 in Appendix \ref{M51_acostuff}. \hyperref[tab:summaryaco]{Table~\ref*{tab:summaryaco}} lists the \aco\ values for M51 using different binnings.

(i) \textit{Galactocentric Radius:} We do not find any significant trend of \aco\ with galactocentric radius in M51. Across the disk of the galaxy, we find an average value of $\langle\aco^{\rm M51}\rangle=3.7\pm{0.6} \rm M_\odot\,pc^{-2}/(\rm K\,km\,s^{-1})$.
This is slightly lower but within the scatter margin for the value found by \citet{Leroy2017}. In that study, the authors performed a slightly different version of the scatter minimization technique: they selected a range in \aco\ that yields a constant $\rm DGR$ trend across the disk of M51. With this technique, they find the following range $\aco^{\rm L17}\approx 4.5-5.0\,\rm M_\odot\,pc^{-2}/(K\,km\,s^{-1})$. 

(ii) \textit{Line Ratio $R_{21}$:} We reiterate that this particular CO ratio is of interest since it sensitive to variations in density, temperature, and the opacity of the molecular gas \citep{Penaloza2017}. As the middle panel of \autoref{fig:M51_M101_comp} shows, M51 covers higher $R_{21}$ values than M101.  Combining the sightlines from both galaxies, we cover a dynamical range in line ratio values of $R_{21}{\sim}0.5-0.9$. This range is similar to the full range of line ratio values commonly found across a sample of nearby spiral galaxies \citep{Yajima2021, Leroy2022}. Given our large uncertainty, we find no significant correlation. The fitted slope is $m=-0.5\pm2$. We note that the predicted slope value based on 3D magnetohydrodynamics galaxy-scale simulations of the cloud-scale ISM as given in \citet{Gong2020} is $-0.87$, which is within our margin of error. In conclusion, despite the range in $R_{21}$, we do not obtain strong constraints from our observations on any possible trends between the line ratio and the conversion factor. This limits the use of $R_{21}$ as a predictor of \aco\ variation for extragalactic studies on kpc scales.

(iii) \textit{Line Ratio $R_{13/12}$:} We reiterate that, assuming optically thin \chem{^{13}CO}{10} emission, $R_{13/12}$ traces a combination of the \chem{^{12}CO} optical depth and abundance variations of the \chem{^{13}CO} species (see \autoref{sec:disc_isoratio}). We expect an optically thin CO line to result in lower \aco\ values \citep{Bolatto2013}. Consequently,  if $R_{13/12}$ is driven by opacity changes, we expect lower line ratios to have low \aco. In \autoref{fig:M51_M101_comp}, we only perform a linear fit to the trend of M51 since we cover a wider range of line ratios for that galaxy ($R_{13/12}{\sim}0.1-0.2$). However, given the uncertainties of our \aco\ values, we do not find any significant trend with $R_{13/12}$ in M51. 

\begin{figure}
    \centering
    \includegraphics[width=0.9\columnwidth]{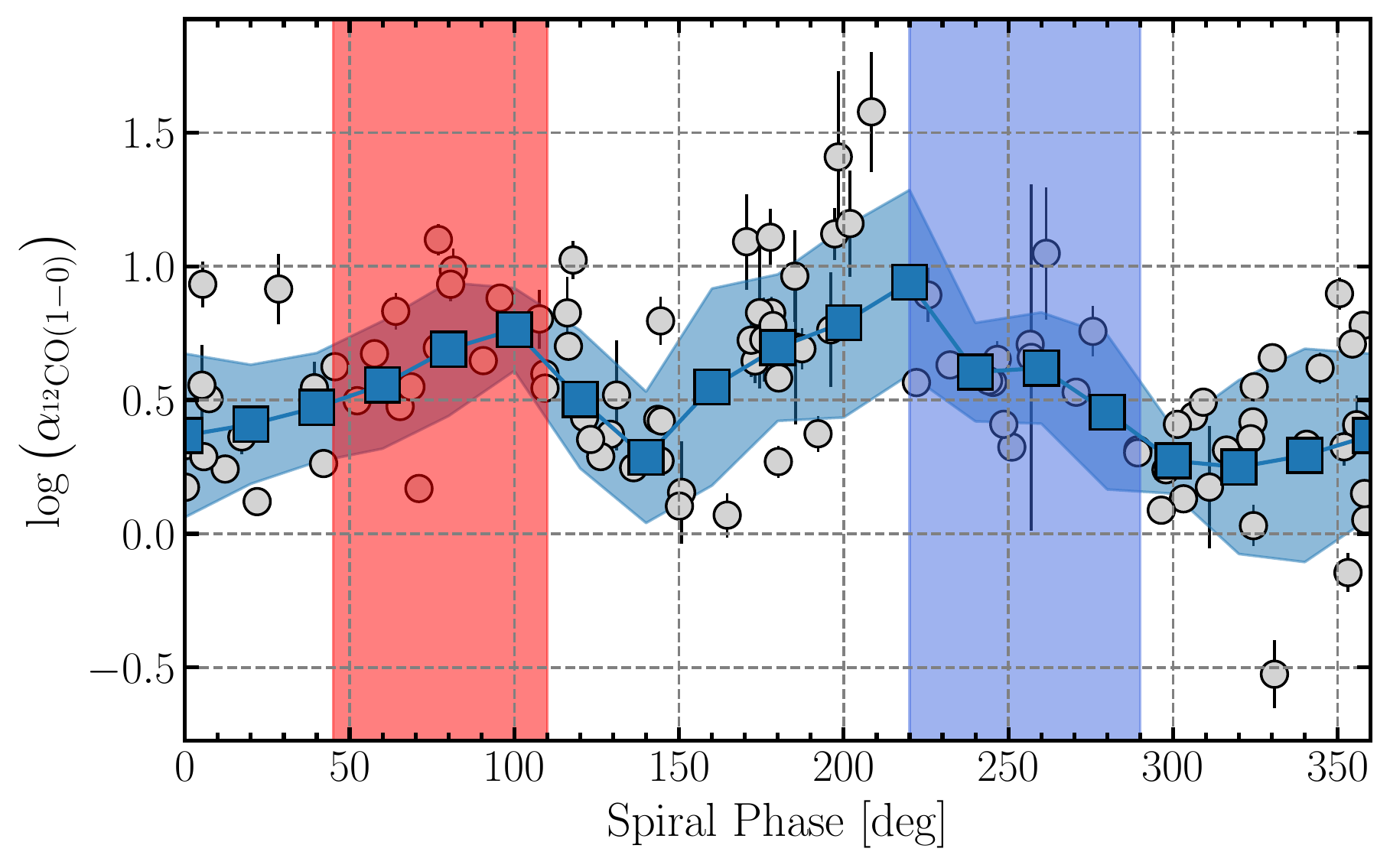}
    \caption{{\bf Arm-interarm variation of \aco in M51} The graph shows the \chem{^{12}CO}{10} intensity binned by spiral phase. The blue points show the stacked line ratio by spiral phases in steps of 20$^\circ$, increasing counter-clockwise. {The vertical red and blue shaded regions show the extent of spiral phases that correspond to the northern (red) and southern (blue) spiral arm of the galaxy.}}
    \label{fig:arm_interarm_m51}
\end{figure}

\begin{figure*}
    \centering
    \includegraphics[width = 0.95\textwidth]{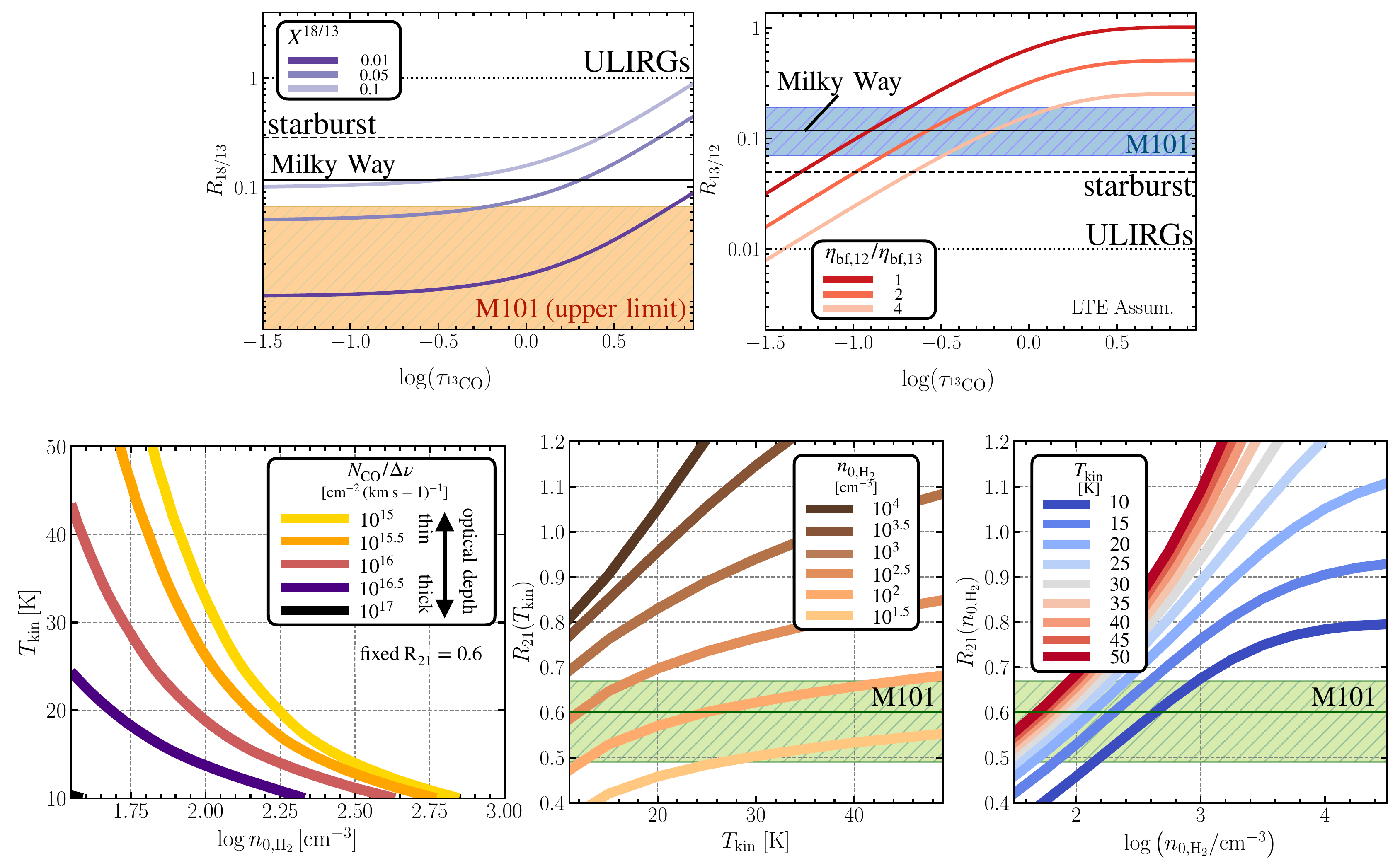}
    \caption{{\bf Effect of \chem{^{13}CO} optical depth on  CO line ratios under LTE (top) and non-LTE (bottom) conditions.} We assume optically thick \chem{^{12}CO} and optically thin \chem{C^{18}O} emission. (\textit{Top left}) Variation of $R_{18/13}$ with $\tau_{\chem{^{13}CO}}$ for different abundance ratios of the \chem{^{13}CO} and \chem{C^{18}O} species. We do not significantly detect \chem{C^{18}O} in the center of the galaxy. The orange shaded region hence shows the region where the line ratio in M101's central 4\,kpc region could fall.  Average MW value (solid line; \citealt{Wouterloot2008}), starburst (dashed line; e.g. \citealt{Tan2011}), and ULIRGs (dotted; e.g., \citealt{Greve2009}) is shown. (\textit{Top right}) Variation of $R_{13/12}$ with $\tau_{\chem{^{13}CO}}$ for different beam filling factor ratios between the \chem{^{12}CO} and \chem{^{13}CO} emission, $\eta_{12}$ and $\eta_{13}$. The blue band shows the range of measured $R_{13/12}$ values in M101.  Average MW value from \citet{RomanDuval2016}, starburst and ULIRGs from \citealt{Sliwa2017b}. (\textit{Bottom panels}) We use the line ratio model calculation provided in \citet{Leroy2022}. These are based on model calculations with RADEX \citep{vanTak2007} and used a lognormal density distributions described in \citet{Leroy2017_density}. (\textit{Bottom left}) We fix the ratio at $R_{21}=0.6$ and use a lognormal density distribution width of $\sigma = 0.6$. The color-coded lines show as a function of H$_2$ volume density ($x$-axis) and kinetic temperature $T_{\rm kin}$ ($y$-axis) the trends for different total CO column densities per line width ($N_{\rm CO}/\Delta\nu$), which roughly correlate  with the optical depth. {(\textit{Bottom center}) The line ratio $R_{21}$ as a function of kinetic temperature for different mean molecular gas densities. (\textit{Bottom right}) The line ratio $R_{21}$ as a function of the mean molecular gas density for different kinetic temperatures. The green shaded region shows the 16$^{\rm th}$ to 84$^{\rm th}$ percentile range for the values found in M101. For the computation of $R_{21}$ as function of temperature and density, we fix $\sigma=0.6\,\rm dex$ and $N_{\rm CO}/\Delta v=10^{16}\,\rm cm^{-2}(km\,s^{-1})^{-1}$. }   We note that we assume a common excitation (density, temperature) for all species and LTE for the top panels.  }
    \label{fig:ratio_analysis}
\end{figure*}

(iv) \textit{Arm-Interarm Variation}: As opposed to M101 (see \autoref{sec:arm_interarm}), we find strong arm-interarm variation in $R_{21}$ \citep{Koda2012, denBrokClaws} in M51, likely reflecting changes in the optical depth or temperature and density of the molecular gas. By decomposing our solution pixels by spiral phase, we can investigate whether \aco\ shows arm-interarm dependence in M51 as well. \hyperref[fig:arm_interarm_m51]{Figure~\ref*{fig:arm_interarm_m51}} shows \aco\ binned by spiral phase (spiral phases belonging to the {northern} spiral arm are shown in red, and blue indicates the spiral phases that define the southern arm; see \citealt{denBrokClaws} for further details). We decompose the spiral bins using a logarithmic spiral to describe the shape. The technique is described in detail in \citet{Koda2012}. We find a significant variation of \aco\ as a function of the spiral phase. The conversion factor is lower in the interarm than the spiral arm region by about ${\sim}0.5$\,dex.
The arm--interarm trend is consistent with the presence of a more prominent diffuse CO component in the interarm region that enhances the CO emissivity. At the same time, it would decrease the required conversion factor to translate the CO intensity to molecular gas mass. The presence of a diffuse component has previously been suggested by \citet{Pety2013}. On the basis of comparing GMC-scale (${\sim}$100\,pc) and large scale (${\sim}1$\,kpc)  observations, they suggest that  ${\sim}$50 \% of the total CO emission could originate from such a diffuse component in this galaxy. Future work using high-resolution observations of the central region of M51 will provide further insight into the mechanism that produces these strong environmental changes in the line ratio and conversion factor (S. Stuber et al., in prep.). In contrast,  we also note that we do not find any clear arm--interarm variation in \aco\ for M101.  We discuss the precise analysis to quantify the arm and interarm regions using logarithmic spirals in \autoref{sec:arm_interarm}. We note that since the spiral structure is less pronounced in M101 than in M51,  we also expect the difference in $R_{21}$ to be smaller.

%%%%%%%%%%%%%%%%%%%%%%%%%%%%%%%%%%%%%%%%%%%%%%%%%%%%%%%%%%%%%%%%%%%%%%%%%%%%%%%%%%%%%%%%%%%%%%%%%%%%

%%%%%%%%%%%%%%%%% DISCUSSION %%%%%%%%%%%%%%%%%%
\section{Discussion}
\label{sec:discuss}
\subsection{Implications from CO isotopologue line ratio trends}
\label{sec:disc_isoratio}

Generally, CO isotopologue line ratio variation across nearby galaxies is either linked to changes in the relative abundances of the isotopologue species or variation in the physical properties of the molecular gas, such as its opacity, temperature, or density \citep[e.g.,][]{Davis2014}. Since the  \chem{^{13}CO}{10} and \chem{C^{18}O}{10} transitions are generally optically thin \citep[see review by ][]{Heyer2015}, they help us to constrain any potential changes in the relative abundances. In more detail, CO isotopologue line ratio variation can generally be explained by the following factors:

\textit{(i) Changes in CO isotopologue abundances}:
Processes that vary the CO isotopologue abundances can be selective nucleosynthesis \citep{Sage1991, Wilson1992}, chemical fractionation \citep{Watson1976, Keene1998} or selective photodissociation \citep{Dishoeck1988}. These three mechanisms either locally enhance the \chem{^{13}CO} abundance (chemical fractionation), increase the \chem{^{12}C} and \chem{^{18}O} isotope abundances (selective nucleosynthesis), or lead to more photodissociation of certain species due to lower shielding and differences in molecular structure (selective photodissociation). Line ratio trends then give us insight into whether any of these mechanisms act as global drivers and, more importantly, whether abundance variations can explain observed CO isotopologue line ratio trends in the first place. The top left panel in \autoref{fig:ratio_analysis} illustrates the effect of relative abundance variations of \chem{^{13}CO} and \chem{C^{18}O} on the observed line ratio. 

\textit{(ii) Optical Depth effects:} 
Because, in particular, the \chem{^{12}CO} emission, and potentially the \chem{^{13}CO} emission is optically thick, changes in the optical depth will then lead to a variation of the observed line ratio. Due to sufficiently low abundance, \chem{C^{18}O} generally remains optically thin. This way, it is possible to assess the optical depth variation of  \chem{^{12}CO} and \chem{^{13}CO}.

In \autoref{fig:ratio_analysis} we show the expected $R_{13/12}$ and $R_{18/13}$ trends with changing \chem{^{13}CO}{10} optical depth, $\tau_{\chem{^{13}CO}}$, under LTE assumption. Since we only derive upper limits for $R_{18/13}$, the top left panel in \autoref{fig:ratio_analysis} highlights the possible line ratio values up to the upper limit. For optically thin \chem{^{13}CO} emission, $R_{18/13}$ traces the abundance ratio between these two CO isotopologues, $X^{18/13}$. Our upper limit of the line ratio suggests an upper limit of the abundance ratio of $X^{18/13}<0.06$.
The top right panel shows the variation of $R_{13/12}$ for different beam filling factor ratios for \chem{^{12}CO}{10} and \chem{^{13}CO}{10}. We note that the observed range in line ratio values found in M101 is in agreement with optically thin \chem{^{13}CO} emission (i.e., $\tau_{\chem{^{13}CO}}<1$).
In the bottom panels of \autoref{fig:ratio_analysis}, we illustrate the dependence of $R_{21}$ on the temperature (kinetic temperature; $T_{\rm kin}$) and density (collider density; $n_{0,\rm H_2}$). We use the model calculations from \citet{Leroy2022}. They employ multiphase RADEX model calculations \citep{vanTak2007} with density layers weighted by a lognormal profile and a common temperature, $T_k$, and column density per line width, $N_{\rm CO}/\Delta\nu$, \citep{Leroy2017_density}. To illustrate the trends, we fix $R_{21}=0.6$ and show the $n_{0,\rm H_2}$-to-$T_{\rm kin}$ degeneracy for different CO column densities per line width.  We expect the column density per line width to decrease toward the center. Consequently, fixing $R_{21}$, would indicate an increase of the temperature (for constant density) and higher density (for constant temperature) toward the central region of M101 (or a combination of both effects). 

\begin{figure*}
    \centering
    \includegraphics[width=0.8\textwidth]{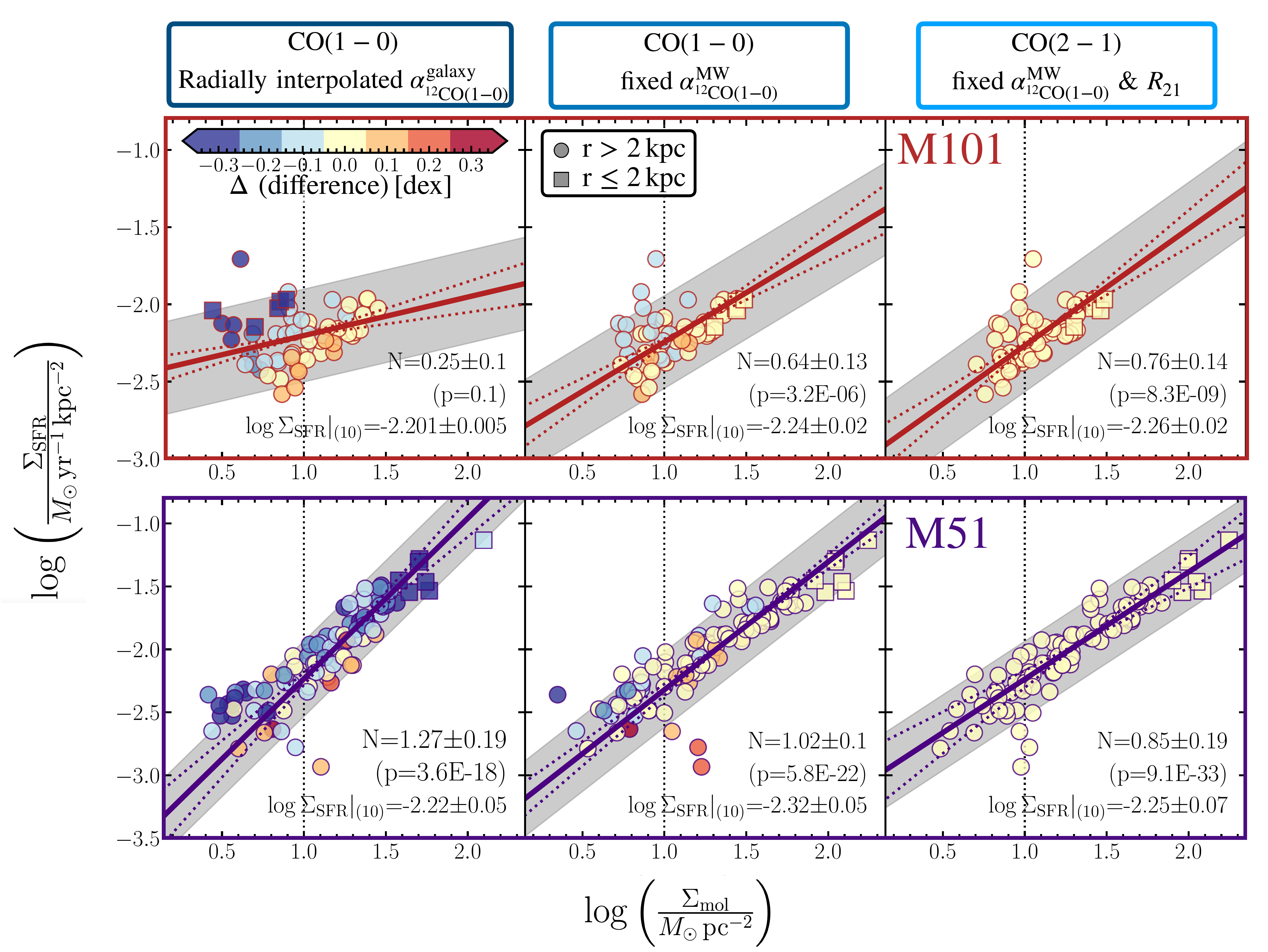}
    \caption{{\bf Implication of varying $R_{21}$ and \aco\ on the Kennicutt-Schmidt relation} We compare the KS relation using three different ways of estimating the molecular gas mass, $\Sigma_{\rm mol}$: (i) \chem{^{12}CO}{10} and a radially interpolated \aco\ derived for each galaxy (ii) \chem{^{12}CO}{10} and a fixed $\aco^{\rm MW}=4.3\,M_\odot\,\rm pc^{-2}/(K\,km\,s^{-1})$, and (iii) \chem{^{12}CO}{21} and a fixed $R_{21}=0.6$ (and $R_{21}=0.9$ in M51; \citealt{denBrokClaws}). The resulting KS index, $N$, is determined using an orthogonal distance regression (fit indicated by solid line) and indicated in each panel with its 1$\sigma$ uncertainty. In addition, each panel also lists the $\Sigma_{\rm SFR}$ value (in log) at $\Sigma_{\rm mol}=10\,M_\odot\,\rm pc^{-2}$, as derived from the linear fit. The KS coefficient, $N$, the Pearson's correlation coefficient, $p$, and as the SFR surface density at a molecular gas mass surface density of 10\,$M_\odot\,\rm pc^{-2}$ or the linear relationship are indicated in each panel as well. The grey shaded region shows the ${\pm}0.30$\,dex dispersion which is expected for a larger sample of galaxies \citep{Bigiel2008, Leroy2013}. The points, which represent the individual solution pixels, are color-coded by their difference (in dex) in $\Sigma_{\rm mol}$ to the molecular gas mass determined using method (iii), which is shown in the right panel. Squares indicate the central points ($r\le2\,\rm kpc$).}
    \label{fig:KS_law}
\end{figure*}

Given optically thin \chem{^{13}CO}, the negative trend we find in $R_{13/12}$ with galactocentric radius either derives from changes in the optical depth of \chem{^{12}CO} or changes in the abundance ratio of [\chem{^{12}CO}/\chem{^{13}CO}] (or a combination of these two factors). An increase of the $X^{13/12}$ abundance ratio toward the center would be consistent with such observed trends in the Milky Way \citep{Milam2005}. Such trends can be explained by selective nucleosynthesis: {i}nside-out star formation scenarios \citep{Tang2019} lead to an increased accumulation of \chem{^{13}CO} sooner towards the center of the galaxy, thus enhancing there the $R_{13/12}$ ratio. Such a scenario is also supported by the increase of $R_{13/12}$ with the star formation rate surface density (see the bottom right panel in \autoref{fig:ratiomaps101}). We note, however, that it remains an open question how precisely the star formation history connects to the abundance of (molecular) gas we observe. To address the connection in the case of M101, higher-resolution observations of the chemical abundance variation on the molecular cloud scale are necessary.

Alternatively, the optical depth of \chem{^{12}CO} is expected to decrease in the presence of diffuse emission or increased turbulence. This would boost the emission of \chem{^{12}CO} relative to \chem{^{13}CO} and lead to a decreasing $R_{13/12}$. In contrast, the optical depth could also increase if the column density increases towards the center of the galaxy. If changes in the optical depth were the main driver for line ratio variation, the increasing trend of $R_{13/12}$ toward the center of M101 would indicate higher optical depth in the center. Given our nearly flat $R_{21}$, higher optical depths would mean less dense or colder {molecular} gas (as can be seen in \autoref{fig:ratio_analysis}). We hence conclude that particularly the trend in $R_{13/12}$ is in part due to changes in the relative abundance of \chem{^{13}CO}. However, for future work to properly disentangle the contribution of abundance variations and optical depth changes to the line ratio, at least another \chem{^{13}CO} rotational $J$ transition is required to perform a non-LTE modeling analysis \citep[e.g.,][]{Teng2022}.

\subsection{Implications of \texorpdfstring{\aco}{Lg} variation on scaling relations}

The Kennicutt-Schmidt (KS) relation \citep{Schmidt1959, Kennicutt1989} links the star formation surface density and the {total} gas surface density, and its slope likely reflects the underlying processes of star formation \citep{Elmegreen2002, Krumholz2005}. The molecular KS law {(which relates the SFR surface density to only the molecular gas surface density)} follows:
\begin{equation}
    \log\left(\frac{\Sigma_{\rm SFR}}{M_\odot\,\rm yr^{-1}\,kpc^{-2}}\right) = N\times \log\left(\frac{\Sigma_{\rm mol}}{M_\odot\,\rm pc^{-2}}\right)+C
\end{equation}
where $N$ indicates the KS slope and $C$ the normalization offset.

\begin{figure*}
    \centering
    \includegraphics[width=0.7\textwidth]{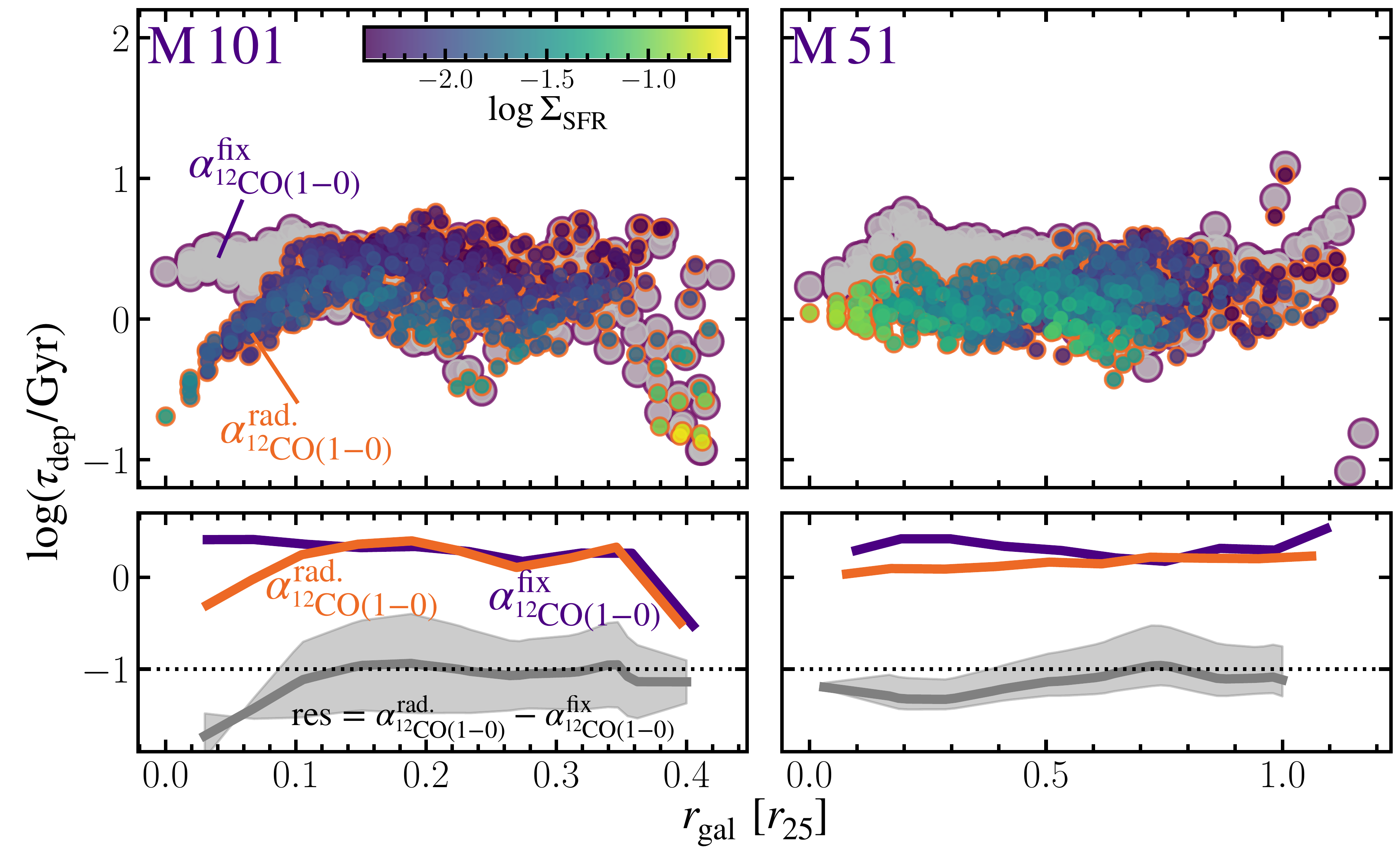}
    \caption{{\bf Implication of varying $R_{21}$ and \aco\ on the molecular gas depletion time} The top panels show the radial trend of the depletion time for M101 (left),  M51 (center), and for the combination of the sightlines from both galaxies (right). The depletion time depends on a measurement of the molecular gas mass. The grey points (with purple edge color) show the measurements using CO(2-1) and fixed $R_{21}$ and \aco\ values. The points color-coded by SFR surface density are from using CO(1-0) and a radially interpolated \aco. The bottom panels compare the radially binned trends of both measurements. The grey line indicates the residual between the two trend lines.}
    \label{fig:t_dep}
\end{figure*}
 
When estimating the KS relation from observations, the molecular gas surface density is most often inferred from CO observations. CO (2-1) and CO (1-0) are often used for this purpose, and translating from CO intensity to molecular gas mass surface density requires adopting some CO-to-H$_2$ conversion factor. This means that for the increasingly common set of high-quality CO (2-1) mapping and integrated data \citep[e.g.,][]{Cicone2017,Noble2019,Pereira2021,Leroy2021_PHANGS}, both the line ratio and adopted \chem{^{12}CO}{10}-to-H$_2$ conversion factor can affect the derived surface density, scaling relation, and {molecular} gas depletion time. In this study, we are uniquely positioned to investigate the impact  of using either a constant or varying line ratio $R_{21}$ (i.e., using either \chem{^{12}CO}{21} and a line ratio, or \chem{^{12}CO}{10} directly) and conversion factor \aco.
 In \autoref{fig:KS_law}, we compare the different derived KS slopes, $N$, in M101 and M51 based on (i) using \chem{^{12}CO}{10} and a radially interpolated \aco\ (ii) \chem{^{12}CO}{10} and a fixed $\aco=4.3\,M_\odot\,\rm pc^{-2}/(K\,km\,s^{-1})$, and (iii) \chem{^{12}CO}{21} and a fixed $R_{21}=0.6$ (and $R_{21}=0.9$ in M51; \citealt{denBrokClaws}) and the same fixed \aco\ value as for (ii). The points, which show the solution pixels, are color-coded by the difference (in dex) compared to the molecular gas mass derived from \chem{^{12}CO}{21} according to (iii).

Each panel also indicates the KS index, $N$ (including the 1$\sigma$ uncertainty shaded in gray). The index is determined using an orthogonal distance regression, which is more robust than the more commonly used linear regression, with $x$ and $y$ as observables with associated uncertainties. The right-most panel shows the relation based on method (iii). We find an index $N<1$ for both galaxies. While the KS relation predicts a close relation between $\Sigma_{\rm SFR}$ and $\Sigma_{\rm mol}$, an overall dispersion from this relation is expected \citep[${\pm}0.30$\,dex;][]{Bigiel2008, Leroy2013} and connected to physical drivers. Regarding free (left and central panels) and fixed (right panels) $R_{21}$, we find a decrease for M101 ($N=0.76$ to $N=0.64$), but an increase in $N$ for M51 ($N=0.85$ to $N=1.02$). We generally expect an increase of the index since $R_{21}$ becomes larger towards the center, leading to an overestimation of the molecular gas mass \citep[${\sim}$ difference of 10-20\%;][]{Yajima2021, denbrok2021, Leroy2022}. We see that in M101, the points with higher $\Sigma_{\rm SFR}$ (connected to the center) also show a negative (blue) difference in molecular mass. 
The impact of $R_{21}$ is limited by its factor of 2 variation because it saturates at 1 and is limited to 0.4 by typical excitation condition \citep[see also. e.g., ][]{Yajima2021, Leroy2022}.

We expect the impact by varying \aco\ to be larger since we observe a variation of the conversion factor of a factor of $5{-}10$. When estimating the molecular gas mass using the radially interpolated \aco\ value (left column in \autoref{fig:KS_law}), we see a further increase in $N$ for M51 and a decrease for M101 compared to using a fixed \aco. Since values with higher surface density seem to show a depression of \aco\ (e.g., the center), we expect that points with higher $\Sigma_{\rm SFR}$ have overestimated molecular gas masses. Correcting this effect by accounting for \aco\ variation will push these points to lower $\Sigma_{\rm mol}$. This can be seen in M101, where the squared points (for which $r\le2\,\rm kpc$) are bluer than for fixed \aco. The other high-SFR points stem from the bright H\textsc{ii} region toward the southeast. 
We overestimate $\Sigma_{\rm mol}$ when not accounting for varying $R_{21}$ and \aco (in particular for the center). We hence expect to find a larger KS slope than if we were accounting for varying \aco\ and $R_{21}$. 
However, we find that a simple KS parameterization is breaking in M101 when accounting for the variation in $R_{21}$ and \aco (left panels of \autoref{fig:KS_law}). This is because (a) the dynamical range in the SFR surface density is small (${\sim}0.5$\, dex) and (b) the points from the center of the galaxy are pushed off the linear relation (c) the \hii\ region (NGC\,5461) at high-SFR shows points offset from the main relation. Hence, a linear fit does not capture the entire trend of the relation. 

\begin{figure*}
    \centering
    \includegraphics[width=\textwidth]{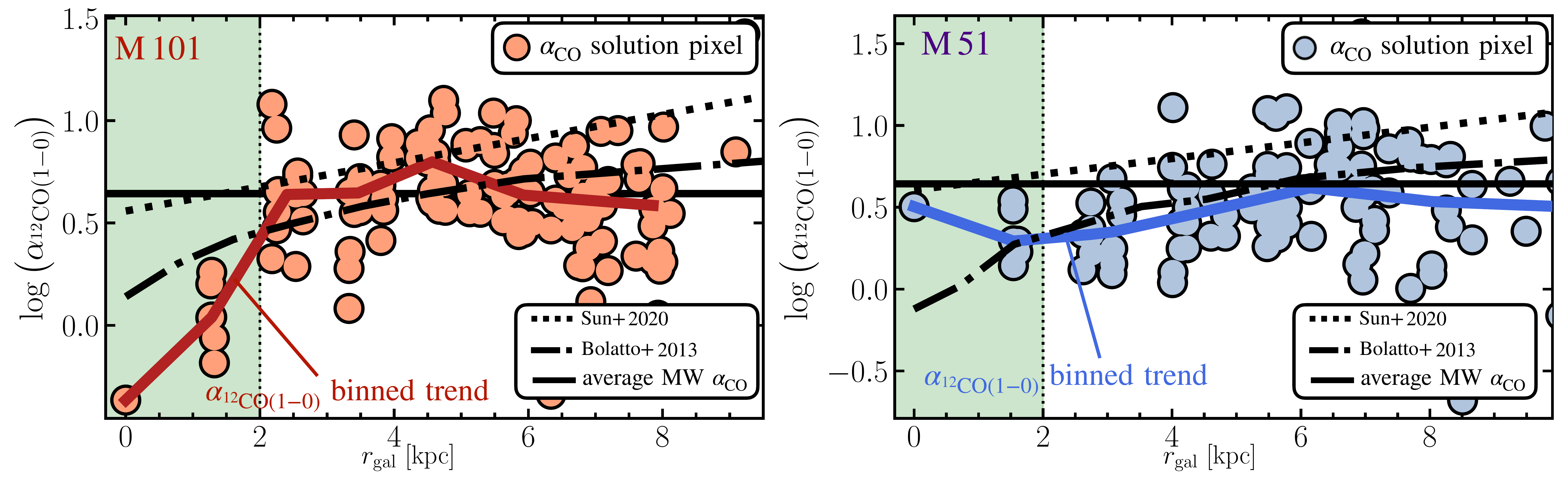}
    \caption{{\bf CO(1-0)-to-H$_2$ conversion factor prescription comparison}  Comparing the radial \aco\ trend derived from binning the data (in color) and the trend derived from applying different prescriptions. The trend for M101 is shown in the left and for M51 on the right.  Comparison of the derived trend when applying the \aco\ prescription used \citet{Sun2022} (dotted line), from \citet{Bolatto2013} (dash-dotted line). The black horizontal line illustrates the average local solar neighborhood \aco\ value. The green-shaded region is an approximate illustration of the central region of the galaxy, where conditions potentially change dramatically to the overall disk.}
    \label{fig:model_comp}
\end{figure*}

Furthermore, to study environmental variation, we investigate how the molecular gas depletion time $\tau_{\rm dep}$ varies as a function of radius across both galaxies. The depletion time is defined as follows:
\begin{equation}
    \tau_{\rm dep}\equiv \frac{\Sigma_{\rm mol}}{\Sigma_{\rm SFR}}
\end{equation}
\hyperref[fig:t_dep]{Figure~\ref*{fig:t_dep}} shows the radial trend of $\tau_{\rm dep}$ for either fixed $R_{21}$ and \aco, or a radially interpolated \aco. Both galaxies show relatively flat depletion times in their discs, independent of the $\Sigma_{\rm mol}$ method. This is in agreement with previous studies of resolved nearby galaxies that also found a relatively narrow distribution of molecular gas depletion time  \citep[e.g.][]{Leroy2008,Bigiel2011,Leroy2013}. Using radial \aco\ and CO(1-0), we find an average depletion time of $\tau_{\rm dep}^{M101}=1.9$\,Gyr and $\tau_{\rm dep}^{M51}=1.5$\,Gyr with a scatter of 0.2 dex for both galaxies.  
We also note that the value for the average depletion time in both galaxies agrees well with the value of 1.6~Gyr found for M51 by \citet{Leroy2017} using PAWS CO(1-0) data and is well within the range expected from varying the choice of SFR tracer. Using the constant \aco\ and $R_{21}$, we find a slightly higher depletion time in M51 by about 500 Myr (which constitutes a ${\sim}$50\% increase). 
Such constant depletion times are thought to represent evidence that averaging over many GMCs on kpc-scales in the disks of nearby spiral galaxies the SFR per unit gas mass is relatively constant \citep[e.g., ][]{Leroy2008,Bigiel2011}. 

While M51 shows a narrow range of depletion time across the galaxy, independent of the method used, we see an apparent decrease of $\tau_{\rm dep}$ toward the center of M101 when we use a radially interpolated \aco. In the center, the constant trend seems to break, and we find $\tau_{\rm dep}\approx 150-300$ Myr, which is an order of magnitude lower than the disk-wide average. Furthermore, M101 shows lower depletion times (again by almost one order of magnitude) in the bright H\textsc{ii} region toward the southeast of the galaxy (NGC 5462). Sightlines within this region incidentally show also high SFR surface densities. 
\citet{Utomo2017}, studying galaxies from the EDGE-CALIFA survey, and \citet{Leroy2013} investigating resolved observations of the HERACLES sample, also found decreased depletion times toward the center of galaxies and suggested that when accounting for \aco\ variation, the extent of the drop in $\tau_{\rm dep}$ will be amplified. A possible explanation for the central depression could be low crossing and free-fall times of the clouds in the central region as opposed to  clouds in the galactic disks \citep[e.g.,][]{Leroy2015}.

\subsection{Comparing prescriptions of \texorpdfstring{\aco}{Lg}}

We compare commonly used \aco-prescriptions to our estimate from the scatter minimization technique in M101 and M51. Such prescriptions are primarily based on first order on the metallicity, $Z$ \citep[e.g.][]{Schruba2012, Accurso2017}, but also incorporate further key parameters such as the surface brightness \citep[e.g.][]{Bolatto2013} or the CO line ratio, $R_{21}$ \citep[e.g.][]{Gong2020}, {which traces to some degree also the temperature dependence of the conversion factor}. In particular, we test the prescription from \citet{Bolatto2013} and prescription from \citet{Accurso2017} (which describes \aco\ in terms of a power-law scaling with metallicity, similar to \citealt{Schruba2012, Amorin2016}). The panels of \autoref{fig:model_comp} show a comparison of the radial trends of \aco\ for the different prescriptions and the trend based on the scatter minimization technique (shown in color) for M101 (left) and M51 (right). 
For M101, in the disk, the prescription based on metallicity alone (dotted line; \citealt{Sun2020}, which is based on the prescription by \citealt{Accurso2017}) describes the range of \aco\ approximately well in M101 and is slightly offset in M51 toward larger values (by about 0.3 dex). However, this prescription does not predict the depression of \aco\ toward the center of the galaxy in M101. In contrast, we see that the prescription by \citet{Bolatto2013}, which accounts for regions with high total mass surface density, also describes a mild depression of \aco\ toward the center of both galaxies. But the extent of the decrease is only ${\sim}0.5$~dex with respect to the average disk value, and not ${\sim}$1 dex, as we see in M101. The prescription by \citet{Bolatto2013} also finds a decreasing trend toward the centre in M51 (by ${\sim}0.5$ dex). This finding suggests that the reason is linked to the peculiarity mentioned above of M51, such as the AGN in the center or its strong interaction with the companion galaxy NGC\,5195.

%%%%%%%%%%%%%%%%%%%%%%%%%%%%%%%%%%%%%%%%%%%%%%%%%%%%%%%%%%%%%%%%%%%%%%%%%%%%%%%%%%%%%%%%%%%%%%%%%%%%

%%%%%%%%%%%%%%%%% CONCLUSION %%%%%%%%%%%%%%%%%%
\section{Conclusions}
\label{sec:conclusion}
This study presents new wide-field IRAM \mbox{30m} low-$J$ CO observations of M101. We address two key aspects of studies of the molecular gas physics in the galaxies M101 and M51: i) How well do CO isotopologue line emissions capture changes in the molecular gas  characteristics, and ii) how does \aco\ vary with environmental parameters across the galaxy.

\noindent Based on our CO isotopologue analysis, we find:
\begin{enumerate}
    \item An average line ratio of $\langle R_{21}\rangle=0.60^{+0.07}_{-0.11}$, which is consistent with previous studies of similar, nearby star-forming galaxies. The ratio stays predominantly flat across the disk of M101, with only a mild increase of 10\% towards the central $1.5\,\rm kpc$ region.
    \item We also detect resolved \chem{^{13}CO}{10} emission across the center, bar end, and spiral arm of M101. We find an average \chem{^{13}CO}/\chem{^{12}CO}{10} line ratio of $\langle R_{13/12}\rangle = 0.11^{+0.03}_{-0.02}$, which suggests optically thin  \chem{^{13}CO} emission throughout the galaxy.
    \item  Using spectral stacking, we can constrain an upper limit for $R_{18/13}<0.07$ for the central 4\,kpc region (by radius). Such low line ratios are more predominantly found in the outskirts of star-forming galaxies and indicate very low relative abundances of the \chem{C^{18}O} species.
    \item Given the observed trend in $R_{13/12}$, which increases toward the center, we  conclude that changes in abundances due to nucleosynthesis are a major driver on galaxy-wide scales. Changes in the opacity of \chem{^{12}CO} do not seem to be the primary driver, since the optical depth generally decreases toward the center, which would result in an opposite $R_{13/12}$ trend.
    
    \end{enumerate}

\noindent Besides this in-depth analysis of the 3mm CO isotopologue line ratios, we investigate the variation of the CO-to-H$_2$ conversion factor, \aco\, across M101. We use a modified version of the scatter minimization technique. The method is based on the dust mass approach, and we use both the \chem{^{12}CO}{10} and \trans{21} emission lines to estimate $\alpha_{\rm CO(1-0)}$. Our main results and conclusion can be summarized as follows:\\

\begin{enumerate}\addtocounter{enumi}{4}

\item We find an average conversion factor of $\acoResSpec $ across the disk of galaxy M101, with an apparent decrease of the value towards the galaxy's center by a factor of ${\sim}$10. The reduction of the conversion factor towards the center of the galaxy follows the qualitative expectation that the turbulence increases, hence decreasing the optical depth, which enhances the $^{12}$CO emission. We note however, that such conditions are usually more expected in the starburst regime than in the center of regular disk galaxies.  For comparison, we also perform a scatter minimization approach in M51. We find a relatively flat \aco\ trend in M51 across the disk and center of the galaxy. 
\item Using the optically thin \chem{^{13}CO} emission, we perform an LTE-based \aco\ estimation in M101. Generally, the conversion factor determined using this approach is lower by a factor $2{-}3$ compared to the scatter minimization technique. The discrepancy is likely due to the simplifying assumption of a similar beam-filling factor of the two lines, using constant excitation temperature and a change in the relative abundance of the species. In general, the variation of all three assumptions is plausible. As a consistency check, we introduce a two-component model of a galaxy and change the conditions in the center and disk. We derive a depression of $\aco^{\rm LTE}$ of similar order as the scatter minimization derived \aco, showing that the depression is plausible under the set of assumptions.
\item Compared to the results one would obtain assuming a fixed $\alpha_{\rm ^{12}CO (2-1)}$, one significantly overestimates the molecular gas mass, particularly in the center of galaxies. We find that, in particular, for M101, the simple linear Kennicutt Schmidt relation breaks if accounting for variable \aco. In addition, we find that the molecular gas depletion time seems to be significantly overestimated in the center by ${\sim}1$ dex with respect to depletion time of 1.9\,Gyr across the disk, if not accounting for \aco\ variation. In contrast, M51 shows a depletion time of 1.5\,Gyr, without a radial trend. 
\item Finding a prescription for \aco\ on kpc-scale remains challenging. For M101 and M51, commonly used prescriptions yield estimates of the conversion factor for the central kpc-regions that are off by $>0.5$~dex. While M101 shows a stronger depression of \aco\ than predicted, M51 does not show any decrease toward the center, despite being predicted by the prescription. 
\end{enumerate}

\noindent Overall, our results shed new light on the degree of variation of \aco\ and the corresponding trends with key galactic properties. In particular, we stress that the points near the galaxy centers need to be treated with care when employing commonly used \aco\ prescriptions, as the depression of the central value is not yet fully captured or understood. With future higher-resolution CO isotopologue observations of molecular clouds in the center and disk of nearby star-forming galaxies, it will become possible to address the physical conditions of the molecular gas that can explain the depression in \aco.

%%%%%%%%%%%%%%%%%%%%%%%%%%%%%%%%%%%%%%%%%%%%%%%%%%%%%%%%%%%%%%%%%%%%%%%%%%%%%%%%%%%%%%%%%%%%%%%%%%%%

%%%%%%%%%%%%%%%%% ACKNOWLEDGEMENTS %%%%%%%%%%%%%%%%%%

\begin{acknowledgements}
      JdB, FB, JP and ATB acknowledge funding from the European Research Council (ERC) under the European Union’s Horizon 2020 research and innovation programme (grant agreement No.726384/Empire). JdB and EWK acknowledges support from the Smithsonian Institution as a Submillimeter Array (SMA) Fellow. JC acknowledges support from ERC starting grant \#851622 DustOrigin. KS was supported by NSF award 2108081. The work of AKL is partially supported by the National Science Foundation under Grants No. 1615105, 1615109, and 1653300. AU acknowledges support from the Spanish grants PGC2018-094671-B-I00, funded by MCIN/AEI/10.13039/501100011033 and by ``ERDF A way of making Europe'', and PID2019-108765GB-I00, funded by MCIN/AEI/10.13039/501100011033. ES and TGW acknowledge funding from the European Research Council (ERC) under the European Union’s Horizon 2020 research and innovation programme (grant agreement No. 694343). ER acknowledges the support of the Natural Sciences and Engineering Research Council of Canada (NSERC), funding reference number RGPIN-2022-03499. IC acknowledges the National Science and Technology Counsil for support through grants 108-2112-M-001-007-MY3 and 111-2112-M-001-038-MY3, and the Academia Sinica for Investigator Award AS-IA-109-M02. TS acknowledges funding from the European Research Council (ERC) under the European Union’s Horizon 2020 research and innovation programme (grant agreement No. 694343). MC gratefully acknowledges funding from the Deutsche Forschungsgemeinschaft (DFG) through an Emmy Noether Research Group, grant number CH2137/1-1. -- COOL Research DAO is a Decentralized Autonomous Organization supporting research in astrophysics aimed at uncovering our cosmic origins. CE acknowledges funding from the Deutsche Forschungsgemeinschaft (DFG) Sachbeihilfe, grant number BI1546/3-1. SCOG acknowledges support from the DFG via SFB 881 “The Milky Way System” (sub-projects B1, B2 and B8) and from the Heidelberg cluster of excellence EXC 2181-390900948 “STRUCTURES: A unifying approach to emergent phenomena in the physical world, mathematics, and complex data,” funded by the German Excellence Strategy. Y-HT acknowledges funding support from NRAO Student Observing Support Grant SOSPADA-012 and from the National Science Foundation (NSF) under grant No. 2108081. 
      
\end{acknowledgements}

% WARNING
%-------------------------------------------------------------------
% Please note that we have included the references to the file aa.dem in
% order to compile it, but we ask you to:
%
% - use BibTeX with the regular commands:
%   \bibliographystyle{aa} % style aa.bst
%   \bibliography{Yourfile} % your references Yourfile.bib
%
% - join the .bib files when you upload your source files
%-------------------------------------------------------------------

\bibliographystyle{aa}
\bibliography{references.bib}

\begin{appendix}
\section{Single Dish Scale Factor Estimation}
\label{sec:hi}

\begin{figure}
    \centering
    \includegraphics[width=0.9\columnwidth]{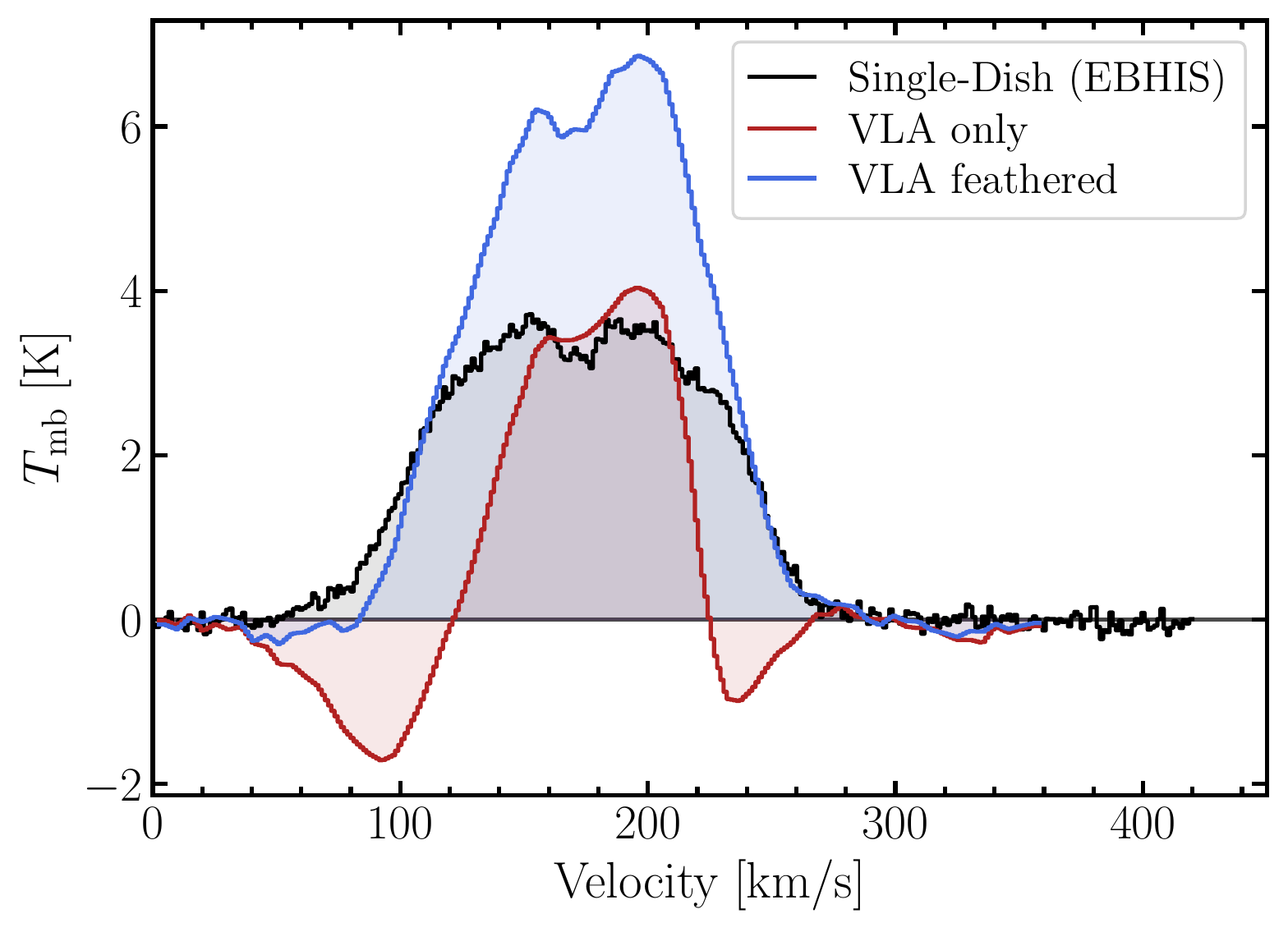}
    \caption{{\bf \hi\ Short Spacing Correction} The THINGS \hi\ data cube for M101 is seriously affected by filtering and bowling artifacts. The red spectrum illustrates the spectrum for an arbitrarily selected line of sight at 650$'$ spatial resolution (angular resolution of the Effelsberg single-dish data). The black spectrum shows the same sightline spectrum obtained from the Effelsberg single-dish data (EBHIS survey). We used the \texttt{uvcombine} package to determine a single-dish scale factor of 1.7. The blue spectrum shows the resulting feathered observation for the selected sightline.  }
    \label{fig:hi_feathere}
\end{figure}

The scatter minimization technique uses total gas mass estimates derived from dust mass measurements. With the help of atomic gas mass estimates via \hi\ observations, we can separate the total gas in to an atomic and a molecular gas component, from which we can deduce \aco. For the analysis, we use \hi\ 21cm cubes from the THINGS survey \citep{Walter2008}. The observations for M101 are, however limited by filtering issues (see \autoref{fig:hi_feathere}; in \autoref{sec:aco_datasets}, the effect when using the \hi\ data that have not been short-space corrected is computed). In order to correct these issues, we feathered the data using \hi\ observations from the Effelsberg-Bonn HI Survey \citep[EBHIS;][]{Winkel2016}.

\hyperref[fig:hi_feathere]{Figure~\ref*{fig:hi_feathere}} illustrates the need for correctly feathering the interferometric VLA data from the THINGS survey. The red spectrum indicates the VLA-only data. Clear bowling on both sides of the spectral line seriously hampers integrated intensity measurements. The black spectrum shows the single-dish data in the figure. Using the Python package \texttt{uvcombine}\footnote{\url{uvcombine.readthedocs.io}} we determine a single-dish factor of $1.7$ by comparing the flux on scales sampled in both the VLA and EBHIS data sets \citep[see Appendix A in][]{Koch2018}. We use the \texttt{casa-feather} tool to feather the data. Not correcting the VLA-only data would significantly underestimate the total \hi\ emission (total intensity lower by 70\% before feathering).

\section{Censored Line Ratio Regions}
\label{sec:CensoredReg}
As a consequence of how we have constructed the line ratio (fainter lines in the numerator), we can also estimate the censored region in the ratio plane. If we observe lines observed with different sensitivity, the noise levels will differ for each line. Since we compare lines of varying brightness, we will obtain many upper limits. We expect to obtain significantly fewer line ratios at lower values since the line in the numerator has reached the sensitivity. Larger line ratios are still possible because this can happen due to either lower line brightness in the denominator (since we have not yet reached the sensitivity limit) or larger brightness of the line in the numerator.
We bin the line ratios by a certain quantity. We then estimate the censored $1\sigma$ (or $3\sigma$) region in the following way: we divide the average rms (or $3{\times}$ this value) of the faint line per bin by the average brightness temperature of the brighter line. We reiterate that this approach is only valid when constructing the line ratio to have the fainter line in the numerator. Since rms and the line brightness vary across the survey field, we expect to find a certain number of significantly detected data points within the censored region.

\section{CO Line Stacks}
\label{app:stacking}
\begin{figure*}
    \centering
    \includegraphics[width=\textwidth]{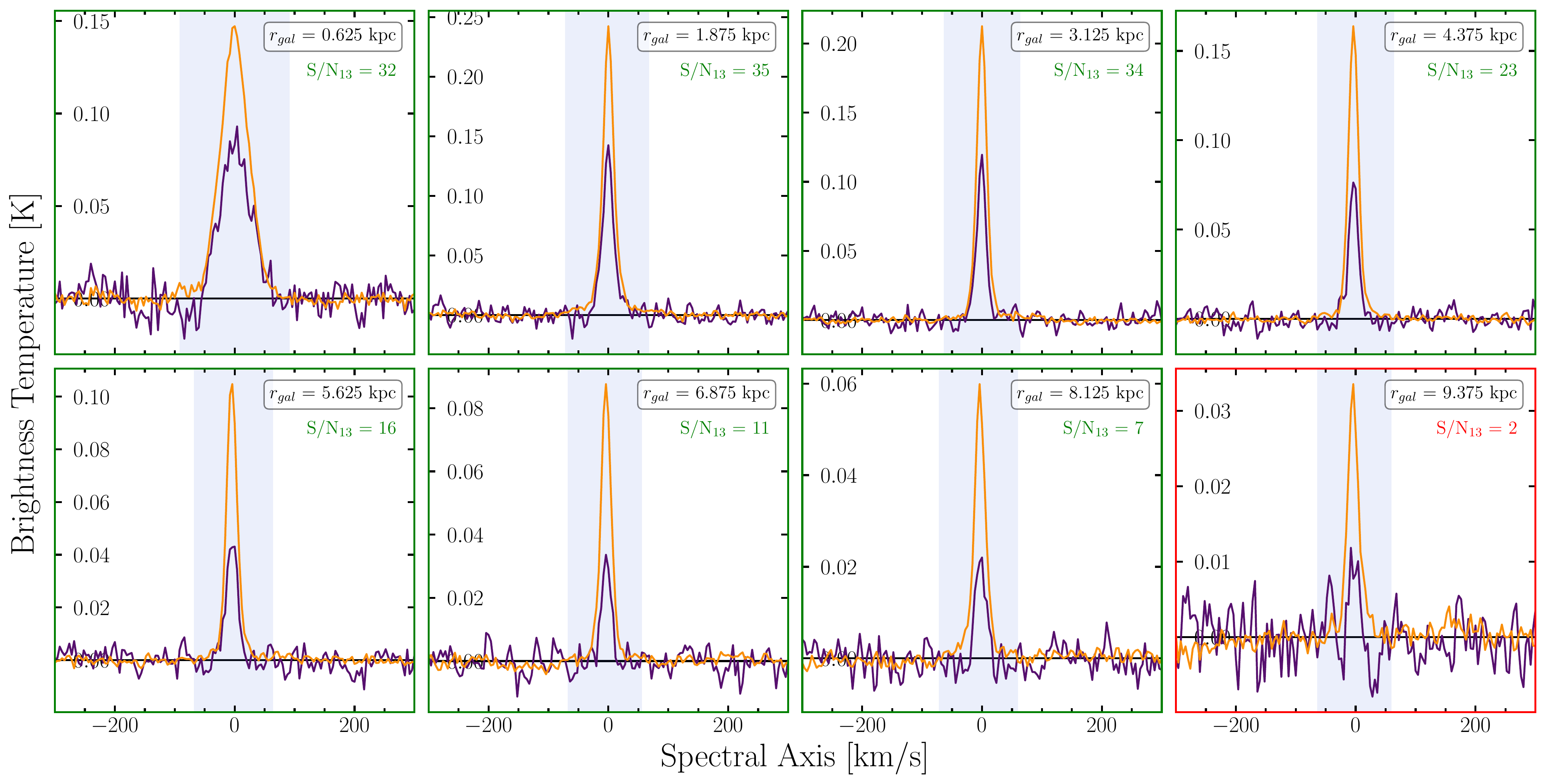}
    \caption{{\bf Radially stacked \chem{^{12}CO}{10} (orange) and \chem{^{13}CO}{10} (purple) spectra.} For a better comparison, we scale the \chem{^{13}CO}{10} brightness temperature up by a factor $5$. The S/N of the \chem{^{13}CO}{10} is indicated in each panel (green indicates spectra where ${\rm S/N_{13}>5}$. We stack in radial bins of size 1.25\,kpc. The shaded region indicates the spectral range over which we integrate the spectra. We detect significant \chem{^{13}CO}{10} emission out to $r_{\rm gal}\sim8\,{\rm kpc}$.}
    \label{fig:stacks13co}
\end{figure*}

\begin{figure*}
    \centering
    \includegraphics[width=\textwidth]{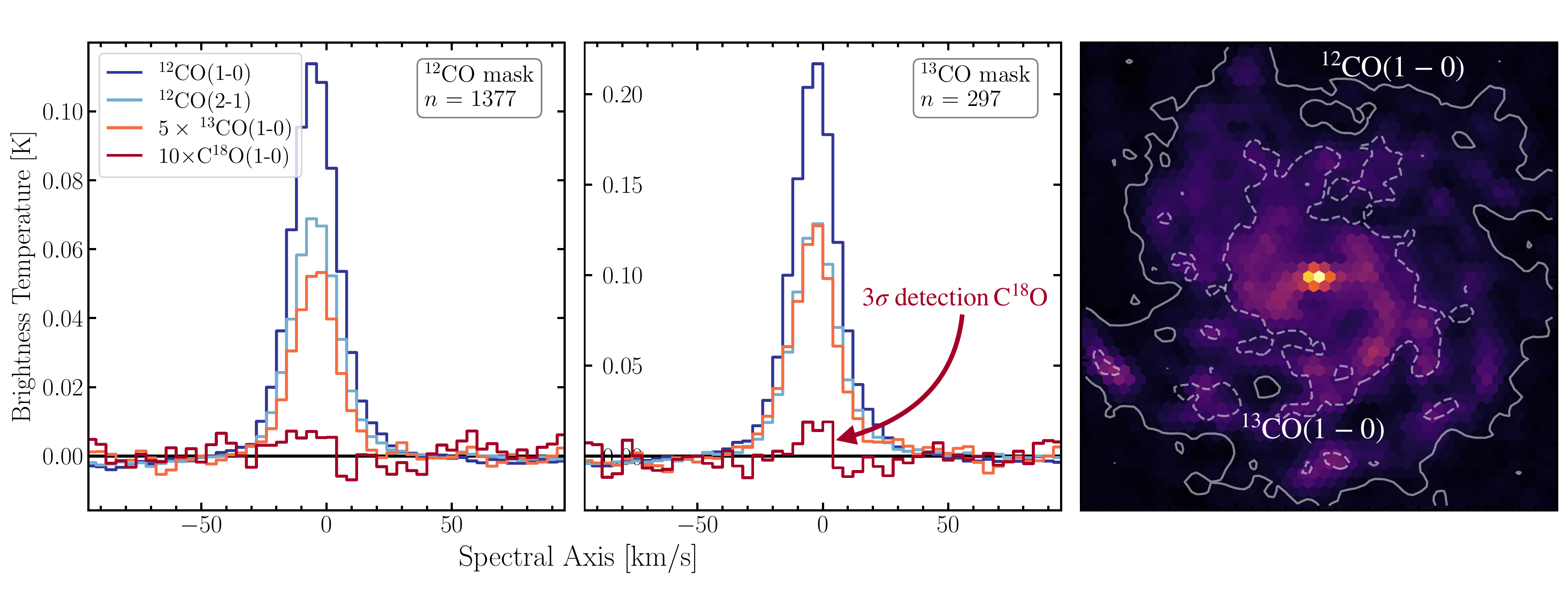}
    \caption{{\bf Stacked CO spectra over the full galaxy} (\textit{Left}) Stacked spectra over the  \chem{^{12}CO}{10} 3$\sigma$ mask. \chem{^{13}CO}{10} spectrum scaled up by a factor of 5, \chem{C^{18}O}{10} by a factor of 10. We mark the number of sightlines per mask with $n$. (\textit{Middle}) Similar to the left panel, but stacked over the  \chem{^{13}CO}{10} 3$\sigma$ mask. We do not detected significant \chem{C^{18}O}{10} emission. (\textit{Right}) moment 0 map of \chem{^{12}CO}{10}. The ${\rm S/N=3}$ contour of \chem{^{12}CO}{10} is illustrated by the solid line, while for \chem{^{13}CO}{10} it is indicated by the dashed line. }
    \label{fig:stacks18co_full}
\end{figure*}

In order to improve the $\rm S/N$ -- which allows for the detection of fainter emission lines -- we stack the spectra after binning by a certain quantity (e.g., radius, star formation rate surface density, etc.). By shifting the spectrum of each line of sight to the zero velocity, we ensure that the spectra are added coherently. In general, the combination of $N$ independent sightlines will enhance the S/N by a factor $\sqrt{N}$.

\hyperref[fig:stacks13co]{Figure~\ref*{fig:stacks13co}} shows the individual radial stacks for the \chem{^{12}CO}{10} and \chem{^{13}CO}{10} molecular transition lines. Each panel indicates the S/N ratio for the integrated \chem{^{13}CO}{10} intensity. We require a detection with $\rm S/N>{3}$ to classify it as \emph{significant}. Significant stacks are shown in green, while non-significant line detections are framed in red. We note that with the help of stacking, we do significantly detect \chem{^{13}CO}{10} out to 8\,kpc (i.e., second to last bin).

When we perform radial stacking (with a bin with of ${\sim}1.3\,\rm kpc$), we still do not recover a significant detection of \chem{C^{18}O}{10}. However, we detect significant emission in our data if we stack over a larger part of the galaxy. When we stack over the full \chem{^{12}CO}{10} mask (illustrated by the solid contour in \autoref{fig:stacks18co_full}), we do not find significant line emission. But in contrast, if we stack over the \chem{^{13}CO}{10} mask (illustrated by the dashed contour line), we detect \chem{C^{18}O}{10} emission with ${\rm S/N}=3$. This detection is valuable since it provides a constraint on the $R_{18/13}{\equiv}\chem{C^{18}O}/\chem{^{13}CO}{10}$ line intensity ratio. Since both these lines are optically thin, that particular line ratio traces the relative abundance ratio of the two CO isotopologues.

\section{Azimuthal Variation in M101}
\label{sec:arm_interarm} 

\cite{Koda2012} provide a prescription of decomposing sightlines by their corresponding spiral phase. We can bin the data using a logarithmic spiral of the following form:
\begin{equation}
    R = e^{k\times\psi}
\end{equation}
where $R$ indicates the galactocentric radius distance of a selected point, $k=\tan\left(\theta\right)$ encapsulates the galaxy's pitch angle $\theta$, and $\psi$ describes the spiral phase. For M51, we use a pitch angle of $\theta = 20^\circ$, which is close to the values found by \citet{Shetty2007} ($\theta = 21^\circ.1$) and \citet{Pineda2020} ($\theta = 18^\circ.5$).

\begin{figure*}
    \centering
    \includegraphics[width=\textwidth]{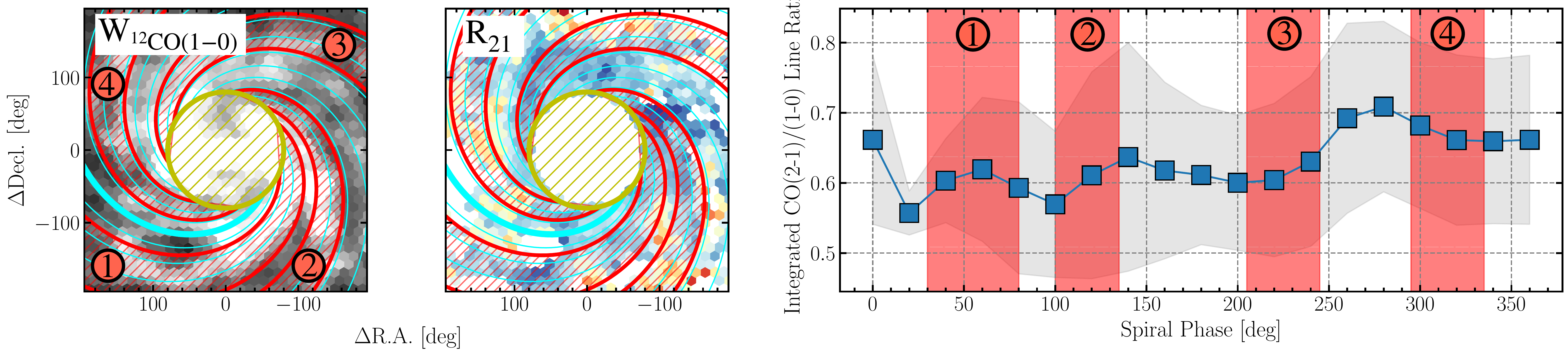}
    \caption{{\bf Azimuthal Variation of $R_{21}$ in M101.} (\textit{Left}) The \chem{^{12}CO}{10} integrated intensity map with the logarithmic spirals with a pitch angle of $\theta=23^\circ$ overlayed. The spirals increment with a spiral phase $\Delta\psi=20^\circ$ and increment in the counter-clockwise direction. The spiral bins have a with of $40^\circ$. The thick cyan line indicates $\psi=0$. The four spiral arms are accordingly labeled and colored in red. For the spiral binning, we exclude the central 80$'$ (in diameter), belonging to the central region of the galaxy (indicated by the golden-hatched region in the middle). (\textit{Center}) The map shows the $R_{21}$ variation across the galaxy. Spiral bins follow the description on the left panel. (\textit{Right}) The line ratio is binned by the spiral phase. The gray-shaded region shows the $1\sigma$ scatter per bin. The red-shaded region indicates the spiral phases of a particular spiral arm. }
    \label{fig:arm_interarm}
\end{figure*}

The spiral arms in M51 could be described using two components: a northern and a southern spiral arm (see \autoref{fig:arm_interarm_m51}). In the case of M101, however, we opted for four spiral arms. We use a pitch angle $\theta^{\rm M101}=23^\circ$ \citep{Abdeen2020}. \autoref{fig:arm_interarm} shows the spiral phases (left and central panel) as well as the decomposition of $R_{21}$ (right panel). 
We bin the data by segments that span over 40$^{\circ}$, and we increment in steps of $\Delta\psi=20^\circ$. The phase angle increase in a counter-clockwise direction.
We find a slightly higher line ratio between spiral arms 3 and 4 ($R_{21}{\sim}0.7$). But generally, we do not find any significant arm or interarm variation.

\section{Different Data Sets}
\label{sec:aco_datasets}

The galaxy M101 is also amongst the galaxies studied by \citet{Sandstrom2013}. Compared to our CO-to-H$_2$ conversion factor estimate in the disk of the galaxy ($\langle\aco\rangle=4.4{\pm}0.9$), they find a lower value of $\aco = 2.3^{+2.6}_{-1.2}$. The value is lower even though they also use the scatter minimization technique. We note that we employed different datasets in this study. To ensure that the discrepancy is not related to our implementation of the scatter minimization technique, we compare the result using different permutations of the different datasets. In particular, we suspect that the discrepancy can stem from
\begin{enumerate}
    \item \emph{Feathered \hi\ data}: as discussed in \hyperref[sec:hi]{Appendix~\ref*{sec:hi}}, the THINGS data cubes are seriously affected by filtering and bowling issues. In this study, we have feathered the data cube to improve the \hi\ data. Using the VLA data without a correction could impact the resulting \aco. We find that the resulting \aco\ value is 0.11\,dex lower if substituting the feathered \hi\ data with the unfeathered ones.
    \item \emph{Different \chem{^{12}CO}{21} Datasets:} as discussed in \cite{denbrok2021}, the mm single-dish datacubes can suffer from flux calibration issues. For observations with HERA on the IRAM 30m telescope, the flux calibration can account up to 20\% difference. We hence compare the result when substituting our \chem{^{12}CO}{21} data to the observations from HERACLES \citep{Leroy2009}. We find that the difference in the derived \aco\ value only differs marginally with 0.05\,dex lower values.
    \item \emph{Fixed $R_{21}$}: in essence, \citet{Sandstrom2013} derive a \chem{^{12}CO}{21}-based \aco, while in this study, we investigate the \chem{^{12}CO}{10}-based \aco. We find that the \chem{^{12}CO}{21}-based \aco\ is 0.2\, dex lower than our combined CO transition approach. 
    %rely on the HERACLES \chem{^{12}CO}{21} observations and apply a constant $R_{21}$ to convert to a \chem{^{12}CO}{10} intensity. The discrepancy could hence also be related to the use of a constant and a variable $R_{21}$.
    
\end{enumerate}

\begin{figure*}
    \centering
    \includegraphics[width=0.8\textwidth]{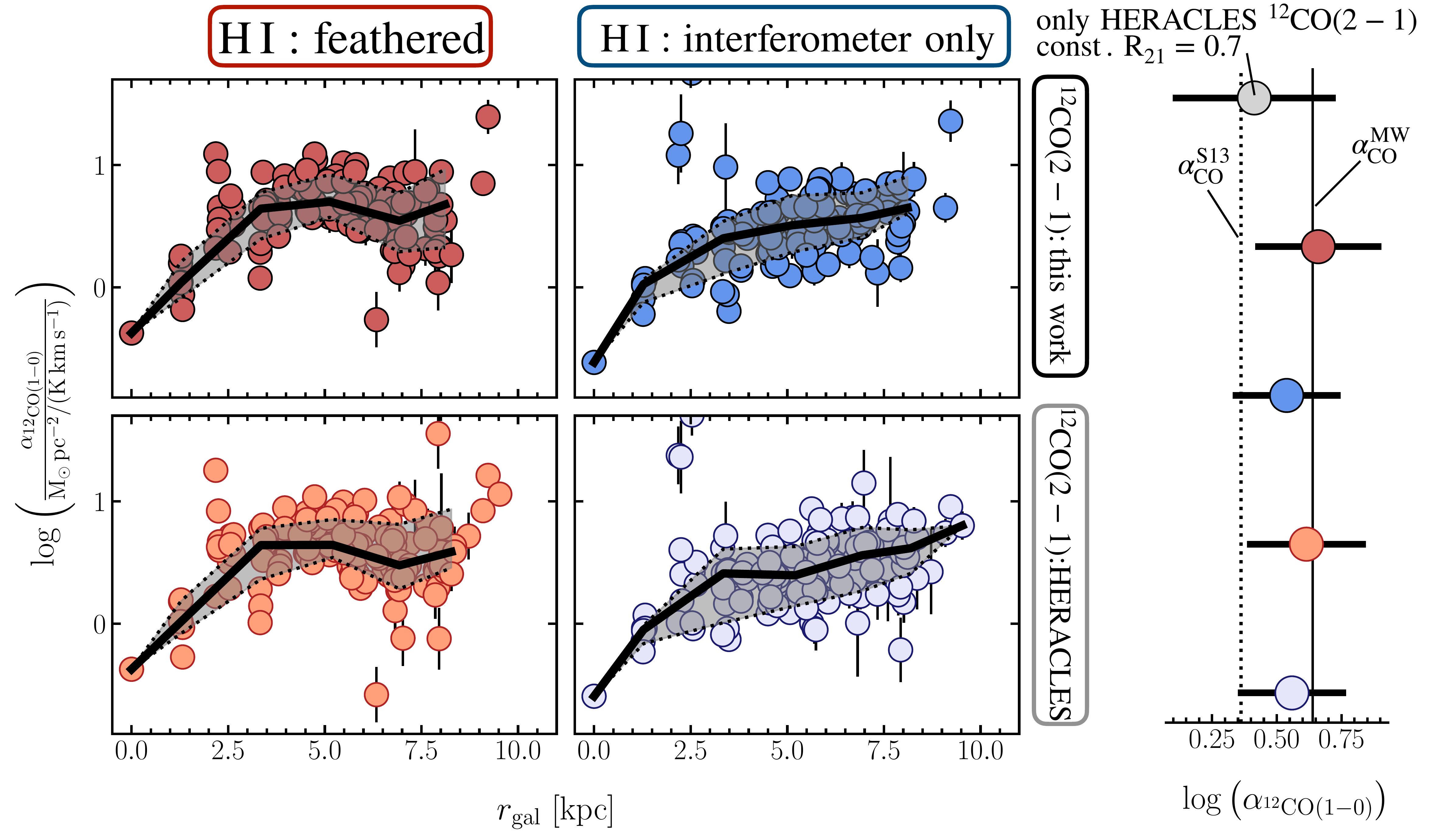}
    \caption{{\bf Comparing the Impact of Different Datasets on \aco\ Estimates.} We compare the results after substituting feathered \hi\ (left column) and non-feathered \hi\ (right column) observations as well as different \chem{^{12}CO}{21} observations (data from this project and HERACLES). The points show the resulting \aco\ value for the different solution pixels. The black line indicates the binned trend, and the shaded region illustrates the $1\sigma$ scatter per bin. The right panel shows the average values of the different permutations (color-coded). The grey point is based on using the \chem{^{12}CO}{21} data only (and deriving the \chem{^{12}CO}{10} data using a fixed $R_{21}$). This approach reproduced the method by \citet{Sandstrom2013}. The solid line indicates the average MW \aco\ value, and the dashed line shows the average value found by \citet{Sandstrom2013} for M101.}
    \label{fig:comp_method}
\end{figure*}

\hyperref[fig:comp_method]{Figure~\ref*{fig:comp_method}} illustrates the comparison for the radial \aco\ trends when using different permutations of dataset. The top row (orange and blue) use the \chem{^{12}CO}{21} observations from this project. The bottom rows (pink and green) use the HERACLES \chem{^{12}CO}{21} data. The columns differ by the use of \hi\ data (the left column shows the results based on the feathered and the right column the interferometric only \hi\ data). The right panel shows the \aco\ mean and scatter for the various data set permutations. The grey point shows when only using \chem{^{12}CO}{21} data from HERACLES and a constant $R_{21}$ value (i.e., reproducing result from \citealt{Sandstrom2013}).

Overall, we find that \aco\ values are ${\sim0.1}$ dex lower when using the non-feathered \hi\ data. Furthermore, only relying on the HERACLES \chem{^{12}CO}{21} data only indeed reproduced an even lower \aco\ value that is in agreement with the finding by \citet{Sandstrom2013}.

\section{Potential Degeneracy for \texorpdfstring{$\rm DGR$}{Lg} and \texorpdfstring{\aco}{Lg} with the Scatter Minimization Technique}
\label{sec:discrep}
\begin{figure}
    \centering
    \includegraphics[width=0.9\columnwidth]{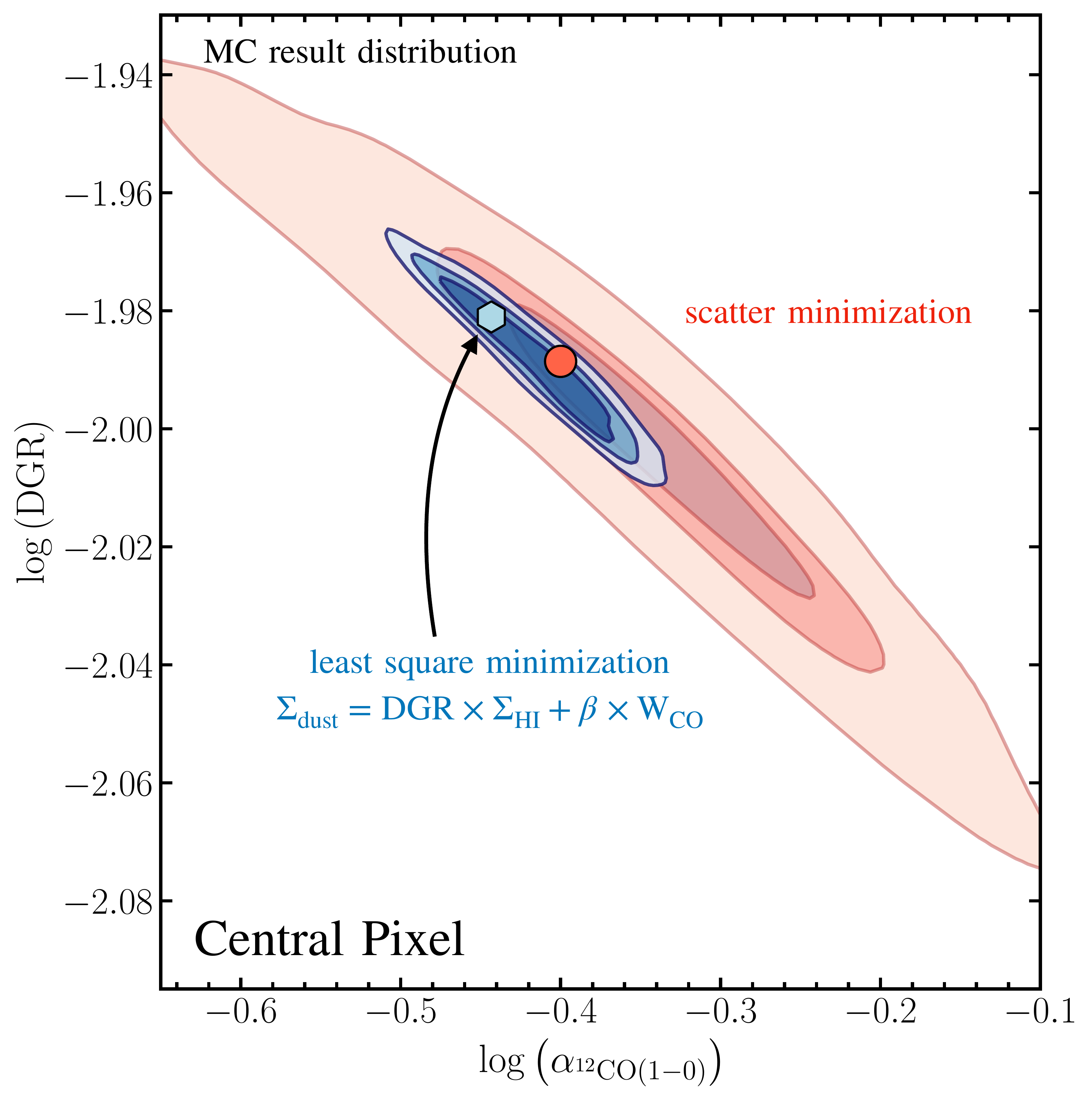}
    \caption{{\bf MC Simulation for Central Solution Pixel.} The red contours show the 10\%, 50\%, and 75\% inclusion of the solution distribution for the scatter minimization technique. Blue shows the result distribution based on the least-square minimization (\autoref{eq:ls_min}). The red and blue points show the solution before adding noise to the input parameters.  }
    \label{fig:mc_sim}
\end{figure}

The scatter minimization technique relies on the presence of a dynamical range of the \hi/CO ratio. However, for instance, in the center of the galaxy, where \hi\ emission becomes weak, the ratio might be dominated by the dynamical range of the CO emission.
In essence, the scatter minimization algorithm is equivalent to a least-square minimization of the following linear equation{, which we derive from \autoref{eq:dust-gas} after multiplying both sides of the equation with the $\rm DGR$ term}:

\begin{equation}
\label{eq:ls_min}
\Sigma_{\rm dust} = {\rm DGR}\times\Sigma_{\hi} + \gamma\times W_{\rm CO},
\end{equation}
where {\rm DGR} and $\gamma \equiv \aco\times {\rm DGR}$ are the two free parameters. If \hi\ is relatively small compared with CO, it is possible to  determine $\gamma$ with good accuracy, but not {\rm DGR}, leading to a degeneracy for the two parameters. We can assess the degree of this degeneracy by performing Monte Carlo iterations. We focus on the central solution pixel, where the CO emission is maximal, while the \hi\ emission is minimal (but still detected with $\rm S{/}N>20$ for the underlying sightlines). We iteratively perform the scatter minimization technique and solve the least-square minimization (\autoref{eq:ls_min}) after adding noise to the input parameters. \hyperref[fig:mc_sim]{Figure~\ref*{fig:mc_sim}} illustrates the solution distribution for the derived \aco\ and ${\rm DGR}$ values based on the two techniques. The red circle and blue hexagon show the solution without adding noise to the input parameters. Indeed, we find a certain degree of degeneracy for both methods, in the sense that lower \aco\ values correspond clearly to higher $\rm DGR$ values. However, for $\rm DGR$, the dynamical range in variation from the MC simulation is about 0.1 dex based on the scatter minimization technique, which is smaller than the scatter we find across M101 (${\sim}$0.2 dex; \autoref{fig:radial_sub}). For \aco, we find a larger dynamical range of ${\sim}0.5$\,dex, which is comparable to the scatter we find across the disk of M101. Nevertheless, we can conclude  that the significant depression of \aco\ by a factor 10 toward the center of M101 is not due to the degeneracy or systematic uncertainty of the scatter minimization technique itself.

Alternatively, we can assess the robustness of the scatter minimization technique by fixing $\rm DGR$ using the {empirical} prescription by \cite{Chiang2018} determined for M101 (for $(12+\log(\rm O/H)){>}8.2$):
\begin{equation}
    \log \rm DGR = (1.9\pm.1)\times (12+\log(\rm O/H))  + (-18.1\pm0.7)
\end{equation}
The fit is derived using a broken emissivity model to determine the dust mass. Using this prescription, we find a dynamical range of ${\sim}1$\,dex in the ${\rm DGR}$, which translates into a dynamical range of 1\,dex for \aco\ between center and disk. So we recover the central depression of \aco\ also when using a ${\rm DGR}$ derived from a prescription instead of treating it as a free parameter in the scatter minimization technique.

\section{\texorpdfstring{$\rm DGR$}{Lg} and \texorpdfstring{\aco}{Lg} in M51}
\label{M51_acostuff}

\begin{figure}
    \centering
    \includegraphics[width=0.7\columnwidth]{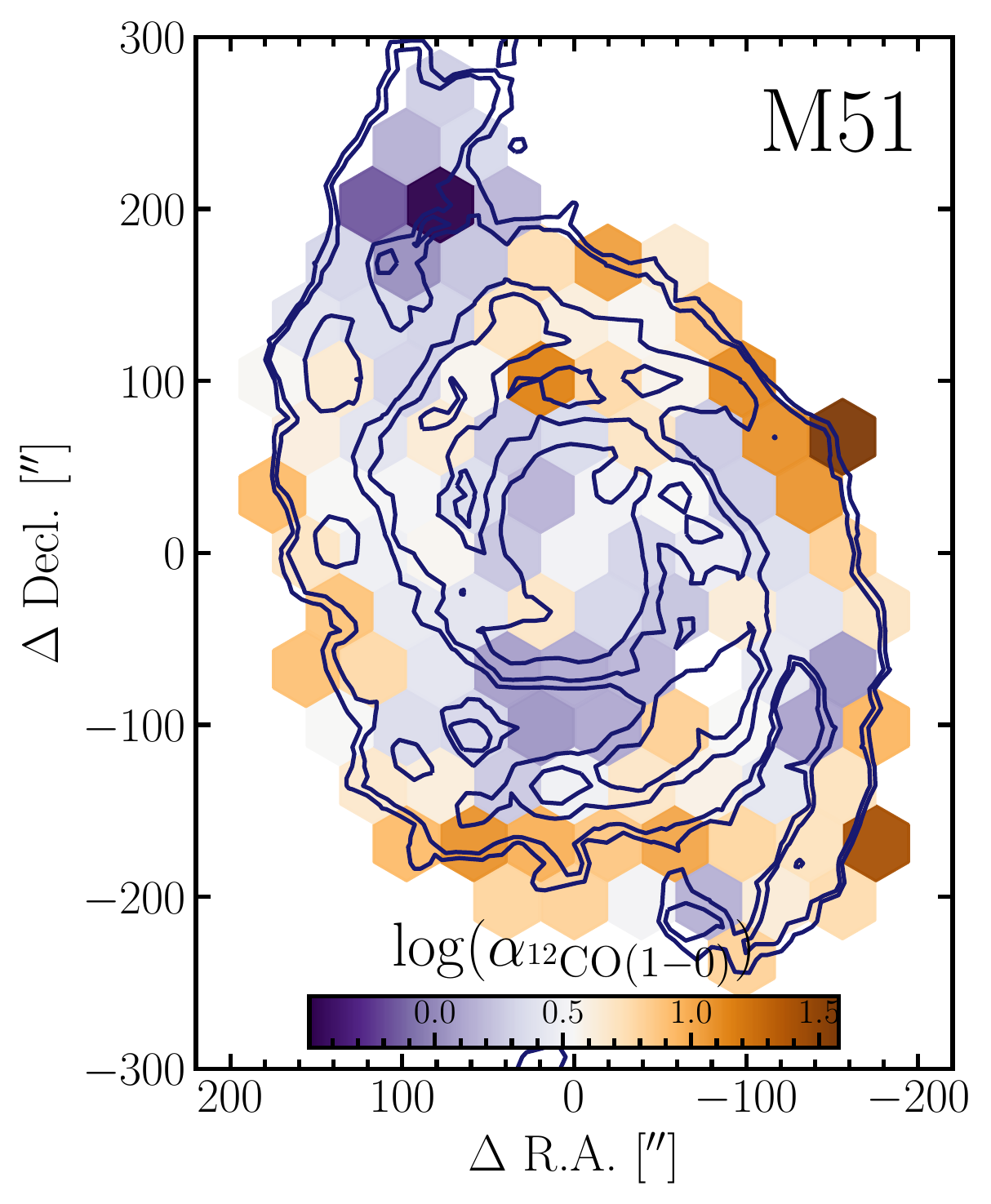}
    \caption{{\bf Distribution of \aco\ across M51} Solution pixels of M51 color-coded by their respective \aco\ value. Contours are drawn at ${\rm S/N}=7,10,30,50,100$ and help visualize the extent and structure of the galaxy. }
    \label{fig:aCO_map_m51}
\end{figure}

In this project, we compare \aco\ estimates across M101 to values and trends we find across M51. \autoref{fig:aCO_map_m51} shows the \aco\ distribution across M51. The solution pixels are color-coded according to their value of \aco. For reference, the \chem{^{12}CO}{10} S/N contours show the extent and morphology of the galaxy.

\end{appendix}

\end{document}